\documentclass[11pt]{article}
\pdfoutput=1 
\usepackage{mathtools}
\usepackage{booktabs}

\usepackage{amsmath,amssymb,amsbsy,amstext, amsthm, simplewick, amsfonts,braket}
\usepackage{graphicx}
\usepackage[small]{caption}
\usepackage{siunitx}
\usepackage{subcaption}
\usepackage{upgreek}
\usepackage{framed}
\usepackage{wrapfig}
\usepackage{multirow}
\usepackage{bbm}
\usepackage[numbers,sort&compress]{natbib}
\usepackage[svgnames,dvipsnames,x11names]{xcolor}
\usepackage[utf8x]{inputenc}
\usepackage{selinput} 
\usepackage{bm}
\usepackage{float}
\usepackage{dsfont}
\usepackage[margin=1cm]{caption}
\usepackage{subcaption}
\usepackage{sidecap}
\usepackage{longtable}
\usepackage{anyfontsize}

\usepackage{epstopdf}
\usepackage{cancel}
\usepackage{tcolorbox}
\usepackage{latexsym,amsmath,amssymb,epsfig}
\usepackage{braket}
\usepackage{tensor}
\usepackage{tocloft}
\usepackage[permil]{overpic}

\usepackage{tcolorbox}
\definecolor{greyish2}{rgb}{.96,.96,.96}

\def\xyma{\xymatrix@M.7em}
\def\xymas{\xymatrix@M.1em}

\newcommand{\Comment}[1]{{}}
\definecolor{darkblue}{rgb}{0.15,0.35,0.55}
\definecolor{reddish}{rgb}{0.65, 0.2, 0.2}
\definecolor{darkgreen}{RGB}{50,150,0}
\definecolor{greyish2}{rgb}{.96,.96,.96}
\usepackage[linktocpage=true]{hyperref}
\hypersetup{
colorlinks=true,
citecolor=darkblue,
linkcolor=darkblue,
urlcolor=darkblue,
pdfauthor={},
pdftitle={},
pdfsubject={}
}

\DeclareFontFamily{OT1}{rsfs10}{}
\DeclareFontShape{OT1}{rsfs10}{m}{n}{ <-> rsfs10 }{}
\DeclareMathAlphabet{\mathscript}{OT1}{rsfs10}{m}{n}
\DeclareMathAlphabet{\mathbbold}{U}{bbold}{m}{n}

\def\gsim{ \lower .75ex \hbox{$\sim$} \llap{\raise .27ex \hbox{$>$}} }
\def\lsim{ \lower .75ex \hbox{$\sim$} \llap{\raise .27ex \hbox{$<$}} }
\def\be{\begin{equation}}
\def\ee{\end{equation}}
\def\bea{\begin{eqnarray}}
\def\eea{\end{eqnarray}}

\newcommand{\baaa}{\begin{eqnarray}}
\newcommand{\eaaa}{\end{eqnarray}}

\usepackage{tikz}
\usetikzlibrary{decorations}
\usetikzlibrary{calc}
\pgfdeclaredecoration{complete sines}{initial}
{
    \state{initial}[
        width=+0pt,
        next state=upsine,
        persistent precomputation={\pgfmathsetmacro\matchinglength{
            \pgfdecoratedinputsegmentlength / int(\pgfdecoratedinputsegmentlength/\pgfdecorationsegmentlength)}
            \setlength{\pgfdecorationsegmentlength}{\matchinglength pt}
        }] {}
    \state{upsine}[width=\pgfdecorationsegmentlength,next state=downsine]{
        \pgfpathsine{\pgfpoint{0.25\pgfdecorationsegmentlength}{0.5\pgfdecorationsegmentamplitude}}
        \pgfpathcosine{\pgfpoint{0.25\pgfdecorationsegmentlength}{-0.5\pgfdecorationsegmentamplitude}}
    }
    \state{downsine}[width=\pgfdecorationsegmentlength,next state=upsine]{
        \pgfpathsine{\pgfpoint{0.25\pgfdecorationsegmentlength}{-0.5\pgfdecorationsegmentamplitude}}
        \pgfpathcosine{\pgfpoint{0.25\pgfdecorationsegmentlength}{0.5\pgfdecorationsegmentamplitude}}
}
    \state{final}{}
}

\definecolor{greyish}{rgb}{.90,.90,.90}
\definecolor{greyish2}{rgb}{.96,.96,.96}
\usepackage{xcolor,colortbl}
\usepackage{tcolorbox}

\usepackage[all]{xy}

\newcommand{\p}{\partial}
\renewcommand{\O}{\mathcal{O}}

\DeclareSymbolFont{matha}{OML}{txmi}{m}{it}

\newcommand{\del}{\nabla}
\renewcommand{\O}{{\mathcal O}}

\newcommand{\bz}{\bar{z}}
\newcommand{\bh}{\bar{h}}
\newcommand{\bP}{\bar{P}}

\newcommand{\half}{\tfrac{1}{2}}

\newcommand{\cp}[1]{\vcenter{\hbox{#1}}}
\newcommand{\id}{\mathbbold{1}}
\newcommand{\modularS}{\mathbb{S}}
\newcommand{\hatS}{\widehat{\mathbb{S}}}
\newcommand{\dilog}{{\rm Li}_2}
\newcommand{\fusionF}[2]{
\mathbb{F}_{#1}
\begin{bsmallmatrix} #2 \end{bsmallmatrix}
}
\newcommand{\hatDelta}{\hat{\Delta}}

\def\centerarc[#1](#2)(#3:#4:#5)
{ \draw[#1] ($(#2)+({#5*cos(#3)},{#5*sin(#3)})$) arc (#3:#4:#5); }

\newcommand{\inlinetetra}{
\vcenter{\hbox{
\begin{tikzpicture}[scale=0.25]
\centerarc[](0,0)(-30:90:1);
\centerarc[](0,0)(90:210:1);
\centerarc[](0,0)(210:335:1);
\draw (0,0) -- (0,1);
\draw (0,0) -- ({cos(30)},{-sin(30)});
\draw (0,0) -- ({-cos(30)},{-sin(30)});
\end{tikzpicture}
}}
}

\usepackage{adjustbox}
\newcommand{\graphS}[5]{\cp{
\adjustbox{margin=-1.5ex 0 -1ex 0}{
\begin{tikzpicture}[scale=0.25]
\coordinate (upperLeft) at (0,1);
\coordinate (lowerLeft) at (0,-1);
\coordinate (vertexLeft) at (0.8,0);
\coordinate (upperRight) at (3.5,1);
\coordinate (lowerRight) at (3.5,-1);
\coordinate (vertexRight) at (2.7,0);
\draw (upperLeft) -- (vertexLeft);
\draw (lowerLeft) -- (vertexLeft);
\draw (vertexLeft) -- (vertexRight);
\draw (upperRight) -- (vertexRight);
\draw (lowerRight) -- (vertexRight);
\node[xshift=-3] at (upperLeft) {\footnotesize \ensuremath{#1}};
\node[xshift=-3] at (lowerLeft) {\footnotesize \ensuremath{#2}};
\node[xshift=3] at (upperRight) {\footnotesize \ensuremath{#3}};
\node[xshift=3] at (lowerRight) {\footnotesize \ensuremath{#4}};
\node at (1.8,0.7) {\footnotesize \ensuremath{#5}};
\end{tikzpicture}
}}}
\newcommand{\graphT}[5]{\cp{
\adjustbox{margin=-1.5ex 0 -1ex 0}{
\begin{tikzpicture}[scale=0.25]
\coordinate (upperLeft) at (-1,3.5);
\coordinate (lowerLeft) at (-1,0);
\coordinate (upperRight) at (1,3.5);
\coordinate (lowerRight) at (1,0);
\coordinate (vertexTop) at (0,2.7);
\coordinate (vertexBottom) at (0, .8);
\draw (upperLeft) -- (vertexTop);
\draw (upperRight) -- (vertexTop);
\draw (lowerLeft) -- (vertexBottom);
\draw (lowerRight) -- (vertexBottom);
\draw (vertexTop) -- (vertexBottom);
\node[xshift=-3] at (upperLeft) {\footnotesize \ensuremath #1};
\node[xshift=-3] at (lowerLeft) {\footnotesize \ensuremath #2};
\node[xshift=3] at (upperRight) {\footnotesize \ensuremath #3};
\node[xshift=3] at (lowerRight) {\footnotesize \ensuremath #4};
\node at (0.7,1.8) {\footnotesize \ensuremath #5};
\end{tikzpicture}
}}}

\newcommand{\torusBlockM}[3]{
\cp{
\adjustbox{margin=-1ex -1ex -1ex -1ex}{
\begin{tikzpicture}[scale=0.25]
\coordinate (circleCenter) at (0,2);
\coordinate (vertex) at (0,1);
\coordinate (bottom) at (0,0);
\draw (bottom)-- (vertex);
\draw (circleCenter) circle (1);
\node at (circleCenter) {\footnotesize #3};
\node[xshift=-6,yshift=2] at (bottom) {\footnotesize \ensuremath{#2}};
\node at (1.4,3.1) {\footnotesize \ensuremath{#1}};
\end{tikzpicture}
}}}
\newcommand{\torusBlockS}[2]{\torusBlockM{#1}{#2}{S}}
\newcommand{\torusBlock}[2]{\torusBlockM{#1}{#2}{}}


\numberwithin{equation}{section}
\newcommand{\vecP}{\mathsf{P}}

\newcommand{\Zvir}{Z_{\rm vir}}
\newcommand{\Ztv}{Z_{\rm CTV}}

\begin{document}
%
\renewcommand{\thefootnote}{\fnsymbol{footnote}}
\vspace{0truecm}
\thispagestyle{empty}

\begin{center}
{
\bf\LARGE
Triangulating quantum gravity in AdS$_3$


\vspace{0.5cm}
}
\end{center}


\begin{center}
{\fontsize{12}{18}\selectfont
Thomas Hartman
}

%

\vspace{.8truecm}

\centerline{{\it  Department of Physics, Cornell University, Ithaca, New York}}

\vspace{0.2cm}

 \centerline{\tt hartman@cornell.edu }

 \vspace{.25cm}

\vspace{.3cm}

\end{center}

\vspace{0.7cm}

\begin{abstract}
\noindent
The path integral of pure 3D gravity with negative cosmological constant is formulated on a finite region of spacetime $M$, with boundary conditions that fix geodesic lengths or dihedral angles on $\p M$. In the dual CFT, this quasi-local amplitude calculates corrections to the Gaussian ensemble of OPE coefficients for black hole states.  By triangulating $M$ with generalized tetrahedra, we develop a general method to construct semiclassical geometries and to calculate the exact gravitational path integral on a fixed hyperbolic topology. The path integral with fixed-length boundary conditions is a Virasoro TQFT amplitude-squared, and with fixed-angle boundary conditions it is a partition function of Conformal Turaev-Viro theory. The two are related by a modular $S$-transform. In addition, we show how to translate the calculation of OPE statistics from Virasoro TQFT to the metric formalism, on general topologies. These results are derived exactly, and some examples are also checked semiclassically, including the geometries dual to the Virasoro $6j$-symbol and the modular $S$-matrix. The classical saddlepoint geometries are finite-volume hyperbolic 3-manifolds ending on pleated Riemann surfaces, which have vanishing extrinsic curvature except on geodesics where they can bend into corners.  The hyperbolic volumes of these geometries match the predictions of Conformal Turaev-Viro theory and the dual CFT.

\end{abstract}

\newpage

\setcounter{page}{2}
\setcounter{tocdepth}{2}
\tableofcontents
\renewcommand*{\thefootnote}{\arabic{footnote}}
\setcounter{footnote}{0}
 
\ \bigskip

\section{Introduction}

The discovery of controlled wormhole effects in quantum gravity \cite{Saad:2018bqo,Saad:2019lba,Penington:2019kki,Almheiri:2019qdq} requires rethinking the gravitational path integral. Pure 2D gravity (i.e., dilaton gravity) is holographically dual to random matrix theory \cite{Saad:2019lba}. This suggests that pure 3D gravity is dual to an ensemble of 2d CFTs.  Indeed, off-shell wormholes give a prediction for the spectral statistics in this ensemble \cite{Cotler:2020ugk} and on-shell wormholes match the statistics of OPE coefficients calculated from the conformal bootstrap \cite{Chandra:2022bqq}. 
The results of \cite{Chandra:2022bqq} build on general arguments that OPE statistics should be related to wormholes \cite{Penington:2019kki,Pollack:2020gfa,Belin:2020hea,Stanford:2020wkf} combined with bootstrap methods for large-$c$ CFTs \cite{Hartman:2013mia,Collier:2019weq}. See e.g.~\cite{Afkhami-Jeddi:2020ezh,Maloney:2020nni,Belin:2023efa,DiUbaldo:2023qli,deBoer:2023vsm,Jafferis:2025vyp,Wang:2025bcx,deBoer:2025rct} for related developments from various perspectives.

In this paper, we study higher topologies in 3D gravity from a new point of view. Classically, hyperbolic geometries with complicated topology can be triangulated by tetrahedra, a fact that has been exploited to great advantage by mathematicians. We will develop a similar approach to the exact path integral of 3D quantum gravity on higher topologies based on gluing together (generalized) tetrahedra, and show that this allows for a direct translation between bulk geometries and CFT statistics. In a companion paper \cite{ctv}, we introduce Conformal Turaev-Viro (CTV) theory, a topological theory that can be used to evaluate the exact gravitational path integral by triangulations. The emphasis in \cite{ctv} is on the TQFT, while in the current paper we apply those results to gravity and compare to the semiclassical description in the metric formulation. The two papers can be read independently, but logically \cite{ctv} comes first.

While the exact TQFT results are specific to low-dimensional gravity, some of the semiclassical methods developed in the current paper can also be applied to gravity in $D>3$ dimensions. In $D=3$, we will study boundary conditions on a finite region of spacetime that fix the lengths of geodesics on the boundary, or their conjugate angles. The analogous boundary conditions in higher dimensions fix the areas or bending angles at codimension-2 extremal surfaces. Recently, related geometries have appeared in the literature on fixed-area or microcanonical states in the gravitational path integral. Examples of these geometries, with corners on extremal surfaces, have been discussed explicitly in 2D gravity \cite{Harlow:2018tqv} and in higher dimensions \cite{Dong:2022ilf,Chua:2023srl,Chua:2023ios,Chandra:2024vhm} and also arise implicitly in e.g.~\cite{Akers:2018fow,Dong:2018seb,Harlow:2019yfa,Chandrasekaran:2022eqq}. 
In these references, the spacetime manifolds are usually assumed to be non-compact, with asymptotic boundary components, whereas in the current paper the boundaries are at finite distance and the semiclassical spacetimes have finite volume.

\subsubsection*{Gravity on a finite region of spacetime}

Our starting point is to formulate pure 3D gravity with negative cosmological constant on compact hyperbolic 3-manifolds with boundaries. These are manifolds without asymptotic conformal boundaries --- only finite boundaries, with prescribed boundary conditions. The on-shell solutions with these boundary conditions have finite volume. 

The path integral on a complete, non-compact hyperbolic 3-manifold, where all the boundaries are asymptotic, is essentially a solved problem. It was solved gradually over several decades, starting with  \cite{Achucarro:1986uwr,Witten:1988hc,Verlinde:1989ua}, with further milestones in e.g.~\cite{, Krasnov:2000ia, Takhtajan:2002cc, Scarinci:2011np,EllegaardAndersen:2011vps,andersen2013new}, and more recently in \cite{Chandra:2022bqq,Collier:2023fwi}. The most complete --- and most convenient --- formulation is in terms of a topological theory known as Virasoro TQFT (VTQFT)  \cite{Collier:2023fwi,Collier:2024mgv}.\footnote{Virasoro TQFT is believed to be the same quantum theory as the Teichmuller TQFT developed in earlier work and formalized precisely by Andersen and Kashaev \cite{EllegaardAndersen:2011vps}, but this is not manifest.}
%
This is not an ordinary TQFT, since not all amplitudes are well defined, but it can be defined rigorously by a set of diagrammatic rules, and used to calculate exact gravity amplitudes on fixed hyperbolic topologies. The results of these calculations match ensemble-averaged observables in the holographic dual, which in many cases can be calculated independently using the conformal bootstrap \cite{Chandra:2022bqq,Collier:2023fwi,Yan:2023rjh,Collier:2024mgv,deBoer:2024mqg,Post:2024itb}. Virasoro TQFT also applies to finite boundaries, but we will need to carefully translate boundary conditions on the metric into the language of the TQFT.

The path integral with finite boundaries is actually much simpler than with asymptotic boundaries. At asymptotic infinity, there is an infinite amount of boundary data, there are Virasoro boundary gravitons to account for, and observables depend on kinematics in a complicated way such that they can be expressed as sums of Virasoro conformal blocks. With finite boundaries, the boundary condition on each genus-$g$ component $\Sigma$ of the boundary is simply a point in the Teichmuller space ${\cal T}(\Sigma)$ \cite{Moncrief:1989dx,Verlinde:1989ua,Krasnov:2005dm,Scarinci:2011np,Kim:2015qoa}, which is $(6g-6)$-dimensional. In terms of the CFT ensemble, the theory of a finite region is a theory of OPE statistics for conformal primaries. There are no kinematics. The asymptotic regions that account for descendants and conformal blocks are stripped off, similar to stripping off trumpets and the corresponding Schwarzian modes in JT gravity \cite{Saad:2019lba}, and they can be glued back on at the end to reproduce averaged CFT observables. From this point of view, the asymptotic AdS boundary is a distraction. 

Consider AdS$_3$ gravity with the central charge parameterized as $c = 1+6(b+b^{-1})^2$, where $b \to 0$ is the semiclassical limit. The gravitational path integral is performed on a fixed topology 3-manifold $M$ with boundary conditions $\gamma(\vecP)$. The boundary conditions are specified as a set $\gamma = (\gamma_1,\dots,\gamma_n)$ of disjoint simple closed curves on $\p M$ labeled by $\vecP \in \mathbb{C}^n$. 
Each boundary component of genus $g$ has $n \leq 3g-3$ such curves. 
We will consider two different boundary conditions for the gravitational path integral, and the corresponding amplitudes
\begin{align}
\mbox{fixed length:} \qquad &Z_L(M, \gamma(\vecP)) \\
\mbox{and fixed angle:} \qquad &Z_A(M, \gamma(\vecP))  \ . 
\end{align}
The fixed-length boundary condition specifies the geodesic lengths $\ell_i$ of the curves $\gamma_i$, with the lengths parameterized as
\begin{align}
\ell_i = 4\pi b P_i \ . 
\end{align}
The interpretation in terms of lengths requires $P_i$ to be real and non-negative, but the fixed-length amplitude can be analytically continued to complex $P_i$. 

The fixed-angle boundary condition fixes the dihedral angles $\psi_i$ on the curves $\gamma_i$, with the angles parameterized as
\begin{align}
\psi_i = 2\pi i b P_i \ . 
\end{align}
In this case, the interpretation in terms of angles requires $\psi_i \in (0,\pi)$, and therefore $P_i$ is purely imaginary.

In fact, each boundary condition has two different interpretations. When the fixed-angle amplitude, defined initially for imaginary $P_i$, is analytically continued to real $P_i$, it fixes the lengths, not of $\gamma$ but of a set of dual cycles $\gamma'$ on a dual manifold $M'$ which generally has different topology from $M$. Therefore the fixed-angle boundary condition is also a \textit{dual length} boundary condition. Similarly, the fixed-length boundary condition at imaginary $P_i$ fixes dual angles. Thus the terminology `fixed-length' and `fixed-angle' refers to the cycles $\gamma$, while the actual geometric interpretation depends on whether $P_i$ is real or imaginary.

By construction, the fixed-length and fixed-angle path integrals are related by a Laplace transform,
\begin{align}\label{introFourierAL}
Z_A(M, \gamma(\vecP')) = \sqrt{2} \int_{\mathbb{R}_+^n} d\vecP (\Pi_{i=1}^n e^{4\pi i P_i' P_i} )Z_L(M, \gamma(\vecP))
\end{align}
The parameterization of lengths and angles by $P_i$ is chosen to match the Liouville parameterization of conformal weights, $h_i = \frac{c-1}{24}+P_i^2$. 

\subsubsection*{Fixed-length path integrals}

A Virasoro TQFT diagram $(M_E, \Gamma(\vecP))$ is a framed trivalent graph $\Gamma$ embedded in a closed 3-manifold $M_E$.  The edges of the graph are labeled by conformal weights $\vecP = (P_1,\dots, P_n) \in \mathbb{C}^n$. We refer to $M_E$ as the \textit{embedding manifold} to distinguish it from the manifold $M$ where gravity lives. 
Amplitudes in Virasoro TQFT are denoted by
\begin{align}
\Zvir(M_E, \Gamma(\vecP)) \ .
\end{align}
Our normalization of the VTQFT vertex differs from \cite{Collier:2023fwi,Collier:2024mgv}, as described below --- we strip off all factors of the OPE variance $C_0$ associated to asymptotic AdS boundaries.

The fixed-length gravitational path integral is directly related to Virasoro TQFT. Given a VTQFT graph $(M_E, \Gamma(\vecP))$, define $M = M_E - N(\Gamma)$ by removing a regular neighborhood of the graph $N(\Gamma)$.\footnote{A regular neighborhood $N(\Gamma)$ is an open neighborhood of $\Gamma$ that deformation retracts to $\Gamma$.} $M$ is the compact 3-manifold with boundary where gravity lives, and we impose boundary conditions $\gamma(\vecP)$ on the meridians of the graph.\footnote{Throughout the paper we are working in pure gravity, so all external weights $\vecP$ in the fixed-length path integral or the VTQFT diagram are above the black hole threshold. For conical defect particles, assuming they are sufficiently heavy \cite{Chandra:2022bqq}, gravity lives on $M = M_E - \Gamma(V)$ where $V$ is the set of vertices. } We will argue by consistency with known results in 3D gravity that the fixed-length gravitational path integral is
\begin{align}\label{summaryLength}
Z_L(M, \gamma(\vecP))  = | \Zvir(M_E, \Gamma(\vecP))|^2
\end{align}
This result is already implicit in \cite{Collier:2023fwi}, so it is not particularly novel. In special cases, versions of this relation also appear in the analysis of Hartle-Hawking wavefunctions \cite{Harlow:2018tqv,Chua:2023srl,Chua:2023ios} and wavefunctions for thin shells \cite{Chandra:2024vhm}.  The new contribution in this paper is to give an expanded discussion of the gravity amplitude on left-hand side, and to work out the semiclassical  geometries in some nontrivial examples, where the core has non-zero volume.

\subsubsection*{Fixed-angle path integrals}

More interesting for our purposes is the gravitational path integral with fixed-angle (or dual length) boundary conditions. This path integral is directly related to triangulations. In the companion paper \cite{ctv} we define and develop Conformal Turaev-Viro (CTV) theory, a dual formulation of Virasoro TQFT in which amplitudes are defined by triangulating $(M_E, \Gamma(\vecP))$ into tetrahedra. The CTV partition function is related to Virasoro TQFT by a modular $S$-transform \cite{ctv},
\begin{align}\label{introFourierVV}
\Ztv(M_E, \Gamma(\vecP')) = \int_{\mathbb{R}_+^n}d\vecP (\Pi_{i=1}^n S_{P_i' P_i})
\left|
\Zvir(M_E, \Gamma(\vecP))
\right|^2
\end{align}
where $S_{P'P} = 2\sqrt{2}\cos(4\pi P'P)$.  A similar result was known previously for discrete spin networks \cite{roberts1995skein,Barrett:2004im}, and in \cite{ctv} we show that there is no essential difficulty in generalizing those results to the Virasoro theory, after making a few changes to the definitions to avoid divergences.

Given the relation between $|\Zvir|^2$ and $Z_L$ in \eqref{summaryLength}, the $S$-transform \eqref{introFourierVV} is precisely the change from fixed-length to fixed-angle boundary conditions. Therefore, the exact fixed-angle path integral is the CTV partition function,
\begin{align}\label{introAngleResult}
Z_A(M, \gamma(\vecP))  = \Ztv(M_E, \Gamma(\vecP))  \ . 
\end{align}
This provides an exact formula for the gravitational path integral in terms of a triangulation of $M_E$. This is not quite the end of the story, because the embedding manifold $M_E$, which is closed, is not the same as $M$, the compact manifold with boundary where gravity lives. Rather, $M$ is the complement of the thickened VTQFT graph $\Gamma \subset M_E$. This means that triangulations of $M_E$ that arise in CTV theory are not  triangulations of $M$. But they can be used to build \textit{generalized} triangulations of $M$, which are cell decompositions into generalized tetrahedra. Generalized tetrahedra are obtained from ordinary tetrahedra by truncating the vertices and dualizing a subset of the edges. 

\begin{figure}[t]
\begin{center}
\begin{overpic}[width=2.5in]{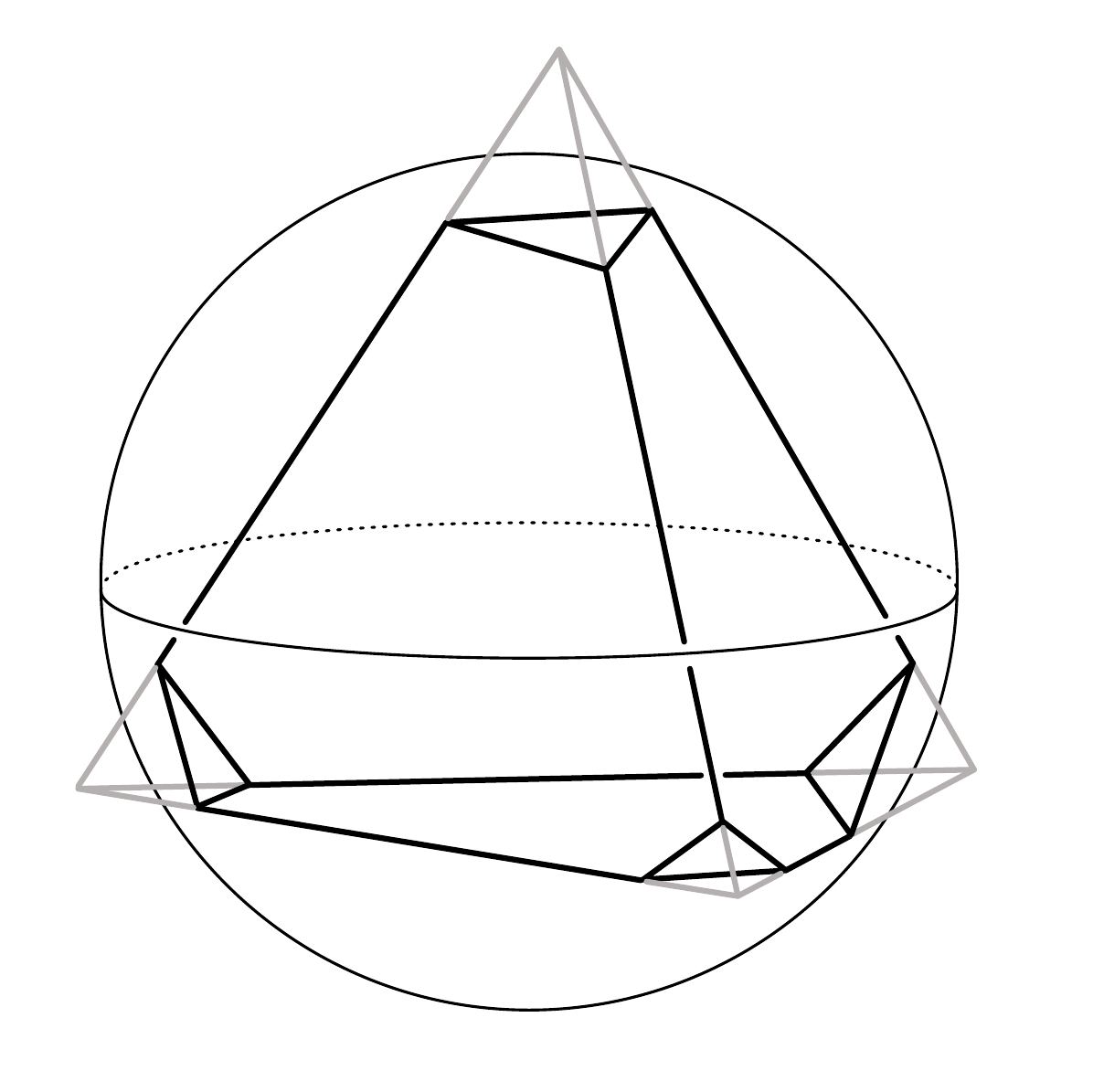}
\end{overpic}
\end{center}
\caption{A generalized hyperbolic tetrahedron, with four hyperideal vertices beyond the conformal boundary of $H_3$. The faces are geodesic hyperplanes, and truncated faces meet their neighbors orthogonally.\label{fig:trunctet} }
\end{figure}

Thus the gravitational path integral is decomposed into topological generalized tetrahedra. The classical saddles are  generalized \textit{hyperbolic} tetrahedra (Figure \ref{fig:trunctet}). Generalized hyperbolic tetrahedra have an elegant description in the embedding space formalism, where $H_3$ (Euclidean AdS$_3$) is realized as a submanifold in $\mathbb{R}^{3,1}$. Truncating the edges corresponds to moving vertices `beyond' the conformal boundary of AdS, into the de Sitter region of embedding space.\footnote{The idea that heavy operators meet beyond the conformal boundary was suggested to me by J. Maldacena.}  Thus we will construct the classical geometries by gluing together polyedra with vertices in the space $H_3 \cup dS_3$, where $H_3$ and $dS_3$ share a conformal boundary. The surprise appearance of de Sitter spacetime here could be just a kinematic curiosity, or something deeper. 

The classical geometries built by gluing together generalized hyperbolic tetrahedra are compact 3-manifolds ending on \textit{pleated surfaces}, which have vanishing extrinsic curvature away from a collection of geodesics where there are corners. Pleated surfaces were introduced by Thurston in his proof of the hyperbolization theorem \cite{thurstonNotes,MR855294}, and the bending locus is a measured geodesic lamination, which figures prominently in the mathematics literature on 3-manifolds. A more elaborate application of the techniques in this paper, which gives a holographic, CFT interpretation of Thurston's hyperbolic knot complements, is described in \cite{knotspaper}. Another approach to 3D gravity on compact manifolds, which also utilizes pleated boundaries, is explored in \cite{Krasnov:2005dm,Krasnov:2006jb,Bonsante:2006tr,Krasnov:2008yq,Krasnov:2009vy,MR3732685,MR4479756}.

\subsubsection*{From Virasoro TQFT to geometry}

Virasoro TQFT \cite{Collier:2023fwi} is the most elegant and efficient way to calculate OPE statistics in the ensemble of 2d CFTs dual to 3D gravity \cite{Chandra:2022bqq}. On the other hand, it obscures the role of the three-dimensional geometry. In some cases, such as some of the calculations in \cite{Post:2024itb}, it is not even clear what topology in 3D gravity corresponds to a given VTQFT calculation, or whether there is a classical saddlepoint.\footnote{There is a calculation in \cite{Post:2024itb} described topologically in VTQFT in terms of a geometry with three torus boundaries. An obstacle in translating this into gravity is that the on-shell geometry may have a singularity that pinches off one of the torus boundaries, leaving two tori and a conical singularity in the geometric description. Thus it is not clear from the VTQFT calculation whether the toroidal boundaries are asymptotic or singular. There are other, similar examples in which the toroidal components are known to develop singularities instead of extending smoothly to conformal boundaries \cite{knotspaper}. For other differences between geometry and VTQFT on off-shell topologies, see \cite{Yan:2025usw}.}    The results summarized above provide the bridge between Virasoro TQFT and geometry in the metric formalism, thereby closing this gap, at least on topologies that admit a hyperbolic metric. The bridge is the inverse of \eqref{introFourierVV}, 
\begin{align}\label{introInverse}
Z_L(M,\gamma(\vecP')) = |\Zvir(M_E, \Gamma(\vecP'))|^2 = \int_{\mathbb{R}_+^n}d\vecP 
 (\Pi_{i=1}^n S_{P_i' P_i}) \Ztv(M_E,\Gamma(\vecP)) \ . 
\end{align}
This expresses the OPE statistics in terms of a triangulated 3-manifold, decomposed into generalized tetrahedra. The CTV partition function on the right-hand side, which translates directly into the metric formalism, assigns a Virasoro $6j$-symbol $\begin{Bsmallmatrix}P_4&P_5&P_6\\P_1&P_2&P_3\end{Bsmallmatrix}$ to each generalized tetrahedron; the $P_i$ correspond to dual lengths (i.e., imaginary angles) of this tetrahedron. The semiclassical limit of \eqref{introInverse} gives the OPE statistics in terms of the volumes of generalized hyperbolic tetrahedra, glued together and integrated over dual lengths. This manifestly agrees with the on-shell action of pure gravity, and gives a practical method to construct the geometries themselves.

\subsubsection*{Semiclassical examples}

We will apply this formalism to two examples, corresponding to the Virasoro TQFT graphs
\begin{align}\label{introVTQFT}
(a)\quad
\vcenter{\hbox{
\begin{tikzpicture}[scale=0.8]
\centerarc[thick](-0.5,0)(65:410:1);
\centerarc[thick](0.5,0)(-115:233:1);
\draw[thick] (-0.5,0) -- (0.5,0);
\end{tikzpicture}
}}
\qquad \ ,  \qquad  \qquad \qquad
(b)\quad
\vcenter{\hbox{
\begin{tikzpicture}[scale=0.8]
\centerarc[thick](0,0)(-30:90:1);
\centerarc[thick](0,0)(90:210:1);
\centerarc[thick](0,0)(210:335:1);
\draw[thick] (0,0) -- (0,1);
\draw[thick] (0,0) -- ({cos(30)},{-sin(30)});
\draw[thick] (0,0) -- ({-cos(30)},{-sin(30)});
\end{tikzpicture}
}}
\end{align}
which contribute to the OPE statistics $\overline{c_{ijk}c_{lmn}}$ and $\overline{c_{ijk}c_{lmn}c_{pqr}c_{stu}}$, respectively. These amplitudes can be calculated with the TQFT, but when the operators are above the black hole threshold, there has been no detailed description of these geometries or comparison to the classical action of pure gravity.

We will show that the geometric description of the handcuff graph (\ref{introVTQFT}a) is a single truncated tetrahedron,
\begin{align}\label{introS}
\vcenter{\hbox{
\begin{tikzpicture}[scale=0.035]
%
\coordinate (a1) at (10,53);
\coordinate (a2) at (9, 27);
\coordinate (a3) at (18, 40);
%
\draw[thick] (a1) -- (a2) node[midway,left] {$A$};
\draw[thick,dotted] (a2) -- (a3);
\draw[thick,dotted] (a3) -- (a1);
%
\coordinate (b1) at (59, 19);
\coordinate (b2) at (45, 5);
\coordinate (b3) at (73, 7);
%
\draw[thick] (b1) -- (b2);
\draw[thick] (b2) -- (b3) node[midway,below] {$B$};
\draw[thick] (b3) -- (b1);
%
\coordinate (c1) at (92,60);
\coordinate (c2) at (86,45);
\coordinate (c3) at (94,34);
%
\draw[thick,dotted] (c1) -- (c2);
\draw[thick,dotted] (c2) -- (c3);
\draw[thick] (c3) -- (c1) node[midway,right] {$A$};
%
\coordinate (d1) at (35,88);
\coordinate (d2) at (50,82);
\coordinate (d3) at (59,90);
%
\draw[thick] (d1) -- (d2);
\draw[thick] (d2) -- (d3);
\draw[thick] (d3) -- (d1) node[midway,above] {$B$};
%
\coordinate (upperLeftE) at (33,61);
\coordinate (lowerLeftE) at (35,26);
\coordinate (lowerRightE) at (71, 30);
\coordinate (upperRightE) at (67,64);
\coordinate (frontE) at (51,50);
\coordinate (backE) at (54,40);
%
%
\draw[thick,red,dashed] plot [smooth,tension=1] coordinates {(a3) (backE) (c2)};
\draw[thick,darkgreen] plot [smooth,tension=1] coordinates {(a1) (upperLeftE) (d1)};
\draw[thick,darkgreen] plot [smooth,tension=1] coordinates {(a2) (lowerLeftE) (b2)};
\draw[thick,darkgreen] plot [smooth,tension=1] coordinates {(b3) (lowerRightE) (c3)};
\draw[thick,darkgreen] plot [smooth,tension=1] coordinates {(c1) (upperRightE) (d3)};
\draw[thick,blue] plot [smooth,tension=1] coordinates {(b1) (frontE) (d2)};
\end{tikzpicture}
}} \ .
\end{align}
On shell, the eight faces are totally geodesic surfaces, and edges meet the triangular faces orthogonally. Gluing this object along the triangular faces $A \leftrightarrow A$ and $B \leftrightarrow B$ produces a finite-volume genus-2 handlebody with pleated boundary, bent on three closed geodesics shown in different colors. The semiclassical gravitational path integral on this topology is the volume of the truncated tetrahedron, plus corner terms, and we will show that this matches the predictions of CTV theory and Virasoro TQFT. With fixed-length boundary conditions, it reproduces the Virasoro modular S-matrix  $|\hatS_{P_1P_2}[P_3]|^2$, which is the Virasoro TQFT amplitude-squared for the handcuff graph.

The second example is a tetrahedron doubled along its truncated faces:
\begin{align}\label{intro6j}
\vcenter{\hbox{
\begin{tikzpicture}[scale=0.035]
%
\coordinate (a1) at (10,53);
\coordinate (a2) at (9, 27);
\coordinate (a3) at (18, 40);
%
\draw[thick] (a1) -- (a2) node[midway,left] {$A$};
\draw[thick,dotted] (a2) -- (a3);
\draw[thick,dotted] (a3) -- (a1);
%
\coordinate (b1) at (59, 19);
\coordinate (b2) at (45, 5);
\coordinate (b3) at (73, 7);
%
\draw[thick] (b1) -- (b2);
\draw[thick] (b2) -- (b3) node[midway,below] {$B$};
\draw[thick] (b3) -- (b1);
%
\coordinate (c1) at (92,60);
\coordinate (c2) at (86,45);
\coordinate (c3) at (94,34);
%
\draw[thick,dotted] (c1) -- (c2);
\draw[thick,dotted] (c2) -- (c3);
\draw[thick] (c3) -- (c1) node[midway,right] {$C$};
%
\coordinate (d1) at (35,88);
\coordinate (d2) at (50,82);
\coordinate (d3) at (59,90);
%
\draw[thick] (d1) -- (d2);
\draw[thick] (d2) -- (d3);
\draw[thick] (d3) -- (d1) node[midway,above] {$D$};
%
\coordinate (upperLeftE) at (33,61);
\coordinate (lowerLeftE) at (35,26);
\coordinate (lowerRightE) at (71, 30);
\coordinate (upperRightE) at (67,64);
\coordinate (frontE) at (51,50);
\coordinate (backE) at (54,40);
%
%
\draw[thick,red,dashed] plot [smooth,tension=1] coordinates {(a3) (backE) (c2)};
\draw[thick,cyan] plot [smooth,tension=1] coordinates {(a1) (upperLeftE) (d1)};
\draw[thick,darkgreen] plot [smooth,tension=1] coordinates {(a2) (lowerLeftE) (b2)};
\draw[thick,orange] plot [smooth,tension=1] coordinates {(b3) (lowerRightE) (c3)};
\draw[thick,purple] plot [smooth,tension=1] coordinates {(c1) (upperRightE) (d3)};
\draw[thick,blue] plot [smooth,tension=1] coordinates {(b1) (frontE) (d2)};
\end{tikzpicture}
\begin{tikzpicture}[scale=0.035]
%
\coordinate (a1) at (10,53);
\coordinate (a2) at (9, 27);
\coordinate (a3) at (18, 40);
%
\draw[thick] (a1) -- (a2) node[midway,left] {$C$};
\draw[thick,dotted] (a2) -- (a3);
\draw[thick,dotted] (a3) -- (a1);
%
\coordinate (b1) at (59, 19);
\coordinate (b2) at (45, 5);
\coordinate (b3) at (73, 7);
%
\draw[thick] (b1) -- (b2);
\draw[thick] (b2) -- (b3) node[midway,below] {$B$};
\draw[thick] (b3) -- (b1);
%
\coordinate (c1) at (92,60);
\coordinate (c2) at (86,45);
\coordinate (c3) at (94,34);
%
\draw[thick,dotted] (c1) -- (c2);
\draw[thick,dotted] (c2) -- (c3);
\draw[thick] (c3) -- (c1) node[midway,right] {$A$};
%
\coordinate (d1) at (35,88);
\coordinate (d2) at (50,82);
\coordinate (d3) at (59,90);
%
\draw[thick] (d1) -- (d2);
\draw[thick] (d2) -- (d3);
\draw[thick] (d3) -- (d1) node[midway,above] {$D$};
%
\coordinate (upperLeftE) at (33,61);
\coordinate (lowerLeftE) at (35,26);
\coordinate (lowerRightE) at (71, 30);
\coordinate (upperRightE) at (67,64);
\coordinate (frontE) at (51,50);
\coordinate (backE) at (54,40);
%
%
\draw[thick,red,dashed] plot [smooth,tension=1] coordinates {(a3) (backE) (c2)};
\draw[thick,purple] plot [smooth,tension=1] coordinates {(a1) (upperLeftE) (d1)};
\draw[thick,orange] plot [smooth,tension=1] coordinates {(a2) (lowerLeftE) (b2)};
\draw[thick,darkgreen] plot [smooth,tension=1] coordinates {(b3) (lowerRightE) (c3)};
\draw[thick,cyan] plot [smooth,tension=1] coordinates {(c1) (upperRightE) (d3)};
\draw[thick,blue] plot [smooth,tension=1] coordinates {(b1) (frontE) (d2)};
\end{tikzpicture}
}} \ . 
\end{align}
This is the geometric description of the tetrahedron graph in (\ref{introVTQFT}b) for heavy operators, and the path integral on this topology reproduces the Virasoro $6j$-symbol, $\left|\begin{Bsmallmatrix} P_1 & P_2 & P_3 \\ P_4 & P_5 & P_6 \end{Bsmallmatrix}\right|^2$. We confirm this by comparing the semiclassical $6j$-symbol to the on-shell action, which is given by the volume plus corner terms.  A similar relation was observed by Teschner and Vartanov \cite{Teschner:2012em} for weights below the black hole threshold, where the geometry is an ordinary (non-truncated) hyperbolic tetrahedron.

\subsubsection*{Related approaches}

A different approach to triangulating AdS$_3$ gravity using a Turaev-Viro-type theory was described recently in an interesting series of papers \cite{Chen:2024unp,Hung:2024gma,Bao:2024ixc,Hung:2025vgs,Geng:2025efs}. Our approach has similar ingredients, but we will use a different definition of the Turaev-Viro partition function and a different mapping to 3D gravity. There are two main distinctions. The first is in how the boundary conditions are implemented. The starting point of \cite{Chen:2024unp} is to triangulate the boundary theory, which requires BCFT techniques, whereas we will only do an ordinary pants decomposition on the boundary, and make no reference to BCFT. The second difference is that we perform the triangulation in terms of fixed-angle rather than fixed-length boundary conditions --- the $S$-transform does not appear in the formalism of \cite{Chen:2024unp}. At the semiclassical level this is not important, but it gives different answers for the exact amplitudes, for example when applied to the handcuff graph (\ref{introVTQFT}a). (The exact result for the tetrahedral graph (\ref{introVTQFT}b) is actually the same in both approaches, thanks to a magical identity stating that the squared $6j$-symbol is self-dual under a Fourier transform \cite{Barrett:2002vi,ctv}.) It would be interesting to incorporate the $S$-transform into BCFT, combining the two approaches.

 Conformal Turaev-Viro theory, developed in \cite{ctv}, is based on the standard Turaev-Viro theory \cite{Turaev:1992hq} for discrete spin networks, which has been studied as a model of dS$_3$ quantum gravity. Indeed, the study of CTV theory in \cite{ctv} borrows heavily from the literature on spin networks \cite{ponzano1968semiclassical,Turaev:1992hq,Barrett:1997gw}, especially Roberts' chain-mail formalism \cite{roberts1995skein} and its extension by Barrett, Garcia-Islas, and Martins to include what are called external edges below \cite{Barrett:2002vi,Garcia-Islas:2004lwa,Barrett:2004im}. (See also \cite{MR1434238,Costantino:2007bnk,MR3073565,Chen:2015wfa}.) There are some important differences: In \cite{Barrett:2002vi,Garcia-Islas:2004lwa,Barrett:2004im} the manifold $M$ is closed, the gravitational theory has positive cosmological constant, the spectrum is discrete, and there is a finite total number of primary states $N$. 
If we tried to repeat their arguments in AdS$_3$ gravity  and Virasoro TQFT directly, we would encounter divergent factors of $N = \int dP \rho_0(P) = \infty$. Some of the definitions and derivations in CTV \cite{ctv} have been modified as compared to \cite{Turaev:1992hq,Barrett:2004im} to avoid these divergences, but the overall structure is similar. It would be very interesting to revisit \cite{Barrett:2004im} and related approaches in light of recent progress on de Sitter quantum gravity (e.g.~\cite{Chandrasekaran:2022cip,Collier:2025lux,Abdalla:2025gzn,Harlow:2025pvj}), since it appears that they may have already described dS$_3$ in a way that is closely analogous to AdS$_3/\overline{\rm{CFT}_2}$\,!

\subsubsection*{Outline}

\noindent In section \ref{s:boundaryconditions} we define fixed-angle and fixed-length boundary conditions on a compact region in 3D gravity. We also briefly discuss the extension to higher dimensions. In section \ref{s:dictionary} we explain how to strip off the asymptotic regions in 3D gravity, and how the normalized OPE statistics are calculated by the fixed-length path integral on a compact region with pleated boundaries. In section \ref{s:pathtri} we discuss the fixed-angle path integral, its connection to the CTV partition function, and how this formalism is naturally organized in terms of triangulations and generalized triangulations. We also describe a general algorithm to convert Virasoro TQFT amplitudes to triangulated geometries. 

The second half of the paper is devoted to the semiclassical limit of the exact results developed in the first half. In section \ref{s:polyhedra}, we review generalized hyperbolic tetrahedra, and the calculation of their volumes and actions. Then, in section \ref{s:classicaltri}, we take the semiclassical limit of the CTV partition function and show that it manifiestly agrees with the metric formalism and decomposes the spacetime into generalized hyperbolic tetrahedra. In sections \ref{s:modulars}-\ref{s:sixj} we study the two semiclassical examples described in the introduction, the geometries dual to the squared Virasoro $6j$-symbol and the squared modular $S$-matrix. In section \ref{s:discussion}, we conclude with a discussion of open questions and future directions.

 In Appendix \ref{app:kernels} we derive the semiclassical limits of the Virasoro crossing kernels, and in appendix \ref{app:cftstatistics} we review the conformal bootstrap prediction of the OPE statistics relevant to the example geometries studied in the paper. Appendix \ref{app:orb} is a brief tutorial on using the software {\tt Orb} to triangulate trivalent graphs, which is not necessary for the examples studied in this paper, but is very useful for more complicated examples.

\subsubsection*{Roadmap for $D \neq 3$}
Readers mainly interested in lessons for gravity in $D \neq 3$ dimensions can take the following shortcut through the paper. Start with the boundary conditions in sections \ref{ss:fixedangles}-\ref{ss:higherdim}, followed by the AdS/CFT dictionary for compact regions in section \ref{s:dictionary}. Skim section \ref{s:polyhedra} to understand the basics of generalized tetrahedra, and then read sections \ref{s:modulars} and \ref{s:sixj} where the classical geometries are constructed in the two examples.

\section{Boundary conditions}\label{s:boundaryconditions}
In this section we define our boundary conditions for gravity on a compact spacetime region. The focus is on AdS$_3$, but we also remark briefly on fixed-area boundary conditions on compact manifolds in higher-dimensional gravity.

\subsection{Fixed angles}\label{ss:fixedangles}
The action of pure 3D gravity on a compact region of Euclidean spacetime is
\begin{align}
I_0 = -\frac{1}{16\pi G} \int_M \sqrt{g}(R+2) - \frac{1}{8\pi G} \int_{\p M} \sqrt{h}K \ ,
\end{align}
with the Euclidean path integral weighted by $e^{-I_0}$. We have included the GHY boundary term ($K=\del_\mu n^\mu$ with $n^\mu$ the outward-pointing normal), but there are no counterterms, because $M$ does not approach the asymptotic AdS boundary. The central charge of the dual CFT is 
\begin{align}
c= \frac{3}{2G} = 1+ 6Q^2 , \qquad Q = b + b^{-1}
\end{align}
 The topology of $M$ is fixed but arbitrary, and the boundary $\p M$ is a union of Riemann surfaces.

We would like to define boundary conditions that fix the dihedral angles $\psi = (\psi_1,\dots \psi_n)$ on a set of disjoint simple closed curves $\gamma = (\gamma_1,\dots,\gamma_n)$ on $\p M$. On each genus-$g$ component of the boundary, we can fix the dihedral angles on up to $3g-3$ cycles. Elsewhere the boundary conditions are free, allowing the metric to fluctuate, so that the classical saddles have totally geodesic boundaries away from $\gamma$ and the curves $\gamma$ are geodesics. 

Classically, this is achieved by the following fixed-angle action:  
\begin{align}\label{IA}
I_A = I_0 + \frac{c}{12\pi}\sum_{i=1}^n (\pi-\psi_i) \int_{\gamma_i} ds \ . 
\end{align}
The variation $\delta I_A$ has a boundary term proportional to the variation of the induced metric, $\delta h_{ij}$, and allowing the boundary metric to vary sets the coefficient of this term to zero. Away from $\gamma$ this imposes $K_{ij} = 0$, and near each $\gamma_i$ it leads to the boundary equation of motion
\begin{align}
K = (\pi - \psi_i) \delta(s)
\end{align}
where $s$ is a normal coordinate measuring the proper distance to $\gamma_i$. This is the equation for the extrinsic curvature at a dimension-1 corner with interior angle $\psi_i$. On shell, the GHY boundary term cancels the contribution from the extra terms in \eqref{IA}, so the action is proportional to the hyperbolic volume,
\begin{align}\label{IAonshell}
I_A^{\rm on-shell} = \frac{c}{6\pi}\mbox{Vol}(M) 
\end{align}

The fixed-angle path integral can be recast as a correlation function of defect operators in the theory with action $I_0$. Let $\sigma \in \p M$ be a closed curve and $P \in \mathbb{C}$. We define the boundary operator
\begin{align}\label{defVp}
V_P(\sigma) = \exp\left[ -\frac{c}{12}(1-2ibP) \int_{\sigma} ds  \right] \ .
\end{align}
With the angles parameterized as $\psi_i = 2\pi i b P_i$, the exponent is the same as the extra term from each $\gamma_i$ in \eqref{IA}. Therefore the fixed-angle partition function can be expressed as
\begin{align}
Z_A(M, \gamma(\vecP)) &= \langle V_{P_1}(\gamma_1) \cdots V_{P_n}(\gamma_n) \rangle_0 \ . 
\end{align}
If too few defects are inserted, or the topology is too simple, then this correlation function will not admit a semiclassical expansion around a saddlepoint. But in the opposite situation there are typically saddles, and we will determine the exact path integral. One of the simplest nontrivial examples is a genus-2 handlebody, where the boundary curves $(\gamma_1,\gamma_2,\gamma_3)$ are drawn in red in the following diagram:
\begin{align}\label{heegS}
\cp{\includegraphics[width=2.2in]{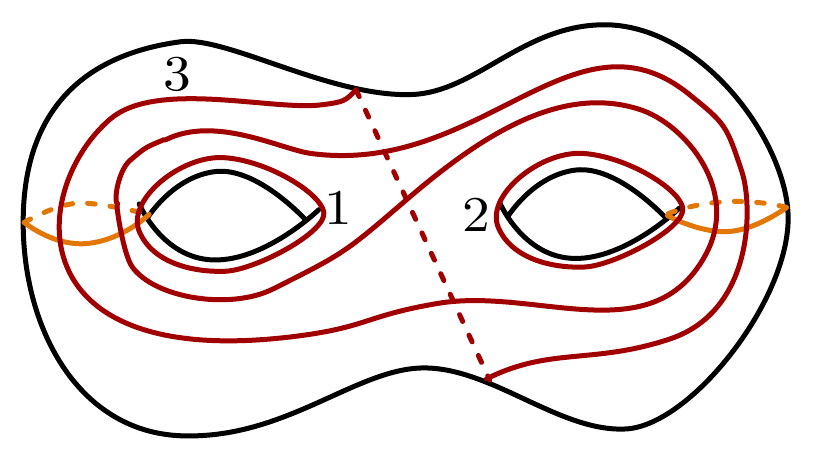}}
\end{align}
Imposing fixed angles on these boundary curves produces a finite-volume saddle that will be constructed in section \ref{s:modulars} below.

In going to the quantum theory, there is an ambiguity in how to define $V_P$ at subleading  orders in $1/c$. We will sidestep this by matching the exact path integral to known results when we glue to asymptotic boundaries, and only insist that it agrees with \eqref{defVp} semiclassically.

We will assume that the exact amplitude is invariant under the reflection $P_i \to - P_i$, taken individually for each $i$. This is a very reasonable assumption given the results in \cite{Wang:2025bcx}, where $V_P$ is interpreted as the holographic dual of a BCFT boundary operator of scaling dimension $h = \frac{c-1}{24} + P^2$ and all amplitudes are reflection-symmetric. In our case, the closed curve $\sigma$ does not hit the AdS boundary, so there is no need to introduce a boundary in the dual CFT. The reflection symmetry is also known to hold for bulk line defects, which are dual to particles below the black hole threshold, and again all physical quantities are functions of $P^2$.

\subsection{Fixed lengths}\label{ss:fixedlengths}
We will now consider fixed-length boundary conditions. The setup is the same, with $M$ a compact 3-manifold and $\gamma$ a set of disjoint simple closed curves on the boundary. Up to a constant shift, the fixed-length action is simply $I_0$, with no additional boundary terms.  Adding a constant changes the normalization. A convenient definition of the fixed-length action is
\begin{align}\label{IL}
I_L = I_0 + \frac{c}{12}\sum_{i=1}^n \ell_i \ . 
\end{align}
A short calculation shows that this action has a good variational principle if we impose $K_{ij} = 0$ away from $\gamma$, fix the lengths $\int_{\gamma_i}ds = \ell_i$, and require the $\gamma_i$ to be extremal. We omit the details, but refer to \cite{Chua:2023srl} for a nearly-identical calculation. 
Classical solutions are again hyperbolic 3-manifolds whose boundaries consist of piecewise-flat hyperplanes glued along dimension-1 corners. The on-shell action has a contribution from the GHY term at the corners, known as Hayward terms in this context, which is
\begin{align}
I_{\rm Hayward} = \left( \frac{\psi}{\pi}-1\right) \frac{\ell}{8G} = \frac{c}{12\pi}\ell(\psi - \pi)
\end{align}
at each corner, where $\psi$ is the interior dihedral angle. Therefore
\begin{align}\label{ILonshell}
I_L^{\rm on-shell} = \frac{c}{6\pi}\left( \mbox{Vol}(M) + \frac{1}{2} \sum_i \ell_i \psi_i \right) \ . 
\end{align}
The reason for using the normalization \eqref{IL} is that it simplifies the gluing rules. If two regions are glued together without creating any internal edges, $I_L$ is additive.  If the gluing creates an internal edge of length $\ell$, then it is additive up to an entropy factor, 
\begin{align}\label{gluingrule}
I_L = -\frac{\ell}{4G} + I_L^{(1)} + I_L^{(2)} \ . 
\end{align}

At the quantum level, we parameterize the lengths by $\ell_i = 4\pi b P_i$, and write the fixed-length path integral as $Z_L(M, \gamma(\vecP))$. By definition, the exact amplitude $Z_L$ is the inverse Laplace transform of the fixed-angle amplitude, with the conventions for the Laplace transform given in \eqref{introFourierAL}. The $P_i \to -P_i$ reflection symmetry of $Z_A$ is automatically inherited by $Z_L$. Written in terms of $\psi$ and $\ell$ instead of $P$ and $P'$, the relation at large $c$ is
\begin{align}
Z_A(\psi) \sim \int_0^{\infty} d^n\ell e^{ \frac{c}{12\pi} \sum_{i=1}^n \psi_i \ell_i } Z_L(\ell) \ ,
\end{align}
which is consistent with the shift in going from \eqref{IAonshell} and \eqref{ILonshell}.  

Complex lengths correspond to independent left and right-moving weights, $\vecP$ and $\bar{\vecP}$. Therefore, it may seem more natural to complexify the lengths and define a fixed-angle-type boundary condition with independent chemical potentials for $\ell$ and $\bar{\ell}$. By contrast, the fixed-angle path integral $Z_A$ that we have defined has only one chemical potential, even if $\psi$ is complex. Indeed, there does appear to be an interesting extension to independent complex $\psi$ and $\bar{\psi}$ that is considered briefly in the discussion section below.  A more formal approach to the problem of classical 3D gravity on compact manifolds, including spin, is pursued in \cite{Krasnov:2005dm,Krasnov:2006jb,Bonsante:2006tr,Krasnov:2008yq,Krasnov:2009vy,MR3732685,MR4479756}.

\subsection{Aside on gravity in $D>3$ dimensions}\label{ss:higherdim}

In higher dimensions, the analogous boundary condition on a compact region $M$ fixes the areas $A_i$ of codimension-2 extremal surfaces $\gamma = (\gamma_1,\dots,\gamma_n)$ on $\p M$, or their conjugate angles. The fixed-angle action in higher dimensions is
\begin{align}
I_{\rm angle} = I_{0} + \frac{1}{8\pi G} \sum_{i=1}^n (\pi - \psi_i) \int_{\gamma_i} dA \ ,
\end{align}
where $I_0$ is the usual Einstein+GHY action in $D$ dimensions. The fixed-area action  (with a constant shift as above) is
\begin{align}
I_{\rm area} &= I_0  + \frac{1}{8G} \sum_{i=1}^n A_i \ ,
\end{align}
and the exact relation between the amplitudes is the Laplace transform
\begin{align}
Z_{\rm angle}(\psi) = \int_0^{\infty} (\Pi_{i=1}^n dA_i) e^{\frac{1}{8\pi G}\Sigma_i \psi_i A_i} Z_{\rm area}(A_i)
\end{align}
There appears to be no problem in principle with studying these boundary conditions in higher dimensions, at least semiclassically. Presumably it is more difficult to construct classical saddlepoints. For motivation and potential comparison, there are some CFT results on the statistics of heavy operators available in the literature \cite{Benjamin:2023qsc}. 

A similar fixed-area path, but with a focus on non-compact manifolds, has been discussed in \cite{Dong:2018seb,Dong:2022ilf}. With a sufficient number of extremal surfaces (or a sufficiently complicated topology), it should be possible to apply the same technique to compact manifolds in order to calculate OPE statistics in the black hole regime.

\section{The AdS/CFT dictionary for a compact region}\label{s:dictionary}
In this section, we will relate the fixed-length path integral on a compact region of spacetime $Z_L(M, \gamma(\vecP))$ to OPE statistics in the dual CFT.  OPE statistics can also be calculated using Virasoro TQFT, which leads to \eqref{summaryLength}. As noted in the introduction, the results of this section are mostly implicit in \cite{Collier:2023fwi,Collier:2024mgv}, but we will make them explicit in geometric language.

\subsection{Background on OPE statistics in AdS$_3$/CFT$_2$}

The OPE coefficients of Virasoro primary operators in a 2d CFT are $c_{ijk} = \langle i|{\cal O}_j |k\rangle$. We assume in this section and throughout most of the paper that all three operators are dual to black hole microstates. This requires each conformal weight to be above the black hole threshold, $P \in \mathbb{R}_+$. The eigenstate thermalization hypothesis suggests the ansatz
\begin{align}\label{eth}
c_{ijk} &= 
f(P_i,\bP_i,P_j,\bP_j) \delta_{ik}
+
g(P_i,\bP_i,P_j,\bP_j,P_k,\bP_k) R_{ijk}
\end{align}
where $f$ and $g$ are smooth functions and $R_{ijk}$ behaves like a random tensor with zero mean and unit variance. The diagonal term $f$ is the thermal 1-point function of ${\cal O}_j$, which is typically subleading in a holographic theory. The ansatz is meant to apply to the chaotic regime. In a general, non-integrable quantum system this may require $\Delta_i, \Delta_k \to \infty$ with some restrictions on $\Delta_j$, but in a holographic theory, it is reasonable to assume that almost all black hole microstates are chaotic and therefore subject to \eqref{eth}.

In a holographic CFT dual to pure 3D gravity, the four-point function of pairwise-identical heavy operators is the Virasoro identity conformal block \cite{Hartman:2013mia}. This statement holds at leading order in a formal sum over topologies. Expanding in the dual channel gives an explicit formula for the leading statistics of heavy OPE coefficients \cite{Collier:2019weq},
\begin{align}\label{c0stat}
\overline{|c_{ijk}|^2} &\approx C_0(P_i, P_j, P_k) C_0(\bP_i, \bP_j, \bP_k) \ , 
\end{align}
where $C_0$ is proportional to the DOZZ structure constants of the Liouville CFT, and once again, this is the leading approximation in a sum over topologies. See appendix \ref{app:cftstatistics} or \cite{Eberhardt:2023mrq} for a review of $C_0$ and other aspects of Virasoro conformal blocks. The average, denoted by a $\overline{\rm bar}$, can be interpreted as either an ensemble average or an average over nearby microstates. In the case that the bulk theory is pure gravity, it should be interpreted as an ensemble average \cite{Chandra:2022bqq} (see \cite{Chandra:2022fwi,Chandra:2023dgq,Chandra:2023rhx} for the other interpretation, in a UV-complete fixed theory). Comparing to \eqref{eth}, we see that $\left[C_0(P_i, P_j, P_k) C_0(\bP_i, \bP_j, \bP_k)\right]^{1/2}$ is the leading approximation to the smooth function $g$ in the eigenstate thermalization hypothesis.

General arguments indicate that OPE statistics should be related to wormhole amplitudes \cite{Penington:2019kki,Pollack:2020gfa,Belin:2020hea,Stanford:2020wkf}, and this was realized concretely in \cite{Chandra:2022bqq}. The conjecture of \cite{Chandra:2022bqq} is that at leading order in a sum over topologies, the ensemble has Gaussian statistics with variance \eqref{c0stat}, and corrections at subleading order can be calculated iteratively by the conformal bootstrap. In addition, the ensemble has leading order density of states given by the Cardy formula,
\begin{align}
\overline{\rho(P,\bP)} = \rho_0(P) \rho_0(\bP) , \qquad 
\rho_0(P) &:= 4\sqrt{2} \sinh(2\pi bP)\sinh(2\pi b^{-1}P)  \approx e^{\pi c b P /3} \ . 
\end{align}
It is unclear whether individual draws from the ensemble are true CFTs, or only the average makes sense as a CFT. Relatedly, while we can study the statistics of CFT data using these methods, it is not known how to formulate a microscopic theory of random 2d CFTs akin to random matrix theory in quantum mechanics, or how to develop a systematic approach to all orders in a topological expansion.\footnote{A proposal in this direction is the random tensor model introduced in \cite{Belin:2023efa}. The authors of \cite{Jafferis:2025vyp} conjectured that an extended version of this model reproduces all of 3D gravity in a topological expansion. The leading terms and certain subleading terms in this model are equivalent to \cite{Cotler:2020ugk,Chandra:2022bqq}, by construction. In principle, the random tensor model also makes systematic predictions for the full series of subleading terms, but they have not been checked vs gravity in the literature. There are potential ambiguities in the random tensor model at subleading orders, so if they match, that would be striking evidence for the strongest version of the correspondence suggested in \cite{Jafferis:2025vyp}. }

\subsection{Stripping off the asymptotic boundaries}
We will now discuss the AdS/CFT dictionary for the statistics of OPE coefficients. We use ${\cal M}$ for a  complete non-compact 3-manifold with asymptotic AdS boundaries, and $M$ for a compact region with (non-asymptotic) boundary.

The standard gravitational path integral in AdS$_3$, on a fixed 3-topology ${\cal M}$ with asymptotic boundary condition $S$,  is written $Z_{\rm grav}({\cal M}; S)$.  The dictionary for the $n$-point statistics of OPE coefficients is the formal sum over 3-manifolds \cite{Chandra:2022bqq}
\begin{align}\label{oldDictionary}
\overline{c_{i_1j_1k_1} \dots c_{i_nj_nk_n}} 
= \sum_{{\cal M}} Z_{\rm grav}({\cal M}; S_{i_1j_1k_1} \sqcup \dots \sqcup S_{i_nj_1n_1})
\end{align}
where $S_{ijk}$ denotes an asymptotic 2-sphere with three operators inserted at $0,1,\infty$. This is a gravitational path integral on a wormhole topology, with an asymptotic boundary for each insertion of $c_{ijk}$. 

The precise definition of the boundary condition $S_{ijk}$ on each component of $\p {\cal M}$ depends on whether the operators are above or below the black hole threshold. The simplest case is the defect regime, where all three operators are below threshold, with imaginary $P$. In the defect regime, the boundary condition $S_{ijk}$ is simply a 2-sphere with three conical defects hitting the boundary. In \cite{Chandra:2022bqq}, with this boundary condition, the dictionary \eqref{oldDictionary} was checked at the level of the Gaussian term and certain subleading corrections.  The simplest example is the 3-point Maldacena-Maoz \cite{Maldacena:2004rf} wormhole, with topology ${\cal M}_0 = S^2 \times \mbox{Interval}$. The amplitude for this wormhole is \cite{Chandra:2022bqq}
\begin{align}\label{c0amplitude}
Z_{\rm grav}({\cal M}_0; S_{ijk}\sqcup S_{ijk}) =  C_0(P_i,P_j, P_k) C_0(\bP_i, \bP_j, \bP_k) \ ,
\end{align}
in agreement with the bootstrap result \eqref{c0stat}.

It is more subtle to define the boundary condition $S_{ijk}$ for heavy operator insertions, above the black hole threshold. This problem was solved in \cite{Abajian:2023jye,Abajian:2023bqv} , where the authors also calculated the leading Gaussian term semiclassically and verified the equation \eqref{c0amplitude} for heavy scalar operators. For heavy operators, ${\cal M}_0$ is a branched cover of (pair of pants)$\times$(interval). The branching structure is such that each asymptotic boundary component is a 2-sphere with heavy operators inserted at $0,1$ and $\infty$. Thus there are two types of $C_0$ wormhole: The defect $C_0$ wormhole considered in \cite{Maldacena:2004rf} and \cite{Chandra:2022bqq}, and the heavy $C_0$ wormhole constructed in \cite{Abajian:2023bqv}. They are related by analytic continuation in the weights. There are also mixed versions, for example with two heavy and one light operator, which has not been discussed in the literature but is a straightforward extension of \cite{Abajian:2023bqv} constructed as a branched cover of a BTZ black hole with a defect particle added.

\begin{figure}[t]
\begin{center}
\includegraphics[width=3.9in]{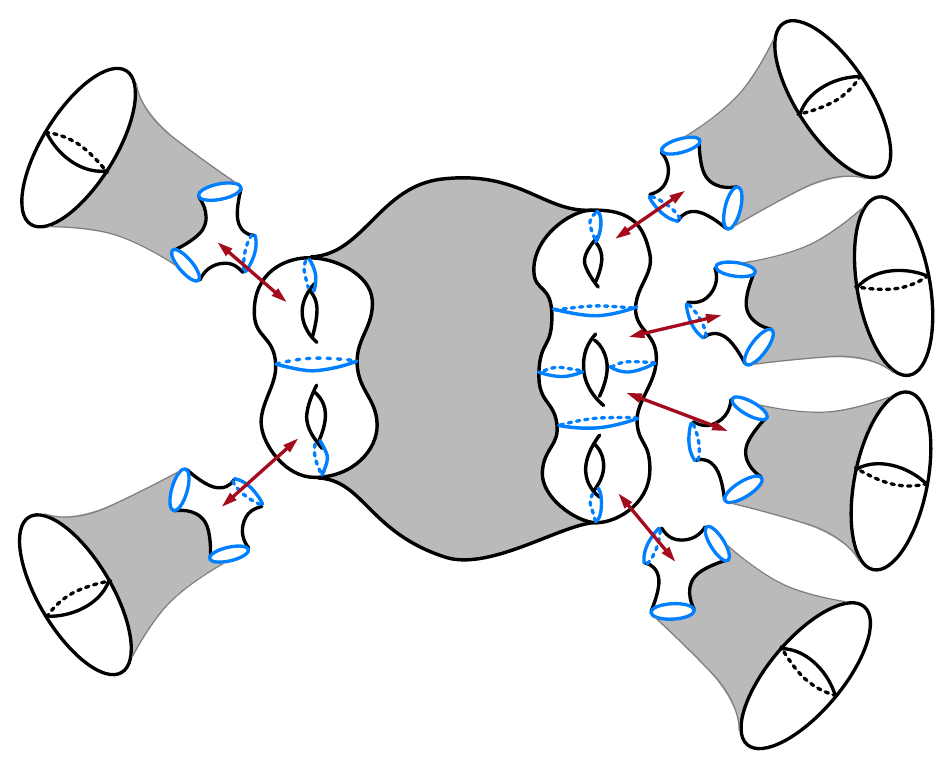}
\end{center}
\caption{\small
Wormhole contributing to the ensemble average $\overline{c_{i_1j_1k_1}\cdots c_{i_6j_6k_6}}$, with all operators above the black hole threshold. Each $C_0$ flare connects an asymptotic boundary $S^2$ with three heavy operator insertions to a totally geodesic pair of pants in the bulk. On-shell, the boundaries of the central region are pleated Riemann surfaces with corners on geodesics (blue). 
}\label{fig:c0flares}
\end{figure}

By imposing the boundary condition from \cite{Abajian:2023bqv} on each component $S_{ijk}$, this extends the dictionary \eqref{oldDictionary} to the $n$-point statistics of heavy operators. The manifolds that contribute to $\overline{c_{i_1j_1k_1} c_{i_2j_2k_2} \dots}$, assuming all operators are heavy, have the structure shown in figure \ref{fig:c0flares}. There is a compact region $M$, with finite boundary $\p M$ consisting of one or more Riemann surfaces. $M$ is glued to the asymptotic boundary by attaching `$C_0$ flares', which play a role similar to the trumpets of JT gravity \cite{Saad:2019lba}. A $C_0$ flare is one half of the $C_0$ wormhole, connecting an asymptotic boundary to a totally geodesic pair of pants. Topologically, the $C_0$ flares are glued to $M$ by choosing a pants decomposition of $\p M$ and gluing each pair of pants on $\p M$ to a flare. The manifold ${\cal M}$ is branched around the separating curves of the pants decomposition in exactly the same way that the $C_0$ wormhole branches around the horizons in \cite{Abajian:2023bqv}. 

\subsection{The dictionary for a compact region}
It is clear from this construction, and also from calculations of OPE statistics in Virasoro TQFT, that for an observable $\overline{c_{i_1j_1k_1}c_{i_2j_2k_2}\cdots}$, each insertion of $c_{ijk}$ always comes with a factor of $\sqrt{C_0 \bar{C}_0}$ associated to the $C_0$ flare. Let us strip off these factors by defining the normalized OPE coefficients
\begin{align}\label{bigC}
C_{ijk} = \frac{1}{\left[C_0(P_i, P_j, P_k) C_0(\bP_i, \bP_j, \bP_k)\right]^{1/2}}  \ c_{ijk} \ ,
\end{align}
which have unit variance at leading order in the sum over topologies.

We would like to rewrite the dictionary for heavy operators in terms of the statistics of $C_{ijk}$. We will first state the dictionary for the compact spacetime region $M$, then check that it agrees with the discussion above by re-attaching the $C_0$ flares. 

Let $M$ be a compact 3-manifold with boundary $\p M = \Sigma_1 \sqcup \dots \sqcup \Sigma_p$. We will assume that each $\Sigma_i$ has genus $g_i \geq 2$. The following prescription can be extended to $g=0,1$ but then we do not expect to have on-shell geometries in pure gravity.

Choose a pants decomposition of $\p M$, specified as a set of separating curves, $\gamma = (\gamma_1,\dots,\gamma_n)$, with $3g-3$ curves on each component of $\p M$. On these curves, we impose the fixed-length boundary condition $\gamma(\vecP)$ described in section \ref{s:boundaryconditions}, with lengths parameterized as $\ell_i = 4\pi b P_i$. 

The pattern in which the pants are glued together defines an abstract trivalent graph $\Gamma(\vecP)$, where the vertices are pairs of pants and the edges are $\gamma_i$. (An abstract graph is a graph defined only by its connectivity; it is not embedded in a 3-manifold.) For real lengths and scalar operators, we conjecture that the normalized OPE statistics are calculated by the following formal sum over topologies:
\begin{align}\label{dictionary}
\overline{C_{i_1j_1k_1}\cdots C_{i_rj_rk_r} }
&=  \sum_{\Gamma}  \delta^{\Gamma}_{i_1j_1k_1\dots i_rj_rk_r} \sum_{M, \gamma} Z_L(M, \gamma(\vecP))
\end{align}
The outer sum in \eqref{dictionary} is over trivalent graphs $\Gamma$ with $n = 3r/2$ edges ($r$ is assumed even; odd moments vanish). Each vertex of $\Gamma$ corresponds to an insertion of $C_{ijk}$.  The tensor structure $\delta^\Gamma$ is a product of Kronecker deltas, one for each edge of the graph, setting equal the states at the two ends of the edge, times an integer counting the number of inequivalent ways that $M$ can be glued to flares. (See examples below.) The inner sum is over distinct\footnote{See e.g. \cite{Collier:2023fwi} for a discussion of what `distinct' means.} $(M,\gamma)$, which are compact topological 3-manifolds such that  $\p M$ has a pants decomposition with graph $\Gamma$ and separating curves $\gamma$. The dictionary is extended to spinning operators by analytically continuing to complex lengths.

The dictionary \eqref{dictionary} means that a given topology $M$ with boundary condition $\gamma(\vecP)$ contributes to the OPE statistics for a particular contraction of OPE coefficients that is specified by the graph. This is similar to light defects, and to Virasoro TQFT.

It is well known that closed geodesics in the bulk are related to primaries in the dual CFT according to $\ell_i = 4\pi b P_i$, so \eqref{dictionary} is a natural proposal. At the semiclassical level, \eqref{dictionary} follows from the standard AdS/CFT dictionary by re-attaching the $C_0$ flares, which immediately reproduces \eqref{oldDictionary}.

\subsection{Comparison to Virasoro TQFT}
In Virasoro TQFT, the statistics of OPE coefficients are calculated by amplitudes 
\begin{align}
|\Zvir(M_E,\Gamma(\vecP))|^2
\end{align} 
where $M_E$ is a closed embedding manifold, and $\Gamma(\vecP)$ is a framed trivalent graph. Our point of view on Virasoro TQFT is that it is defined rigorously by a set of diagrammatic rules for calculating amplitudes. The rules were stated in \cite{Collier:2023fwi,Collier:2024mgv} and summarized, with our conventions, in \cite{ctv}. Our conventions differ from \cite{Collier:2023fwi,Collier:2024mgv} in one respect: We normalize the trivalent vertex by the condition
\begin{align}\label{thetanorm}
\Zvir\left(S^3, \cp{\includegraphics[width=.75in]{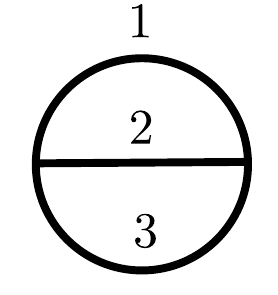}}\right)
&\quad=\quad  1
\end{align}
This removes a factor of $\sqrt{C_0}$ from each vertex as compared to \cite{Collier:2023fwi,Collier:2024mgv}, and as a result, no factors of $C_0$ appear in VTQFT amplitudes when they are expressed in terms of the $6j$-symbols and modular $S$-matrix $\hatS_{P_1P_2}[P_3]$ in Racah-Wigner normalization. With the normalization \eqref{thetanorm}, $|\Zvir|^2$ calculates statistics of the normalized OPE coefficients $C_{ijk}$ defined in \eqref{bigC}.

Therefore, the dictionary \eqref{dictionary} for the fixed-length path integral on compact $M$ is equivalent to 
\begin{align}
Z_L(M, \gamma(\vecP)) = |\Zvir(M_E,\Gamma(\vecP))|^2 \ . 
\end{align}
As explained in the introduction, gravity lives on $M$, while the Virasoro TQFT amplitude is associated to a graph $\Gamma$ embedded in a closed manifold $M_E$. The two manifolds are related by
\begin{align}
M = M_E  - N(\Gamma)
\end{align}
where $N(\Gamma)$ is a regular neighborhood of the graph. Thus gravity lives on the complement of the thickened graph, the boundary is
\begin{align}
\p M = \p N(\Gamma) \ ,
\end{align} 
and the cycles $\gamma$ in the gravity boundary condition are the meridians of the graph.

\subsection{The degenerate core of the $C_0$ wormhole}\label{ss:c0core}
As the first example let us reproduce the heavy $C_0$ wormhole \cite{Abajian:2023bqv}, with the $C_0$ flares stripped off. The $C_0$ flare is by definition half of the $C_0$ wormhole, so stripping off the two flares gives a core $M$ with zero volume. That is, $M$ is a degenerate 3-manifold that can be visualized as $(\mbox{pair of pants})\times(\mbox{interval})$ in the limit where the length of the interval goes to zero:
\begin{align}
M \qquad = \qquad 
\cp{\includegraphics[width=2.2in]{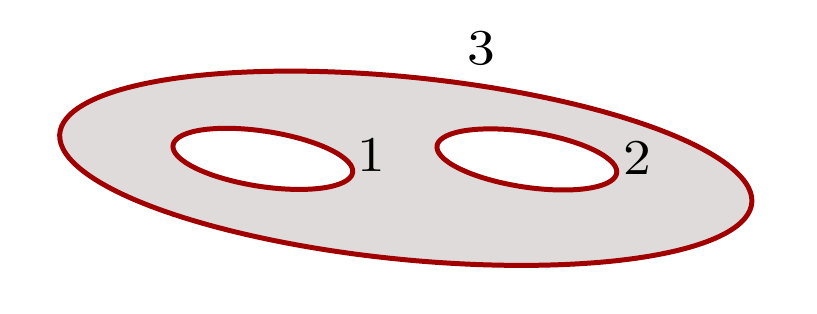}}
\end{align}
Thus $M$ is a flattened genus-2 handlebody.  Its volume is zero, but its boundary $\Sigma = \p M$, consisting of the top of the shaded region glued to the bottom of the shaded region along the three geodesics with dihedral angles $\psi_1=\psi_2=\psi_3 = 0$, is a genus-2 Riemann surface with perfectly regular intrinsic geometry.

The fixed-length on-shell action is \eqref{ILonshell}. The volume vanishes, and for $\psi_i = 0$, the corner contribution drops out, so $I_L=0$. Now we apply the dictionary \eqref{dictionary}, and find to this order
\begin{align}
\overline{C_{ijk }^2} = 1 \ .
\end{align}
This is of course the expected answer because of the normalization \eqref{bigC}. This example is trivial by design, and we now turn our attention to subleading contributions with a nontrivial core.

\section{The path integral by triangulations}\label{s:pathtri}

\subsection{The exact fixed-angle path integral}
Given (i) the results of the previous section on the fixed-length path integral, (ii) the Laplace transform \eqref{introFourierAL} relating fixed-length to fixed-angle boundary conditions, and (iii) the result \eqref{introFourierVV}, proved in \cite{ctv}, we can immediately determine the exact path integral of gravity with fixed-angle boundary conditions. It is given by the partition function of Conformal Turaev-Viro theory,
\begin{align}\label{zactv}
Z_A(M, \gamma(\vecP)) &= \Ztv(M_E, \Gamma(\vecP))  \ .
\end{align}
The CTV partition function is defined as follows \cite{ctv}. Let $T$ be a triangulation of $(M_E, \Gamma(\vecP))$ into tetrahedra, such that the graph $\Gamma$ is part of the triangulation, $\Gamma \subset T$. Edges and vertices of $T$ are called \textit{external} if they below to $\Gamma$, and \textit{internal} otherwise, and we denote by $V_{int}$ the set of internal vertices. A triangulation is called \textit{large} if $H_2(M_E - V_{int}) = 0$; large triangulations do not allow any internal vertices whose neighborhood is a ball. This means that it is not possible to make a large triangulation arbitrarily fine-grained by subdividing the tetrahedra. The external edges are labeled by $\vecP$ and we introduce labels $\tilde{\vecP} = (\tilde{P}_1,\dots,\tilde{P}_m)$ on the internal edges. The CTV partition function is
\begin{align}\label{defzctv}
\Ztv(M_E, \Gamma(\vecP))  &= \int_{\mathbb{R}_+^m} (\Pi_{i=1}^m d\tilde{P}_i \rho_0(\tilde{P}_i) )
\prod_{\Delta \in T} W(\Delta)
\end{align}
where $T$ is any large triangulation of $(M_E, \Gamma(\vecP))$, and each tetrahedron $\Delta$ in the triangulation is assigned the amplitude
\begin{align}\label{tetW}
W\left(
\cp{
\begin{tikzpicture}[scale=0.9]
\centerarc[very thick](0,0)(-30:90:1);
\centerarc[very thick](0,0)(90:210:1);
\centerarc[very thick](0,0)(210:335:1);
\draw[very thick] (0,0) -- (0,1);
\draw[very thick] (0,0) -- ({cos(30)},{-sin(30)});
\draw[very thick] (0,0) -- ({-cos(30)},{-sin(30)});
\node at (1,1) {$P_1$};
\node at (-1,1) {$P_2$};
\node at (-0.3,.55) {$P_3$};
\node at (-.55,.0) {$P_4$};
\node at (0.55,0) {$P_5$};
\node at (0,-1.25) {$P_6$};
\end{tikzpicture}
}
\right) 
\quad=\quad 
\begin{Bmatrix}P_4 & P_5 & P_6 \\ P_1 & P_2 & P_3 \end{Bmatrix} \ . 
\end{align}
The fixed-length gravitational path integral is the Fourier transform,
\begin{align}\label{ZLtri}
Z_L(M, \gamma(\vecP'))
&= \int_{\mathbb{R}_+^n}d\vecP (\Pi_{i=1}^n S_{P_i'P_i}) \Ztv(M_E, \Gamma(\vecP)) \\
&=\int_{\mathbb{R}_+^{n+m}}d\vecP d\tilde{\vecP} (\Pi_{i=1}^n S_{P_i'P_i}) (\Pi_{i=1}^m \rho_0(\tilde{P}_i) )
\prod_{\Delta \in T} W(\Delta)
\end{align}
This is one of our central results. It expresses OPE statistics of primary states in terms of a triangulation. This will allow us to easily describe the geometries that underly various calculations in VTQFT, and to match to the on-shell action of pure gravity in the semiclassical limit. 

\subsection{Generalized triangulations}\label{ss:gentri}
We must be careful to distinguish between triangulations of the embedding manifold $M_E$ vs the manifold $M = M_E - N(\Gamma)$ where gravity lives. In the CTV partition function, $T$ is a triangulation of $M_E$, and it is this triangulation that appears in \eqref{ZLtri}. We will now use the triangulation $T$ of $(M_E, \Gamma(\vecP))$ into ordinary tetrahedra to engineer a \textit{generalized} triangulation $\hat{T}$ of $(M, \gamma(\vecP))$ into \textit{generalized} tetrahedra. This is perhaps easiest to understand by looking at examples in the semiclassical limit, which is done below in sections \ref{ss:gentriS} and \ref{ss:vtqftsixj}, but first we will describe the procedure in topological terms. We restrict to triangulations with no internal vertices.

Let $T$ be a large triangulation of $(M_E, \Gamma(\vecP))$, with external edges labeled by $\vecP$ and internal edges labeled by $\tilde{\vecP}$. The triangulation $T$ consists of a set of (ordinary) tetrahedra, with labeled edges, and face pairings describing how they are glued together. Each tetrahedron $\Delta \in T$ can be drawn by flattening it onto the plane, as  
\begin{align}
\cp{\includegraphics[width=1in]{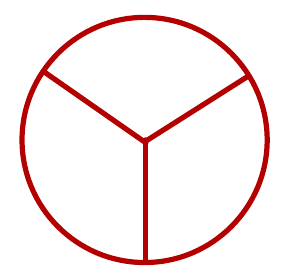}}
\end{align}
where one of the faces is the region outside the circle. The edges are implicitly labeled by weights.

The generalized triangulation $\hat{T}$ is defined by the following procedure. First, we truncate all the vertices,
\begin{align}
\cp{\includegraphics[width=1in]{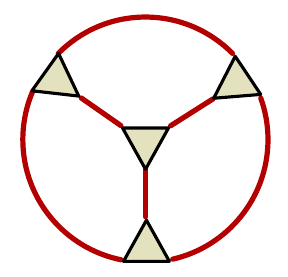}}
\end{align}
Then, we dualize every external edge by replacing
\begin{align}
\cp{\includegraphics[width=0.3in]{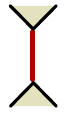}}
\qquad \longrightarrow \qquad
\cp{\includegraphics[width=0.5in]{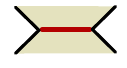}}
\end{align}
This procedure associates to each tetrahedron $\Delta$ a generalized tetrahedron $\hat{\Delta}$, which is a polyhedron together with an assignment of shaded faces. Each $\hat{\Delta}$ has 8 faces (of which 4 are shaded) and 18 edges (of which 6 are labeled by weights). The generalized triangulation $\hat{T}$ is a CW complex where the cells are generalized tetrahedra. The face pairings of $T$ induce face pairings of all the unshaded faces of $\hat{T}$.

The generalized triangulation $\hat{T}$ is a cell decomposition of $M$. The boundary $\p M$ consists of the shaded faces. External edges of $T$ become simple closed curves $\gamma \subset \p M$, so the labels $\vecP$ on the original graph $\Gamma(\vecP)$ become labels on cycles on $\p M$, consistent with our notation for the gravitational boundary condition $\gamma(\vecP)$. 

For example, consider a tetrahedron $\Delta \in T$ with one external edge and five internal edges. First truncating, then dualizing, produces a generalized tetrahedron $\hat{\Delta} \in \hat{T}$ as follows:
\begin{align}
\cp{\includegraphics[width=1in]{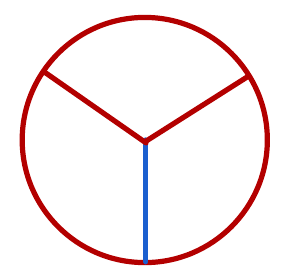}}
\qquad \longrightarrow \qquad
\cp{\includegraphics[width=1in]{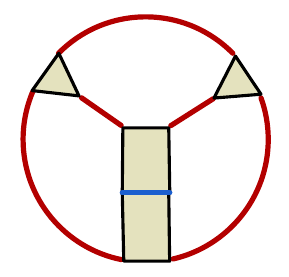}}
\end{align}
where the external edge is shown in blue. The unshaded faces are glued according to the face pairings of $T$, and the shaded faces lie on $\p M$. 

As another example, consider a tetrahedron $\Delta \in T$ where all six edges are external. Truncating, then dualizing (and rearranging the figure), gives
\begin{align}\label{dualtetEx}
\cp{\includegraphics[width=1in]{figures/gentri1.pdf}}
\qquad \longrightarrow \qquad
\cp{\includegraphics[width=1in]{figures/gentri2.pdf}}
\qquad \longrightarrow \qquad
\cp{\includegraphics[width=1in]{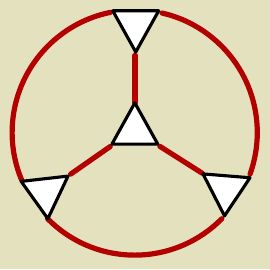}}
\end{align}
Hence the tetrahedron becomes a truncated tetrahedron, with gluings on the truncated faces.

In general, the cell decomposition $\hat{T}$ will consist of various types of generalized tetrahedra similar to these examples, glued together on half of the faces.

Let us now return to the gravitational path integral. Define the amplitude $\hat{W}(\hat{\Delta})$ of a generalized tetrahedron $\hat{\Delta}$ as the $6j$-symbol associated to the ordinary tetrahedron $\Delta$ from which it was obtained, 
\begin{align}
\hat{W}(\hat{\Delta}) := W(\Delta) \ . 
\end{align}
Now we can express the path integral entirely in terms of the manifold $M$ where gravity lives, with no reference the embedding manifold $M_E$. The fixed-angle and fixed-length amplitudes \eqref{zactv} and \eqref{ZLtri} are
\begin{align}\label{exactZAhat}
Z_A(M, \gamma(\vecP)) = \int_{\mathbb{R}_+^m}(\Pi_{i=1}^m d\tilde{P}_i \rho_0(\tilde{P}_i)) \prod_{\hat{\Delta} \in \hat{T}} \hat{W}(\hat{\Delta})
\end{align}
and
\begin{align}\label{exactZLhat}
Z_L(M, \gamma(\vecP'))
&= \int_{\mathbb{R}_+^{n+m}}d\vecP d\tilde{\vecP} (\Pi_{i=1}^n S_{P_i'P_i}) (\Pi_{i=1}^m \rho_0(\tilde{P}_i) )
\prod_{\hat{\Delta} \in \hat{T} } \hat{W}(\hat{\Delta})
\end{align}
This way of writing the amplitudes will make it easy to compare to the semiclassical limit, as we will see that each $6j$-symbol $\hat{W}(\hat{\Delta})$ agrees with the on-shell action of a generalized hyperbolic tetrahedron in the semiclassical limit.

The procedure to generate $\hat{T}$ from $T$ has a natural physical interpretation in the classical limit, in terms of on-shell hyperbolic geometries. Imagine that we start with a VTQFT amplitude for light defect operators, with $\sum \psi < \pi$ at every vertex. The corresponding on-shell geometry (if there is one) is simply $(M_E, \Gamma(\vecP))$ with a hyperbolic metric, and conical defects on the graph. Thus for light enough particles, gravity simply lives on the embedding manifold $M_E$. Since this manifold is closed, there does not appear to be any interpretation of this geometry in AdS/CFT. Now we gradually increase the conformal weights, i.e. the masses of the defect particles. When the total angle $\sum \psi$ at a vertex surpasses $\pi$, the vertex gets truncated by a geoedesic hyperplane, as described in more detail below. Therefore the first step in the procedure above, truncating all the vertices, corresponds to studying VTQFT in the regime $P_i \in i (0, \frac{b+b^{-1}}{4})$, where all states are conical defects and all vertices are truncated. (This is the regime considered in most of the paper \cite{Chandra:2022bqq}.) The second step, dualizing the edges, corresponds to analytically continuing the weights above the black hole threshold to the regime $P_i \in \mathbb{R}_+$.

\subsection{From Virasoro TQFT to geometry}\label{ss:vtqftGeom}

The procedure to turn a triangulation $T$ into a generalized triangulation $\hat{T}$ also gives a practical recipe to translate results from Virasoro TQFT to the metric formalism in pure gravity, where all weights are above the black hole threshold. The recipe is:
\begin{enumerate}
\item Triangulate the VTQFT graph, $(M_E, \Gamma(\vecP))$, with ordinary tetrahedra.
\item Follow the procedure above to find a generalized triangulation of $M$.
\item To solve the Einstein equations, interpret each $\hat{\Delta}$ as a generalized hyperbolic tetrahedron, and extremize over the internal edge lengths $\tilde{\ell}$.
\end{enumerate}
This procedure constructs the saddlepoint geometry corresponding to the VTQFT amplitude on $(M_E, \Gamma(\vecP))$, if there is one.

The first step --- triangulating the closed manifold $M_E$ with an embedded trivalent graph $\Gamma$ --- is probably not familiar to many physicists, but it is straightforward and completely algorithmic. For a pedagogical introduction, see  \cite{thurstonNotes} and \cite{MR4249621}. A comprehensive treatment of 3-manifolds on trivalent graph complements can be found in \cite{heard2005computation} and \cite{MR2676749}, where the authors also discuss the conditions for a trivalent graph to admit a hyperbolic metric, define an ordering of graphs by `complexity', appropriately defined, and analyze the first 129 graphs. These papers have, in effect, calculated many of the leading corrections to OPE statistics in 3D gravity. There is also a freely available software package, {\tt Orb} \cite{heard2005computation}, with a point-and-click interface to find triangulations of trivalent graphs embedded in general 3-manifolds, and to calculate the hyperbolic volumes of graphs in the conical defect regime numerically. A brief tutorial on using {\tt Orb} is included in appendix \ref{app:orb}.

\section{Generalized tetrahedra in hyperbolic space}\label{s:polyhedra}

Thus far, most of the results that we have discussed are exact. In the rest of the paper we focus on the semiclassical limit. The building blocks for classical geometries that are naturally suited to the formalism described above are generalized hyperbolic tetrahedra. 

Generalized polyhedra are polyhedra with (zero or more) truncated vertices. A truncated vertex is removed from the polyhedron and replaced by a face that meets all neighboring faces orthogonally. These are the on-shell geometries associated to the generalized tetrahedra $\hat{\Delta}$ described topologically in section \ref{ss:gentri}. 

In this section we review the properties of truncated polyhedra in hyperbolic space following \cite{thurstonNotes,MR1435975,ushijimaMR2191251} and calculate their classical gravitational action.

\subsection{Hyperplanes in embedding space}

Define the embedding space $\mathbb{R}^{n,1}$ with the metric
\begin{align}\label{ds2embed}
ds^2 = -X_0^2 + X_1^2 + \cdots + X_n^2 \ .
\end{align}
Hyperbolic space $H_n$ is the component of the hyperboloid $X^2 = -1$ with $X_0>0$.  De Sitter space $dS_{n}$ is the hyperboloid $X^2 = 1$. It is natural to view $H_n$ as a subset of $RP_n$, the space of lines through the origin in embedding space.  The upper half of the de Sitter hyperboloid, with $X_0>0$, is also a subset of $RP_n$. Globally we have
\begin{align}
RP_n \cong H_n \cup {\cal C} \cup (dS_n  / \mathbb{Z}_2)
\end{align}
where $\mathbb{Z}_2$ acts by $X \to -X$. The component ${\cal C}$, defined as the set of lines on the lightcone $X^2=0$, is the conformal boundary of hyperbolic space and the future and past conformal boundaries of de Sitter, which are all identified in $RP_n$. 

Define the Klein patch $K_n \subset RP_n$ as the subset of lines with non-zero slope. We have
\begin{align}\label{kleinPatch}
K_n \cong H_n \cup {\cal C} \cup dS_n^+
\end{align}
where $dS_n^+$ is the future half of de Sitter, with $X_0>0$. $K_n$ is canonically mapped to the hyperplane $X_0 =1$ by rescaling $X$. The full projective space is $RP_n = K_n \cup K_n^\infty$ with $K_n^\infty$ the sphere at infinity identified antipodally. When we discuss the geometry of the Klein patch, we use the metric inherited from \eqref{kleinPatch}, which has constant negative curvature on the AdS region and constant positive curvature on the de Sitter region.\footnote{The mathematics literature on the Klein model \cite{thurstonNotes} does not usually assign a metric to the outside region as we have done here, but see \cite{MR2994035}. } 

Given a point $X \in \mathbb{R}^{n,1}$, define the half-space $R_X$ and the hyperplane $X^\perp$ by
\begin{align}
R_X &= \{ Y| X\cdot Y \leq 0 \} , \qquad
X^\perp = \p R_X = \{ Y | X \cdot Y = 0\} \ . 
\end{align}
If $X^2>0$, then $X^\perp$ intersects $H_n$, and thus determines a geodesic hyperplane $\Pi_X = X^\perp \cap H_n$ in $H_n$. Any point with $X^2>0$ can be projected onto the de Sitter hyperboloid by rescaling, so points in de Sitter correspond to geodesic hyperplanes in hyperbolic space. The correspondence between points in $dS_n$ and half-spaces in $H_n$ bounded by geodesic hyperplanes is one-to-one.

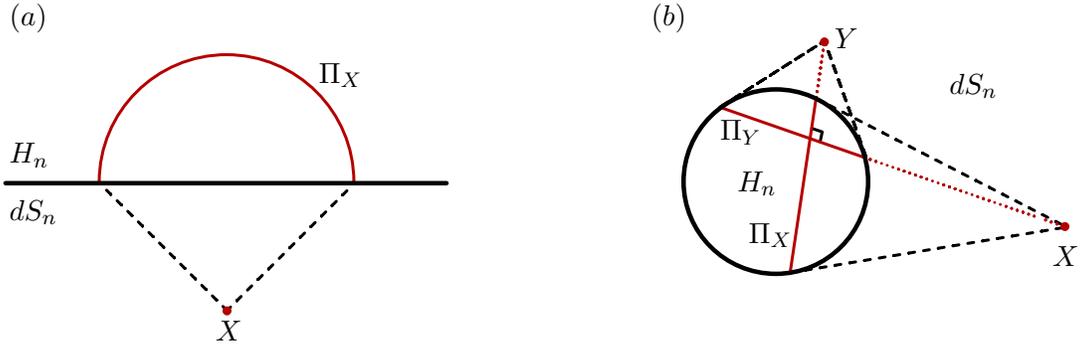
\begin{figure}
\begin{center}
\begin{overpic}[grid=false,width=6.0in]{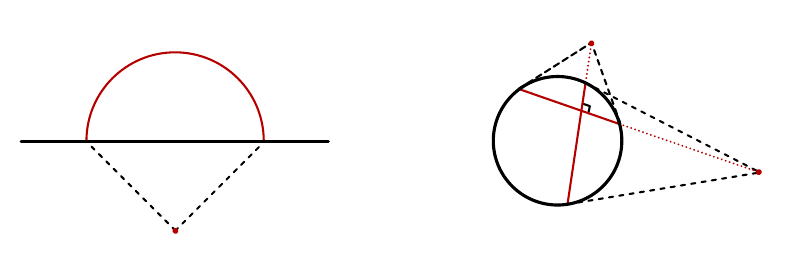}
\put (30,310) {$(a)$}
\put (30,190) {$H_n$}
\put (30,140) {$dS_n$}
\put (210,35) {$X$}
\put (300,260) {$\Pi_X$}
\put (590,310) {$(b)$}
\put (940,100) {$X$}
\put (750,290) {$Y$}
\put (675,120) {$\Pi_X$}
\put (650, 210) {$\Pi_Y$}
\put (665,165) {$H_n$}
\put (850,250) {$dS_n$}
\end{overpic}
\end{center}
\caption{The Klein patch $K_n = H_n \cup {\cal C} \cup dS_n^+$, in $(a)$ Poincare coordinates and $(b)$ Klein (projective) coordinates. Geodesic hyperplanes in hyperbolic space correspond to points in de Sitter. Dashed black lines are de Sitter lightcones and red lines are geodesic hyperplanes. If $X \in Y^\perp$, then $\Pi_X \perp \Pi_Y$ as shown in $(b)$.}\label{fig:hyperplanes}
\end{figure}

Given a point $X \in dS_n$ and $Y \in \Pi_X$, we have $(X-Y)^2 = -2X \cdot Y =0$, so $X$ and $Y$ are null separated. It follows that $\Pi_X$ and the de Sitter lightcone emanating from point $X$ meet the conformal boundary ${\cal C}$ on the same locus. This is illustrated in Poincar\'e coordinates in figure \ref{fig:hyperplanes}a.

Given two points $X,Y \in dS_n$ with $X \neq \pm Y$, the hyperplanes $\Pi_X$ and $\Pi_Y$ intersect in $H_n$ iff $|X \cdot Y | < 1$, with equality when they intersect at the conformal boundary. If they intersect, the hyperbolic angle $\psi$ between them measured in $R_X \cap R_Y \cap H_n$ is given by
\begin{align}\label{piAngle}
\cos \psi = - X \cdot Y \ . 
\end{align}
Therefore any hyperplane through the origin that contains $X$ is orthogonal to $\Pi_X$ in $H_n$. This is shown in figure \ref{fig:hyperplanes}b. 
If $\Pi_X$ and $\Pi_Y$ do not intersect, then the geodesic distance $\ell$ between them satisfies 
\begin{align}\label{piLen}
\cosh \ell = |X \cdot Y| \ . 
\end{align}
Comparing to \eqref{piAngle}, we see that $\psi \to \pi - i \ell$ or $\psi \to i \ell$ under analytic continuation when the intersection is moved outside of $H_n$.

A convenient coordinate system on the Klein patch $K_n$
is obtained by setting
\begin{align}\label{kleinembedding}
X = \frac{1}{\sqrt{|1-x\cdot x|}}(1,x)
\end{align}
with $x \in \mathbb{R}^n$ and $x\cdot y := \delta_{ij}x^i y^j$. For $x \cdot x < 1$, this solves the equation $X^2 = - 1$, so $x$ is a coordinate on $H_n$, while for $x\cdot x >1$ it solves $X^2 = 1$, so $x$ is a coordinate on a patch of $dS_n$. The shared conformal boundary ${\cal C}$ is the unit sphere, $x\cdot x = 1$. This is called the Klein model or projective model. The coordinate $x$ labels a point on the hyperplane $X_0 = 1$, which is mapped to $K_n$ by rescaling.

In the de Sitter region of the $x$-plane, time runs radially inward, and lightcones meet the unit sphere tangentially.   Given a point $x \in dS_n$, the corresponding hyperplane is $x^\perp = \{y|y\cdot x =1\}$, which is the equation for an ordinary Euclidean hyperplane in $\mathbb{R}^n$. Therefore by working in Klein coordinates, we can trivially lift Euclidean hyperplanes to geodesic hyperplanes in $K_n$. See figure \ref{fig:hyperplanes}b.

\subsection{Generalized polyhedra}\label{ss:genpoly}

Let $(v_1, \dots v_r)$ be the vertices of a polyhedron in $\mathbb{R}^n$ that contains the origin. By viewing the vertices as points in $K_n \cong H_n \cup {\cal C} \cup dS_n^+$ and faces as geodesic hypersurfaces in $K_n$, we obtain a polyhedron $\hat{P} \subset K_n$. Each vertex is either \textit{finite}, \textit{ideal}, or \textit{hyperideal}, depending on whether it lies in $H_n$, ${\cal C}$, or $dS_n$, respectively. 
We denote the vertices lifted to embedding space by $V_i = X(v_i)$ with $X(x)$ defined in \eqref{kleinembedding}. 

A \textit{generalized polyhedron} $P$ in $H_n$ is defined from $\hat{P}$ by truncating each hyperideal vertex $V$ at the geodesic hyperplane $\Pi_V$. That is, 
\begin{align}
P = H_n \cap \hat{P} \bigcap_{i \in I_h} R_{V_i}
\end{align}
with $I_h$ the set of hyperideal vertices. A truncation is called \textit{deep} if the truncated face intersects one or more other truncated faces in the interior of $H_n$, and \textit{mild} otherwise. By construction, a mildly truncated face $\Pi_V$ meets all neighboring faces orthogonally. From the remarks around \eqref{piAngle}, all truncations are mild if and only if
\begin{align}
| V_i \cdot V_j | > 1 , \qquad \forall \  i, j \in I_h, \quad i\neq j \ . 
\end{align}
 Some examples of generalized triangles in $H_2$ are as follows (in Poincar\'e disk coordinates):
\begin{align}\label{triangles}
\includegraphics[width=4in]{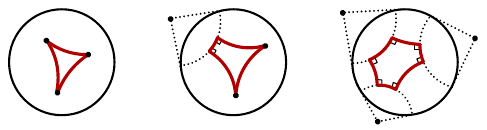}
\end{align}
These examples have zero, one, and three mildly truncated faces, respectively. 
The last example, with three truncations, is the all-right hexagon, which can be doubled along three non-adjacent edges to make a hyperbolic 3-holed sphere.

The description of a generalized polyhedron in terms of vertices $v_i$ is not unique. A generalized simplex with vertices $(V_1, \dots, V_{n+1})$ has a dual description $(W_1,\dots, W_{n+1})$ where $W_i \cdot V_j = \delta_{ij}$. For instance, the first triangle in \eqref{triangles} is drawn with three finite vertices, but it can also be realized by three deeply truncated hyperideal vertices.

The advantage of working with generalized polyhedra is that many of their properties are related by analytic continuation \cite{thurstonNotes}. As a simple example, in $H_2$, the hyperbolic law of cosines applies to both finite and generalized triangles under an appropriate continuation of dihedral angles to geodesic lengths of truncated edges, $\psi \to i \ell $. The continuation from light to heavy operators in $H_3$ is similar, as we will see.

\subsection{The truncated tetrahedron}
We now specialize to a generalized tetrahedron $T$ in $H_3$ with all four vertices mildly truncated.
The vertices $(V_1,\dots, V_4)$ are points in embedding space normalized to lie on the de Sitter hyperboloid, $V_i^2 = 1$. Labeling conventions for the edges and angles are shown in figure \ref{fig:tetrahedron} (before truncation). After truncation, $T$ has four triangular faces and four hexagonal faces, with the triangular faces meeting their neighbors orthogonally. Figure \ref{fig:trunctet} in the introduction shows an example, drawn in Klein coordinates. 

The tetrahedral symmetry acts on the edge tuple by column permutations, and by swapping any two entries in the top row with the corresponding entries in the bottom row, generated by
\begin{align}\label{tetGenerators}
\begin{pmatrix} 
e_1 & e_2 & e_3 \\ 
e_4 & e_5 & e_6 
\end{pmatrix}
 = 
 \begin{pmatrix}
 e_2 & e_1 & e_3 \\
 e_5 & e_4 & e_6 
 \end{pmatrix}
 =
 \begin{pmatrix}
 e_1 & e_3 & e_2 \\
 e_4 & e_6 & e_5 
 \end{pmatrix}
 =
\begin{pmatrix}
e_4 & e_5 & e_3 \\
e_1 & e_2 & e_6 
\end{pmatrix}\ . 
\end{align}
Swapping the two rows dualizes the tetrahedron, exchanging faces with vertices.  

\begin{figure}
\begin{center}
\begin{tikzpicture}[scale=5]
\coordinate (v3) at (0, 0.3); 
\coordinate (v2) at (0.75, 0); 
\coordinate (v1) at (1, 0.5); 
\coordinate (v4) at (0.5, 1); 

\draw[thick] (v3) -- (v2) node[midway, below] {$e_6$};
\draw[thick,dashed] (v1) -- (v3) node[pos=0.65,above] {$e_5$};
\filldraw[white,fill=white] (0.64,0.43) circle (0.035);
\draw[thick] (v3) -- (v4) node[midway,left] {$e_1$};
\draw[thick] (v2) -- (v1) node[midway,right] {$e_4$};
\draw[thick] (v2) -- (v4) node[pos=0.7,left] {$e_2$};
\draw[thick] (v1) -- (v4) node[midway,right] {$e_3$};

\node[left] at (v3) {$v_3$};
\node[below] at (v2) {$v_2$};
\node[right] at (v1) {$v_1$};
\node[above] at (v4) {$v_4$};

\node[right,text width=6.5cm] at (1.3,0.5) {\small
\textit{Dihedral angles}
\begin{align}\notag
\begin{pmatrix}\psi_1 & \psi_2 & \psi_3 \\
\psi_4 & \psi_5 &\psi_6 
\end{pmatrix}
&\equiv
\begin{pmatrix} 
\psi_{12} & \psi_{13} & \psi_{23} \\
\psi_{34} & \psi_{24} & \psi_{14}
\end{pmatrix}
\end{align}
\vspace{0.3cm}

\textit{Lengths}
\begin{align}\notag
\begin{pmatrix}\ell_1 & \ell_2 & \ell_3 \\
\ell_4 & \ell_5 &\ell_6 
\end{pmatrix}
&\equiv
\begin{pmatrix} 
\ell_{34} & \ell_{24} & \ell_{14} \\
\ell_{12} & \ell_{13} & \ell_{23}
\end{pmatrix}
\end{align}
};
\end{tikzpicture}
\end{center}
\caption{\small Tetrahedron labeling conventions. 
 \label{fig:tetrahedron}}
\end{figure}

The vertices $(V_1, V_2, V_3, V_4)$ are a basis for $\mathbb{R}^{n,1}$. The dual basis $V_i^*$ defined by $V_i^* \cdot V_j = \delta_{ij}$ has $(V_i^*)^2 > 0$, so we can rescale the dual vectors to define $U_i = V_i^* / \sqrt{(V_i^*)^2}$ which are on the de Sitter hyperboloid. The hexagonal faces of $T$ lie on the hyperplanes $\Pi_{U_i}$. Define the \textit{Gram matrix} $G_{ij}$, which encodes the dihedral angles $\psi_i$, 
\begin{align}\label{angleGram}
G_{ij} = U_i \cdot U_j = - \cos \psi_{ij} , \quad 
G = \begin{pmatrix}
1 & -\cos \psi_1 & -\cos \psi_2 & -\cos \psi_6\\
-\cos \psi_1 & 1 & -\cos \psi_3 & -\cos \psi_5 \\
-\cos \psi_2 & -\cos \psi_3 & 1 & -\cos \psi_4 \\
-\cos \psi_6 & -\cos \psi_5 & -\cos \psi_4 & 1 
\end{pmatrix} \ .
\end{align}
Let $M$ be the diagonal matrix
\begin{align}
M_{ij} = V_i \cdot U_j = \frac{1}{\sqrt{G_{ii}^{-1}}}\delta_{ij} \ . 
\end{align}
We have the relations
\begin{align}\label{UVrel}
V = M G^{-1} U , \qquad U = G M^{-1} V \ . 
\end{align}
The inner products of the vertices define the \textit{vertex Gram matrix} (or dual Gram matrix), which encodes the edge lengths $\ell_i$, 
\newcommand{\GL}{\widetilde{G}}
\begin{align}\label{lengthGram}
\GL_{ij} = V_i \cdot V_j   = -\cosh \ell_{ij}  \ , \qquad
\GL &= \begin{pmatrix}
1 & -\cosh \ell_4 & -\cosh \ell_5 & -\cosh \ell_3 \\
-\cosh \ell_4 & 1 & -\cosh \ell_6 & -\cosh \ell_2 \\
-\cosh \ell_5 & -\cosh \ell_6 & 1 & -\cosh \ell_1 \\
-\cosh \ell_3 & -\cosh \ell_2 & -\cosh \ell_1 & 1
\end{pmatrix} \ . 
\end{align}
The Gram matrix and its dual are related by $\GL = M G^{-1} M$. Using $M_{ii} = \frac{1}{\sqrt{G_{ii}^{-1}}} =  \frac{1}{\sqrt{\GL_{ii}^{-1}}}$ this gives formulae for the edge lengths as a function of dihedral angles,
\begin{align}
\cosh \ell_{ij} = - \ \frac{G^{-1}_{ij}}{\sqrt{G^{-1}_{ii} G^{-1}_{jj}}} \ ,
\end{align}
and the dihedral angles as a function of edge lengths,
\begin{align}\label{anglesFromLengths}
\cos \psi_{ij} = - \ \frac{\GL^{-1}_{ij}}{ \sqrt{ \GL^{-1}_{ii} \GL^{-1}_{jj} }} \ . 
\end{align}
%
%
%
%
Given an allowed set of dihedral angles or edge lengths, one can reconstruct the tetrahedron from \eqref{angleGram}-\eqref{UVrel} or \eqref{lengthGram} respectively. Thus the set of six lengths or six angles determines the tetrahedron uniquely, if it exists. 

Necessary and sufficient conditions for a generalized tetrahedron to exist, in terms of angles, can be found in \cite{ushijimaMR2191251}. 
We will need the following characterization of allowed edge lengths:
A generalized tetrahedron with four mildly truncated hyperideal vertices and edge lengths $(\ell_1,\dots,\ell_6)$ exists if and only if the vertex Gram matrix $\GL$ defined by the last equality in \eqref{lengthGram} has $\det \GL < 0$. \textit{Proof:}  By Descartes' rule, the matrix $\GL$ has at most two negative eigenvalues, so $\det \GL< 0$ is equivalent to $\GL$ having signature $(-1,1,1,1)$. If $\mbox{sig}(G) = (-1,1,1,1)$ then we can find vertices $V_i$ with $\GL_{ij} = V_i \cdot V_j$
and $V_i^0 > 0$. The diagonal elements of \eqref{lengthGram} are $V_i^2=1$ so all four vertices are hyperideal,  and the off-diagonals are negative $V_i \cdot V_j < 0$ for $i \neq j$, so all truncations are mild. Conversely, if $\mbox{sig}(G) \neq (-1,1,1,1)$ then there are no $V_i$ with $V_i^2=1$ that span $\mathbb{R}^{n,1}$ and solve $\GL_{ij} = V_i \cdot V_j$.

\subsubsection{Volume}\label{sss:volume}
The problem of finding an explicit formula for the volume of a finite hyperbolic tetrahedron is quite difficult. It was solved by Cho and Kim in 1999 \cite{MR1706606}, recast in a form respecting the tetrahedral symmetry by Murakami and Yano \cite{MR2154824}, and extended to mildly truncated generalized tetrahedra by Ushijima \cite{ushijimaMR2191251}.

\newcommand{\boldpsi}{\boldsymbol\psi}
\newcommand{\boldell}{\boldsymbol\ell}

Denote the volume of a generalized hyperbolic tetrahedron $T$ as a function of dihedral angles  by
\begin{align}
\mbox{Vol}(T) = V_A(\psi) , \qquad \psi := \begin{psmallmatrix} \psi_1 & \psi_2 & \psi_3 \\ \psi_4 & \psi_5 & \psi_6 \end{psmallmatrix}  \ ,
\end{align}
and as a function of edge lengths,
\begin{align}
\mbox{Vol}(T) =  V_L(\ell) , \qquad \ell :=  \begin{psmallmatrix} \ell_1 & \ell_2 & \ell_3 \\ \ell_4 & \ell_5 & \ell_6 \end{psmallmatrix}  \ . 
\end{align}
Following \cite{ushijimaMR2191251}, the key step in the derivation of the volume formula is to use the Schl\"afli identity for the derivative with respective to dihedral angles:
\begin{align}
\frac{\p}{\p \psi_i} V_A(\psi) = - \frac{1}{2} \ell_i \ . 
\end{align}
The volume as a function of edge lengths satisfies the corresponding identity
\begin{align}\label{schlafliL}
\frac{\p}{\p \ell_i} \left( V_L(\ell) + \frac{1}{2}\sum_{j=1}^6 \ell_j \psi_j \right)  = \frac{1}{2}\psi_i \ . 
\end{align}
An explicit formula for $V_A(\psi)$ in terms of dilogarithms appears in \cite{MR2154824, ushijimaMR2191251}. We will not use it or repeat it here; a different but equal formula in terms of edge lengths is derived below. For a finite tetrahedron, a formula for $V_L(\ell)$ was derived in  \cite{MR2193233} and can be generalized to allow for mild truncations with some minor modifications. 

Given a candidate formula for the volume function, it suffices to check the Schl\"afli identity and a special case, such as a degenerate limit or ideal tetrahedron, to fix the constant of integration. This is how the volume formulae are proved in \cite{ushijimaMR2191251} and below. For reference we record Milnor's formula for the volume of an ideal tetrahedron, which is specified by two independent dihedral angles $(\alpha,\beta)$:
\begin{align}
V_{\rm ideal}(\alpha,\beta) = V_A\begin{psmallmatrix} \alpha & \beta & \pi - \alpha - \beta \\ \alpha & \beta & \pi - \alpha - \beta \end{psmallmatrix}
= \Lambda(\alpha) + \Lambda(\beta) + \Lambda(\pi - \alpha - \beta)
\end{align}
with $\Lambda(x) = -\int_0^x dy \log|2 \sin y|$ the Lobachevsky function.

\subsection{Gravitational action}\label{ss:tetaction}

Truncated tetrahedra are the natural building blocks for hyperbolic 3-manifolds when the path integral is evaluated using Virasoro TQFT or Conformal Turaev-Viro theory, as discussed in section \ref{s:pathtri}. With fixed-angle boundary conditions, the on-shell action \eqref{IAonshell} of a single truncated tetrahedron is simply
\begin{align}
I_A^{tet}(\psi) = \frac{c}{6\pi}V_A(\psi) \ . 
\end{align}
With fixed-length boundary conditions, the on-shell action was given in \eqref{ILonshell} and generally includes a contribution $\ell_i \psi_i$ at every edge. However, it is convenient to modify the definition slightly for a truncated tetrahedron, and leave out the Hayward terms for the edges coming from truncation. The dihedral angles at truncated faces are always $\frac{\pi}{2}$, so this modification is a choice of normalization. Thus we define
\begin{align}\label{defILtet}
I_L^{tet}(\ell) = \frac{c}{6\pi}\left( V_L(\ell) + \frac{1}{2}\sum_{i=1}^6 \ell_i \psi_i \right) \ . 
\end{align}
The advantage of this definition is that the action is additive when tetrahedra are glued on the truncated faces.

\renewcommand{\cp}[1]{\vcenter{\hbox{#1}}}

\section{Semiclassical triangulations}\label{s:classicaltri}

In section \ref{s:pathtri},  the exact fixed-length and fixed-angle gravitational path integrals were expressed in terms of triangulations using Conformal Turaev-Viro theory. We will now discuss the semiclassical limit, always assuming all external weights are above the black hole threshold, as is the case in pure gravity. 

In section \ref{s:pathtri} we showed that the calculation can be phrased in terms of a triangulation $T$ of the VTQFT graph $(M_E, \Gamma)$ by ordinary tetrahedra $\Delta$, or in terms of a generalized triangulation $\hat{T}$ of the gravity manifold $M = M_E - N(\Gamma)$ by generalized tetrahedra $\hat{\Delta}$. Throughout this section, we will use the language of the generalized triangulation $\hat{T}$.

\subsection{Without internal edges}
Let us first consider the case where the triangulation has no internal edges. The exact fixed-angle path integral on $M$ \eqref{exactZAhat} reduces to 
\begin{align}\label{zanoint}
Z_A(M, \gamma(\vecP)) &= \prod_{\hat{\Delta} \in \hat{T}} W(\hat{\Delta}) 
\end{align}
where $\hat{T}$ is a generalized triangulation of $M$.
To discuss the semiclassical limit, parameterize the weights in terms of angles by $\psi_i = 2\pi i b P_i$. 
The semiclassical limit is $b \to 0$ with $\psi_i$ held fixed. In this limit, the classical solution is a collection of truncated tetrahedra glued together on their truncated faces. The Virasoro $6j$-symbol can be expressed in terms of the volume of a generalized hyperbolic tetrahedron \cite{Teschner:2012em},
\begin{align}\label{sixjAA}
\begin{Bmatrix}P_4 & P_5 & P_6 \\ P_1 & P_2 & P_3 \end{Bmatrix}
\approx 
\exp\left[ - \frac{c}{6\pi} V_A
\begin{pmatrix} \psi_4 & \psi_5 & \psi_6 \\
\psi_1 & \psi_2 &  \psi_3 \end{pmatrix}
\right]
\end{align}
as reviewed in appendix \ref{app:kernels}. The right-hand side is $\exp[-I_A^{tet}(\hat{\Delta})]$. When truncated tetrahedra are glued together on their truncated faces, the action is additive. Therefore the on-shell action of the full manifold $M$ is
\begin{align}
I_A^{\rm on-shell}(M) = \sum_{\hat{\Delta} \in \hat{T}} I_A^{tet}(\hat{\Delta}) \ ,
\end{align}
and the semiclassical limit of \eqref{zanoint} is
\begin{align}
Z_A(M, \gamma(\vecP)) \approx e^{-I_A^{\rm on-shell}(M)} \ . 
\end{align}
This is of course the expected result from the metric point of view. The conclusion is that in the semiclassical limit, the CTV partition function is \textit{manifestly} equal to the gravity path integral with fixed-angle boundary conditions. Each generalized tetrahedron in the CTV partition function is a  hyperbolic generalized tetrahedron in the geometry, and the weight $W$ agrees with the classical action.

To transform to fixed-length boundary conditions, we use \eqref{ZLtri}. With no internal edges ($m=0$) this formula in the semiclassical limit reads
\begin{align}\label{zlgga}
Z_L(M, \gamma(\vecP')) \approx \int d^n \psi \exp\left[ -\frac{c}{12\pi}\sum_{i=1}^n \psi_i \ell_i'  - \frac{c}{6\pi} \sum_{\hat{\Delta}\in \hat{T}}V_A(\psi^{(\hat{\Delta})}) \right]
\end{align}
where $\psi_j^{(\hat{\Delta})}$ and $\ell_j^{(\hat{\Delta})}$ are the dihedral angles and lengths of tetrahedron $\hat{\Delta}$, and we have parameterized $\ell_i' = 4\pi b P_i'$.  
Let $E(i; \hat{\Delta})$ be the set of edges in tetrahedron $\hat{\Delta}$ with angle $\psi_i$. Using the Schl\"afli formula, the saddlepoint equation from extremizing with respect to $\psi_i$ is
\begin{align}
\ell_i' = \sum_{\hat{\Delta}\in \hat{T}} \sum_{j \in E(i;\hat{\Delta})} \ell_j^{(\hat{\Delta})}
\end{align}
That is,  the total length of the closed cycle with angle $\psi_i$ is set equal to $\ell_i'$. Evaluating the right-hand side of \eqref{zlgga} at the extremum gives
\begin{align}\label{ZLnoInt}
Z_L(M, \gamma(\vecP'))
 = |\Zvir(M_E, \Gamma(\vecP'))^2
\approx \exp\left( - I_L^{\rm on-shell}(M) \right)
\end{align}
where
\begin{align}
I_L^{\rm on-shell}(M) = 
\sum_{\hat{\Delta}\in\hat{T}}I_L^{\rm tet}(\hat{\Delta})
\end{align}
with $I_L^{tet}$ defined in \eqref{defILtet}. The conclusion is that the exact fixed-length path integral  --- and therefore the OPE statistics, and Virasoro TQFT amplitude $|\Zvir|^2$ --- also manifestly agree with the classical action of gravity, when expressed in terms of the CTV partition function.

This result also gives a practical way to find the classical geometry associated to any Virasoro TQFT calculation on a hyperbolic topology. First we triangulate the graph $(M_E, \Gamma)$ using the method described in section \ref{ss:vtqftGeom} (or, sidestep the triangulation by calculating the chain-mail invariant \cite{ctv}). Then, follow the procedure in section \ref{ss:gentri} to convert the triangulation to a generalized triangulation, transforming each topological tetrahedron $\Delta$ into a generalized hyperbolic tetrahedron $\hat{\Delta}$. The face-pairing of the triangulation dictates how the truncated faces of the $\hat{\Delta}$'s are glued together to construct the semiclassical geometry. We will work out two examples in sections \ref{s:modulars}-\ref{s:sixj}. 

\subsection{With internal edges}
In the presence of internal edges, the CTV partition function has tetrahedra with two types of labels on the edges: external weights $\vecP$ corresponding to the fixed-angle boundary conditions, and internal weights $\tilde{\vecP}$ which are integrated with the measure $\rho_0(\tilde{P}_i)$. In the semiclassical limit, with boundary conditions corresponding to real angles, the saddles in $Z_A$ have imaginary $P_i$ and real $\tilde{P}_i$. Thus the $P_i$ naturally correspond to angles and the $\tilde{P}_i$ to lengths. It is cumbersome to treat them differently, so instead, we will parameterize all the weights by lengths
\begin{align}
\ell_i = 4\pi b P_i, \qquad \tilde{\ell}_i = 4\pi b \tilde{P}_i
\end{align}
with $\ell_i$ purely imaginary. When all six weights are parameterized in terms of lengths, the semiclassical $6j$-symbol, in the range of parameter space where all the hyperbolic tetrahedra are non-degenerate,\footnote{This is the answer in the range of $P_i$ where the generalized hyperbolic tetrahedron exists and has finite volume. There are degenerate regimes where the answer simplifies.  See appendix \ref{app:kernels} for details.} is
\begin{align}\label{sixjRR}
\begin{Bmatrix}P_4 & P_5 & P_6 \\ P_1 & P_2 & P_3 \end{Bmatrix}
\approx 
\exp\left[ - I_L^{tet}
\begin{pmatrix} \half \ell_1 & \half \ell_2 & \half \ell_3 \\
\half \ell_4 & \half \ell_5 & \half \ell_6 \end{pmatrix}
\right]
\end{align}
where $I_L^{tet}$ was defined in \eqref{defILtet}. 
This is derived in appendix \ref{app:kernels}. 
The CTV partition function \eqref{exactZAhat} in the semiclassical limit gives the fixed-angle amplitude as
\begin{align}\label{zauu}
Z_A(M, \gamma(\vecP)) \approx \int d^m \tilde{\ell} \exp\left[ \frac{c}{12} \sum_{i=1}^m \tilde{\ell}_i  - \frac{c}{6\pi}\sum_{\hat{\Delta}\in\hat{T}}\left(V_L(\ell^{(\hatDelta)})  + \frac{1}{2}\sum_{i=1}^6 \ell_i^{(\hatDelta)}\psi_i^{(\hatDelta)} \right)\right]
\end{align}
Let $\tilde{E}(i, \hat{\Delta})$ be the set of edges in tetrahedron $\hat{\Delta}$ with length $\frac{1}{2}\tilde{\ell}_i$. The saddlepoint condition from extremizing with respect to $\tilde{\ell}_i$ is
\begin{align}\label{sadpp}
\sum_{\hat{\Delta}\in\hat{T}}\sum_{j \in \tilde{E}(i,\hatDelta)} \psi_i^{(\hatDelta)} = 2\pi \ . 
\end{align}
That is, the total angle around each internal edge is equal to $2\pi$. At the extremum, the first and last terms in \eqref{zauu} cancel. Therefore the semiclassical fixed-angle path integral is
\begin{align}
Z_A(M, \gamma(\vecP)) \approx \exp\left[ - \frac{c}{6\pi} \mbox{Vol}(M) \right] \
\end{align}
where the volume is calculated on the solution to \eqref{sadpp}. 

The transformation to fixed-angle boundary conditions is straightforward, following the same steps that led to \eqref{ZLnoInt} above. The final result is the same,
\begin{align}
Z_L(M, \gamma(\vecP')) = |\Zvir(M_E, \Gamma(\vecP')|^2 \approx \exp\left[ - I_L^{\rm on-shell}(M)\right]
\end{align}
with the lengths of the external edges fixed by the boundary condition $\vecP'$, and the lengths of the internal edges determined by the smoothness condition \eqref{sadpp}. Equivalently, since the smoothness conditions also follows from extremizing the internal lengths, the classical action can be calculated from
\begin{align}\label{tildeIcl}
I_L^{\rm on-shell}(M) = 
\mathop{\mathrm{ext}}\limits_{\tilde{\ell}} \sum_{\hatDelta \in \hat{T}}\tilde{I}_L(\hatDelta)
\end{align}
where  in this expression each $\hatDelta$ is a generalized tetrahedron parameterized by edge lengths, and  $\tilde{I}_L(\hatDelta)$ is a function of $(\ell, \tilde{\ell})$ defined as
\begin{align}
\tilde{I}_L(\hat{\Delta}) = \frac{c}{6\pi} \left( \mbox{Vol}(\hat{\Delta}) + \frac{1}{2}\sum_{i \rm{\ external}} \psi_i^{(\hatDelta)} \ell_i^{(\hatDelta)} \right) \ . 
\end{align}
The on-shell action in the form \eqref{tildeIcl} follows from the CTV partition function by doing first the $\ell$ integral, then the $\tilde{\ell}$ integral, by saddlepoints.

\section{Geometry of the modular $S$-matrix}\label{s:modulars}
We will now apply the techniques developed in this paper to two examples: The three-dimensional geometries dual to the Virasoro crossing kernels, the $6j$-symbol and the modular $S$-matrix. The exact CTV partition functions for these examples are worked out in \cite{ctv}. In what follows, we will summarize the exact results briefly but the focus is on the semiclassical geometries. In each example, we compare the fixed-length path integral to Virasoro TQFT amplitude-squared, and the fixed-angle path integral to the CTV partition function.

\subsection{Fixed lengths}

Consider the 2-point OPE statistics, $\overline{C_{ijk}C_{lmn}}$. As in section \ref{s:dictionary}, upper-case $C_{ijk}$ is the OPE coefficient normalized to have unit variance in the Gaussian approximation to the CFT ensemble. According to the dictionary \eqref{dictionary}, the statistics of scalar primaries are calculated by a formal sum over topologies,
\begin{align}
\overline{C_{ijk}C_{lmn}} = \sum_{\Gamma} \delta^{\Gamma}_{ijklmn} \sum_{(M,\gamma)}Z_L(M, \gamma(\vecP))
\end{align}
where $Z_L$ is the fixed-length gravitational path integral on a compact spacetime, $M$.
The sum over trivalent graphs $\Gamma$ with two vertices has 15 terms, corresponding to contractions of indices on the OPE coefficients. They come in two types, theta graphs and barbell graphs:
\begin{align}\label{tpgraphs}
\cp{ 
\begin{tikzpicture}[scale=1]
\draw (0,0) circle [radius=0.5];
\draw (-0.5,0) -- (0.5,0);
\node at (-0.5,0.4) {\footnotesize $i$};
\node at (-0.2,0.15) {\footnotesize $j$};
\node at (-0.5,-0.4) {\footnotesize $k$};
\node at (0.5,0.4) {\footnotesize $l$};
\node at (0.2,0.15) {\footnotesize $m$};
\node at (0.5,-0.4) {\footnotesize $n$};
\end{tikzpicture}
}
\to \delta^\Gamma= \delta_{il}\delta_{jm}\delta_{kn}
  , \qquad
\cp{
\begin{tikzpicture}[scale=1]
\draw (-1,0) circle [radius=0.5];
\draw (1,0) circle [radius=0.5];
\draw (-0.5,0) -- (0.5,0);
\node at (-0.6,0.5) {\footnotesize $i$};
\node at (-0.35,-0.2) {\footnotesize $j$};
\node at (-0.7,-0.6) {\footnotesize $k$};
\node at (0.6,0.5) {\footnotesize $l$};
\node at (0.3,-0.2) {\footnotesize $m$};
\node at (0.7,-0.6) {\footnotesize $n$};
\end{tikzpicture}
}
\to \delta^\Gamma= 2\delta_{ik}\delta_{jm}\delta_{ln} \ . 
\end{align}
There are in total 6 permutations of the theta graph plus 9 permutations of the barbell graph. The factor of 2 in the tensor structure for the barbell graph is the number of inequivalent ways of the gluing the core to $C_0$ flares. Note that these graphs are purely combinatorial objects to keep track of the index structure; they are not embedded into a 3-manifold like the graphs of Virasoro TQFT, but as abstract graphs they are the same. 

For the theta graph, the leading term in the sum over topologies is the $C_0$ wormhole \cite{Chandra:2022bqq,Abajian:2023bqv}. The compact core has zero volume and we confirmed in section \ref{ss:c0core} that it reproduces the expected leading-order statistics.

The barbell graph contributes to $\overline{C_{113}C_{223}}$. We assume 1,2,3 label distinct scalar primary operators. The dictionary \eqref{dictionary} states that there is a contribution
\begin{align}\label{barbellD}
\overline{C_{113}C_{223}}  \quad \supset  \quad 2\,   Z_L(M, 
\cp{
\begin{tikzpicture}[scale=0.5]
\draw (-1,0) circle [radius=0.5];
\draw (1,0) circle [radius=0.5];
\draw (-0.5,0) -- (0.5,0);
\node at (-1,-1) {\footnotesize $\ell_1$};
\node at (1,-1) {\footnotesize $\ell_2$};
\node at (0,0.5) {\footnotesize $\ell_3$};
\end{tikzpicture}
}
)
\end{align}
where $\ell_i =4\pi b P_i$. The boundary condition in \eqref{barbellD} indicates that we perform the path integral with $\p M$ given by a genus-2 Riemann surface with three fixed lengths,
\begin{align}\label{bba}
\p M \quad = \quad 
\cp{
\begin{overpic}[grid=false,width=1.5in]{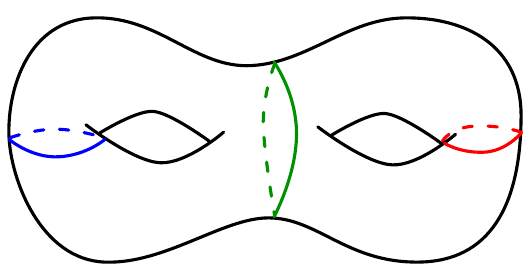}
\put (100,300) {$\ell_1$}
\put (900,300) {$\ell_2$}
\put (460,250) {$\ell_3$}
\end{overpic}
} \qquad \ ,
\end{align}
and $K_{ij}=0$ away from the marked geodesics. In principle there can be a contribution from any topology $M$ that satisfies this boundary condition. If we fill in a handlebody as drawn in \eqref{bba}, then there is no hyperbolic solution. There is a hyperbolic solution on a different topology given by a truncated tetrahedron glued to itself along its truncated faces. The semiclassical geometry is
\begin{align}\label{MclS}
M_{cl} \qquad =  \qquad
\vcenter{\hbox{
\begin{tikzpicture}[scale=0.035]
%
\coordinate (a1) at (10,53);
\coordinate (a2) at (9, 27);
\coordinate (a3) at (18, 40);
%
\draw[thick] (a1) -- (a2) node[midway,left] {$A$};
\draw[thick,dotted] (a2) -- (a3);
\draw[thick,dotted] (a3) -- (a1);
%
\coordinate (b1) at (59, 19);
\coordinate (b2) at (45, 5);
\coordinate (b3) at (73, 7);
%
\draw[thick] (b1) -- (b2);
\draw[thick] (b2) -- (b3) node[midway,below] {$B$};
\draw[thick] (b3) -- (b1);
%
\coordinate (c1) at (92,60);
\coordinate (c2) at (86,45);
\coordinate (c3) at (94,34);
%
\draw[thick,dotted] (c1) -- (c2);
\draw[thick,dotted] (c2) -- (c3);
\draw[thick] (c3) -- (c1) node[midway,right] {$A$};
%
\coordinate (d1) at (35,88);
\coordinate (d2) at (50,82);
\coordinate (d3) at (59,90);
%
\draw[thick] (d1) -- (d2);
\draw[thick] (d2) -- (d3);
\draw[thick] (d3) -- (d1) node[midway,above] {$B$};
%
\coordinate (upperLeftE) at (33,61);
\coordinate (lowerLeftE) at (35,26);
\coordinate (lowerRightE) at (71, 30);
\coordinate (upperRightE) at (67,64);
\coordinate (frontE) at (51,50);
\coordinate (backE) at (54,40);
%
\node[xshift=-5,yshift=5] at (upperLeftE) {\small $\ell_3$};
\node[xshift=5,yshift=7] at (frontE) {\small $\ell_1$};
\node[gray,xshift=-13,yshift=5] at (backE) {\small $\ell_2$};
%
\draw[thick,red,dashed] plot [smooth,tension=1] coordinates {(a3) (backE) (c2)};
\draw[thick,darkgreen] plot [smooth,tension=1] coordinates {(a1) (upperLeftE) (d1)};
\draw[thick,darkgreen] plot [smooth,tension=1] coordinates {(a2) (lowerLeftE) (b2)};
\draw[thick,darkgreen] plot [smooth,tension=1] coordinates {(b3) (lowerRightE) (c3)};
\draw[thick,darkgreen] plot [smooth,tension=1] coordinates {(c1) (upperRightE) (d3)};
\draw[thick,blue] plot [smooth,tension=1] coordinates {(b1) (frontE) (d2)};
\end{tikzpicture}
}}
\end{align}
where $\ell_i = 4\pi b P_i$ is the total length of each closed curve (so each green segment has length $\ell_3/4$), and the triangular faces are glued $A \leftrightarrow A$ and $B \leftrightarrow B$. The same geometry was drawn differently in \eqref{heegS}. The generalized triangulation of $M_{cl}$ therefore consists of a single truncated hyperbolic tetrahedron $\hat{\Delta}$, with edge lengths
\begin{align}
\begin{pmatrix} \frac{1}{4}\ell_3 & \frac{1}{4}\ell_3 & \ell_1 \\
\frac{1}{4}\ell_3 & \frac{1}{4}\ell_3 & \ell_2 
\end{pmatrix} \ . 
\end{align}
At the level of the classical action, the contribution to the gravitational path integral is
\begin{align}\label{smatrixZL}
Z_L(M, \gamma(\vecP)) &\approx \exp\left[-I_L^{tet}
\begin{pmatrix} \frac{1}{4}\ell_3 & \frac{1}{4}\ell_3 & \ell_1 \\
\frac{1}{4}\ell_3 & \frac{1}{4}\ell_3 & \ell_2 
\end{pmatrix}
\right]\\
&\approx \exp\left[-\frac{c}{6\pi}\left(
V_L\begin{pmatrix} \frac{1}{4}\ell_3 & \frac{1}{4}\ell_3 & \ell_1 \\
\frac{1}{4}\ell_3 & \frac{1}{4}\ell_3 & \ell_2 
\end{pmatrix}
+ \frac{1}{2}\sum_{i=1}^3\ell_i \psi_i 
\right)
\right]
\end{align}
Now let us compare to Virasoro TQFT and the dual CFT ensemble. In Virasoro TQFT, the corresponding diagram is 
\begin{align}\label{handcuffjj}
\vcenter{\hbox{\footnotesize
\begin{tikzpicture}[scale=0.8]
\centerarc[very thick](-0.5,0)(65:410:1); 
\centerarc[very thick](0.5,0)(-115:233:1);
\draw[very thick] (-0.5,0) -- (0.5,0);
\node at (-1.3,1) {$1$};
\node at (1.3,1) {$2$};
\node at (0,.35) {$3$};
\end{tikzpicture}
}} \  .
\end{align}
The VTQFT amplitude for this diagram embedded in $S^3$ evaluates to $|\hatS_{P_1P_2}[P_3]|^2$ where $\hatS$ is the modular S-matrix in Racah-Wigner normalization \cite{Collier:2024mgv}. (See \cite[Appendix A]{ctv} for our VTQFT conventions and a review of this calculation.)  The same prediction for the OPE statistics can be derived from the conformal bootstrap in the dual CFT, as reviewed in appendix \ref{app:cftstatistics}. In appendix \ref{app:kernels} we analyze the $S$-matrix in the semiclassical limit and show that it satisfies
\begin{align}
\log |\hatS_{P_1P_2}[P_3]|^2 \sim  
-I_L^{tet}
\begin{pmatrix} \frac{1}{4}\ell_3 & \frac{1}{4}\ell_3 & \ell_1 \\
\frac{1}{4}\ell_3 & \frac{1}{4}\ell_3 & \ell_2  
\end{pmatrix} \ ,
\end{align}
in agreement with \eqref{smatrixZL}. Thus we have verified the match between gravity, Virasoro TQFT, and the dual CFT, 
\begin{align}
Z_L(M, \gamma(\vecP))  =  2|\hatS_{P_1P_2}[P_3]|^2 \ ,
\end{align}
at the level of the classical action.


\subsection{Fixed angles}
The analysis of the semiclassical fixed-angle action is even simpler. Recall that the fixed-angle gravitational path integral is exactly equal to the CTV partition function. In the present case, the triangulation of the graph \eqref{handcuffjj} in $S^3$ has a single tetrahedron, so the exact path integral is
\begin{align}
Z_A(M, \gamma(\vecP)) &= \Ztv(S^3, \Gamma(\vecP)) = \begin{Bmatrix} P_3 & P_3 & P_1 \\ P_3 & P_3 & P_2 \end{Bmatrix} \ . 
\end{align}
Semiclassically, using the results of \cite{Teschner:2012em} reproduced in \eqref{tvsixj} below, this predicts
\begin{align}
Z_A(M, \gamma(\vecP))
\approx \exp\left[
-\frac{c}{6\pi}V_A\begin{pmatrix}
\psi_3 & \psi_3 & \psi_1 \\
\psi_3 & \psi_3 & \psi_2
\end{pmatrix}
\right]
\end{align}
where $\psi_i = 2\pi i b P_i$. This agrees with the on-shell action of the truncated tetrahedron \eqref{MclS} with fixed-angle boundary conditions. 

The relation between the fixed-angle result and the fixed-length result is of course the Laplace transform. Underlying this relation is the following exact identity for the Virasoro modular $S$-matrix, derived in \cite{ctv}:
\begin{align}\label{introTransformS}
\begin{Bmatrix}
P_3' & P_3' & P_1' \\
P_3' & P_3' & P_2' 
\end{Bmatrix}
&=
\int_{0}^{\infty} ( \Pi_{i=1}^3 dP_i \, S_{P_i' P_i} )
\left| \hatS_{P_1P_2}[P_3] \right|^2 \ .
\end{align}

\subsection{From Virasoro TQFT to geometry}\label{ss:gentriS}
The triangulation of $(S^3, \Gamma)$ is not the same as the generalized triangulation of the gravity manifold, $M$.\footnote{In this example, it does little harm to confuse the two, because 
$
\begin{Bsmallmatrix}
P_3 & P_3 & P_1\\
P_3 & P_3 & P_2
\end{Bsmallmatrix}
=
\begin{Bsmallmatrix}
P_3 & P_3 & P_2\\
P_3 & P_3 & P_1
\end{Bsmallmatrix}
$
by tetrahedral symmetry. This is an accident of this example, and more generally the distinction is important.
}
 Let us explain how they are related, and in the process, give an example of the general procedure discussed in section \ref{ss:gentri} to convert a triangulation of the VTQFT graph $(M_E, \Gamma)$ to a generalized triangulation of $M = M_E - N(\Gamma)$. 

We are interested in the case of heavy operators $P_i \in \mathbb{R}$, but the procedure starts with conical defect operators, which have $h_i \in (0, \frac{c-1}{24})$ and $P_i$ purely imaginary. In the semiclassical limit, the relation between the conformal weight $h$ and the total angle $\theta$ around the defect is $\theta = 2\pi(1-2\eta)$, with $h = \frac{c}{6}\eta(1-\eta)$. When three defects meet at a vertex, the vertex type is
\begin{align}
\mbox{finite} &\mbox{ if $\sum_i\eta_i < 1$} \\
\mbox{ideal} & \mbox{ if  $\sum_i\eta_i = 1$} \notag\\
\mbox{hyperideal} & \mbox{ if $\sum_i \eta_i > 1$.}\notag
\end{align}
Suppose the weights are such that both vertices are finite. Then we simply place a conical defect on each edge of the graph, and interpret the graph, embedded in $S^3$, as the 3-manifold $M$. This 3-manifold admits a hyperbolic metric, and it can be triangulated by a single tetrahedron:
\begin{align}\label{tetcuff}
\cp{
\begin{tikzpicture}[scale=0.8]
\centerarc[thick,blue](-0.5,0)(65:410:1);
\centerarc[thick,red](0.5,0)(-115:233:1);
\draw[thick,darkgreen] (-0.5,0) -- (0.5,0);
\draw (0,0) circle (2.5);
\node at (-2,-1) {$S^3$};
\node at (-1.1,0) {\footnotesize $1$};
\node at (1.1,0) {\footnotesize 2};
\node at (0,.3) {\footnotesize 3};
\end{tikzpicture}
}
\quad &= \quad
\cp{
\begin{tikzpicture}[scale=3]
\coordinate (v3) at (0, 0.3); 
\coordinate (v2) at (0.75, 0); 
\coordinate (v1) at (1, 0.5); 
\coordinate (v4) at (0.5, 1); 
\draw[thick,darkgreen] (v3) -- (v2) node[midway, below,black] {\footnotesize 3};
\draw[thick,red] (v1)--(v3) node[pos=0.65,above,black] {\footnotesize 2};
\draw[white,fill=white] (0.64,0.43) circle (0.1em);
%
\draw[thick,darkgreen] (v3) -- (v4) node[midway,left,black] {\footnotesize 3};
\draw[thick,darkgreen] (v2) -- (v1) node[midway,right,black] {\footnotesize 3};
\draw[thick,blue] (v2) -- (v4) node[pos=0.7,left,black] {\footnotesize 1};
\draw[thick,darkgreen] (v1) -- (v4) node[midway,right,black] {\footnotesize 3};
\end{tikzpicture}
}
\end{align}
where the two front faces and the two back faces of the tetrahedron are glued together pairwise, producing conical defects on the edges. The manifold is compact and finite volume.  The dihedral angles of the tetrahedron are
\begin{align}
\psi_1 = 2\pi(1 - 2\eta_1) , \quad
\psi_2 = 2\pi(1 - 2\eta_2) , \quad
\psi_3 = \frac{\pi}{2}(1-2\eta_3) \ . 
\end{align}

If we increase the weights so that $2\eta_1+\eta_3 \to 1$ or $2\eta_2+\eta_3 \to 1$, then the corresponding vertex becomes ideal. A tetrahedron with an ideal vertex is non-compact but still has finite volume.

Now let us increase the weights further, such that both vertices are hyperideal. (But still in the defect regime, $\eta_i \in (0,\frac{1}{2})$.)  The corresponding geometry is the truncated tetrahedron reviewed in section \ref{s:polyhedra}, with the dihedral angles set by the weights $P_1, P_2, P_3$.  As we go from finite to hyperideal vertices, the 1-skeleton changes as follows:
\begin{align}
\Delta = \cp{
\begin{tikzpicture}[scale=1]
\centerarc[thick,darkgreen](0,0)(-30:90:1);
\centerarc[thick,darkgreen](0,0)(90:210:1);
\centerarc[thick,red](0,0)(210:330:1);
\draw[thick,blue] (0,0) -- (0,1);
\draw[thick,darkgreen] (0,0) -- ({cos(30)},{-sin(30)});
\draw[thick,darkgreen] (0,0) -- ({-cos(30)},{-sin(30)});
\end{tikzpicture}
}
\qquad \longrightarrow
\qquad
\cp{
\begin{tikzpicture}[scale=1]
\centerarc[thick,darkgreen](0,0)(-20:80:1);
\centerarc[thick,darkgreen](0,0)(100:200:1);
\centerarc[thick,red](0,0)(224:317:1);
\draw[thick,blue] (0,0) -- (0,0.96);
\draw[thick,darkgreen] (0,0) -- ({0.96*cos(30)},{-0.96*sin(30)});
\draw[thick,darkgreen] (0,0) -- ({-0.96*cos(30)},{-0.96*sin(30)});
\filldraw[thick,black,fill=white] (0.2,0.98)-- (0,0.7) --  (-0.2,0.98) -- cycle;
\filldraw[thick,black,fill=white] (-0.94,-0.33) -- (-0.6,-0.35) -- (-0.72,-0.68) -- cycle;
\filldraw[thick,black,fill=white] (0.94,-0.33) -- (0.6,-0.35) -- (0.72,-0.68) -- cycle;
\filldraw[thick,black,fill=white] (0,.2)-- (.2,-.1) -- (-.2,-.1) -- cycle;
\end{tikzpicture}
}
\end{align}
Increasing the weights even further, to the black hole regime $\eta = \frac{1}{2} + i \mathbb{R}$, results in deep truncations and dualizes the tetrahedron. The effect on the 1-skeleton is
$
\cp{
\adjustbox{margin=-1.5ex 0 -1ex 0}{
\begin{tikzpicture}[scale=0.25]
\coordinate (upperLeft) at (-1,3.5);
\coordinate (lowerLeft) at (-1,0);
\coordinate (upperRight) at (1,3.5);
\coordinate (lowerRight) at (1,0);
\coordinate (vertexTop) at (0,2.7);
\coordinate (vertexBottom) at (0, .8);
\draw (upperLeft) -- (vertexTop);
\draw (upperRight) -- (vertexTop);
\draw (lowerLeft) -- (vertexBottom);
\draw (lowerRight) -- (vertexBottom);
\draw[thick,blue] (vertexTop) -- (vertexBottom);
\end{tikzpicture}
}}
\to
\cp{
\adjustbox{margin=-1.5ex 0 -1ex 0}{
\begin{tikzpicture}[scale=0.25]
\coordinate (upperLeft) at (0,1);
\coordinate (lowerLeft) at (0,-1);
\coordinate (vertexLeft) at (0.8,0);
\coordinate (upperRight) at (3.5,1);
\coordinate (lowerRight) at (3.5,-1);
\coordinate (vertexRight) at (2.7,0);
\draw (upperLeft) -- (vertexLeft);
\draw (lowerLeft) -- (vertexLeft);
\draw[thick,blue] (vertexLeft) -- (vertexRight);
\draw (upperRight) -- (vertexRight);
\draw (lowerRight) -- (vertexRight);
\end{tikzpicture}
}}
$
and the same for the other colored edges, which results in the transformation
\begin{align}
\cp{
\begin{tikzpicture}[scale=1]
\centerarc[thick,darkgreen](0,0)(-20:80:1);
\centerarc[thick,darkgreen](0,0)(100:200:1);
\centerarc[thick,red](0,0)(224:317:1);
\draw[thick,blue] (0,0) -- (0,0.96);
\draw[thick,darkgreen] (0,0) -- ({0.96*cos(30)},{-0.96*sin(30)});
\draw[thick,darkgreen] (0,0) -- ({-0.96*cos(30)},{-0.96*sin(30)});
\filldraw[thick,black,fill=white] (0.2,0.98)-- (0,0.7) --  (-0.2,0.98) -- cycle;
\filldraw[thick,black,fill=white] (-0.94,-0.33) -- (-0.6,-0.35) -- (-0.72,-0.68) -- cycle;
\filldraw[thick,black,fill=white] (0.94,-0.33) -- (0.6,-0.35) -- (0.72,-0.68) -- cycle;
\filldraw[thick,black,fill=white] (0,.2)-- (.2,-.1) -- (-.2,-.1) -- cycle;
\node at (-0.83,0.83) {\footnotesize 3};
\node at (0.83,0.83) {\footnotesize 3};
\node at (0,-1.2) {\footnotesize 2};
\node at (.2,.4) {\footnotesize 1};
\node at (.5,-.1) {\footnotesize 3};
\node at (-.5,-.1) {\footnotesize 3};
\end{tikzpicture}
} \qquad \longrightarrow \qquad
\cp{
\begin{tikzpicture}[scale=1]
\begin{scope}[rotate=180]
\centerarc[thick,darkgreen](0,0)(-20:80:1);
\centerarc[thick,darkgreen](0,0)(100:200:1);
\centerarc[thick,blue](0,0)(224:317:1);
\draw[thick,red] (0,0) -- (0,0.96);
\draw[thick,darkgreen] (0,0) -- ({0.96*cos(30)},{-0.96*sin(30)});
\draw[thick,darkgreen] (0,0) -- ({-0.96*cos(30)},{-0.96*sin(30)});
\filldraw[thick,black,fill=white] (0.2,0.98)-- (0,0.7) --  (-0.2,0.98) -- cycle;
\filldraw[thick,black,fill=white] (-0.94,-0.33) -- (-0.6,-0.35) -- (-0.72,-0.68) -- cycle;
\filldraw[thick,black,fill=white] (0.94,-0.33) -- (0.6,-0.35) -- (0.72,-0.68) -- cycle;
\filldraw[thick,black,fill=white] (0,.2)-- (.2,-.1) -- (-.2,-.1) -- cycle;
\node at (-0.83,0.83) {\footnotesize 3};
\node at (0.83,0.83) {\footnotesize 3};
\node at (0,-1.2) {\footnotesize 1};
\node at (.2,.4) {\footnotesize 2};
\node at (.5,-.1) {\footnotesize 3};
\node at (-.5,-.1) {\footnotesize 3};
\end{scope}
\end{tikzpicture}
}
 = \hat{\Delta}
\end{align}
The faces of the original tetrahedron are now the vertices, so the identifications glue the truncated faces of the final tetrahedron pairwise. Thus the generalized tetrahedron $\hatDelta$ at the end of this procedure is exactly the same hyperbolic geometry \eqref{MclS} that we used to calculate the gravitational path integral.

\section{Geometry of the $6j$-symbol}\label{s:sixj}
In the next example we will apply the dictionary for a compact region to the calculation of $\overline{C_{ijk}C_{lmn}C_{pqr}C_{stu}}$. 
This is known to have a contribution proportional to the Virasoro $6j$-symbol-squared, with tetrahedral index contraction, a result which has been derived from both the dual CFT and Virasoro TQFT as reviewed in appendix \ref{app:cftstatistics}. Our goal is to reproduce this result from a gravity calculation, in metric language, thereby describing the geometric dual of the Virasoro $6j$-symbol for heavy operators.\footnote{The gravity dual of the $6j$-symbol for global conformal blocks, in any number of dimensions, is a tetrahedral Witten diagram \cite{Krasnov:2004gm,Krasnov:2005fu,Liu:2018jhs}. The discussion section of \cite{Krasnov:2004gm} also hints at the connection to Thurston's decomposition using truncated hyperbolic tetrahedra. }

The corresponding calculation for light defect operators was already described in \cite{Teschner:2012em,Collier:2023fwi,Collier:2024mgv}. In that case, the three-dimensional geometry is an ordinary hyperbolic tetrahedron with finite vertices. The geometry below can be viewed as the analytic continuation to heavy weights.

\subsection{Fixed lengths}

The dictionary \eqref{dictionary} for the compact region $M$,  applied to the 4-point statistics, states that
\begin{align}
\overline{C_{ijk}C_{lmn}C_{pqr}C_{stu}} 
&=  \sum_G \delta^G_{ijklmnpqrstu} \sum_{(M,\gamma)} Z_L(M; \gamma(\vecP))
\end{align}
where the outer sum is over trivalent graphs with four vertices. We are interested in the contribution from the tetrahedron graph, $\Gamma = \inlinetetra$. This graph contributes to the connected statistics of the 4-point function as 
\begin{align}\label{fpdict}
\overline{C_{123}C_{156}C_{246}C_{345}} \supset
\sum_{(M,\gamma)} Z_L(M, 
\cp{\footnotesize
\begin{tikzpicture}[scale=0.7]
\draw (0,0) circle (1);
\draw (0,0) -- (0,1) node[midway,right,xshift=-3] {$\ell_3$};
\draw (0,0) -- ({cos(30)},{-sin(30)}) node[midway,below,yshift=4] {$\ell_5$};
\draw (0,0) -- ({-cos(30)},{-sin(30)}) node[midway,below,yshift=4,xshift=3] {$\ell_4$};
\node[left,below] at (1,1) {$\ell_1$};
\node[right,below] at (-1,1) {$\ell_2$};
\node[below] at (0,-1) {$\ell_6$};
\end{tikzpicture}
}
)
\end{align}
where the boundary condition in this expression means 
\begin{align}\label{genus3loops}
\p M \quad = \quad 
\cp{
\begin{overpic}[grid=false,width=1.5in]{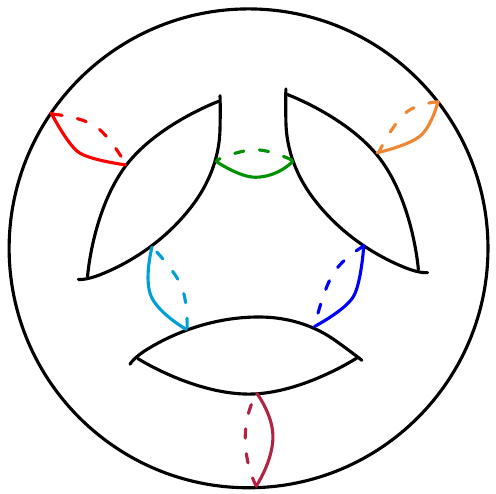}
\put (900,800) {$\ell_1$}
\put (30,800) {$\ell_2$}
\put (480,720) {$\ell_3$}
\put (230,350) {$\ell_4$}
\put (720,350) {$\ell_5$}
\put (410,80) {$\ell_6$}
\end{overpic}
} \qquad  
\end{align}
with fixed lengths on the geodesics and $K_{ij}=0$ elsewhere. 
There is a genus-3 handlebody satisfying this boundary condition constructed by gluing together two tetrahedra along their truncated ends (which is the not the handlebody drawn in \eqref{genus3loops} --- it fills in a different set of cycles). The classical saddlepoint consists of two truncated hyperbolic tetrahedra,
\begin{align}\label{fpres}
M_{cl} \qquad =  \qquad 
\vcenter{\hbox{
\begin{tikzpicture}[scale=0.035]
%
\coordinate (a1) at (10,53);
\coordinate (a2) at (9, 27);
\coordinate (a3) at (18, 40);
%
\draw[thick] (a1) -- (a2) node[midway,left] {$A$};
\draw[thick,dotted] (a2) -- (a3);
\draw[thick,dotted] (a3) -- (a1);
%
\coordinate (b1) at (59, 19);
\coordinate (b2) at (45, 5);
\coordinate (b3) at (73, 7);
%
\draw[thick] (b1) -- (b2);
\draw[thick] (b2) -- (b3) node[midway,below] {$B$};
\draw[thick] (b3) -- (b1);
%
\coordinate (c1) at (92,60);
\coordinate (c2) at (86,45);
\coordinate (c3) at (94,34);
%
\draw[thick,dotted] (c1) -- (c2);
\draw[thick,dotted] (c2) -- (c3);
\draw[thick] (c3) -- (c1) node[midway,right] {$C$};
%
\coordinate (d1) at (35,88);
\coordinate (d2) at (50,82);
\coordinate (d3) at (59,90);
%
\draw[thick] (d1) -- (d2);
\draw[thick] (d2) -- (d3);
\draw[thick] (d3) -- (d1) node[midway,above] {$D$};
%
\coordinate (upperLeftE) at (33,61);
\coordinate (lowerLeftE) at (35,26);
\coordinate (lowerRightE) at (71, 30);
\coordinate (upperRightE) at (67,64);
\coordinate (frontE) at (51,50);
\coordinate (backE) at (54,40);
%
\node[xshift=-5,yshift=5] at (upperLeftE) {\small $\ell_4$};
\node[xshift=5,yshift=7] at (frontE) {\small $\ell_5$};
\node[xshift=5,yshift=5] at (upperRightE) {\small $\ell_6$};
\node[xshift=5,yshift=-5] at (lowerRightE) {\small $\ell_1$};
\node[gray,xshift=-13,yshift=5] at (backE) {\small $\ell_2$};
\node[xshift=-5,yshift=-5] at (lowerLeftE) {\small $\ell_3$};
%
\draw[thick,red,dashed] plot [smooth,tension=1] coordinates {(a3) (backE) (c2)};
\draw[thick,cyan] plot [smooth,tension=1] coordinates {(a1) (upperLeftE) (d1)};
\draw[thick,darkgreen] plot [smooth,tension=1] coordinates {(a2) (lowerLeftE) (b2)};
\draw[thick,orange] plot [smooth,tension=1] coordinates {(b3) (lowerRightE) (c3)};
\draw[thick,purple] plot [smooth,tension=1] coordinates {(c1) (upperRightE) (d3)};
\draw[thick,blue] plot [smooth,tension=1] coordinates {(b1) (frontE) (d2)};
\end{tikzpicture}
\begin{tikzpicture}[scale=0.035]
%
\coordinate (a1) at (10,53);
\coordinate (a2) at (9, 27);
\coordinate (a3) at (18, 40);
%
\draw[thick] (a1) -- (a2) node[midway,left] {$C$};
\draw[thick,dotted] (a2) -- (a3);
\draw[thick,dotted] (a3) -- (a1);
%
\coordinate (b1) at (59, 19);
\coordinate (b2) at (45, 5);
\coordinate (b3) at (73, 7);
%
\draw[thick] (b1) -- (b2);
\draw[thick] (b2) -- (b3) node[midway,below] {$B$};
\draw[thick] (b3) -- (b1);
%
\coordinate (c1) at (92,60);
\coordinate (c2) at (86,45);
\coordinate (c3) at (94,34);
%
\draw[thick,dotted] (c1) -- (c2);
\draw[thick,dotted] (c2) -- (c3);
\draw[thick] (c3) -- (c1) node[midway,right] {$A$};
%
\coordinate (d1) at (35,88);
\coordinate (d2) at (50,82);
\coordinate (d3) at (59,90);
%
\draw[thick] (d1) -- (d2);
\draw[thick] (d2) -- (d3);
\draw[thick] (d3) -- (d1) node[midway,above] {$D$};
%
\coordinate (upperLeftE) at (33,61);
\coordinate (lowerLeftE) at (35,26);
\coordinate (lowerRightE) at (71, 30);
\coordinate (upperRightE) at (67,64);
\coordinate (frontE) at (51,50);
\coordinate (backE) at (54,40);
%
%
\draw[thick,red,dashed] plot [smooth,tension=1] coordinates {(a3) (backE) (c2)};
\draw[thick,purple] plot [smooth,tension=1] coordinates {(a1) (upperLeftE) (d1)};
\draw[thick,orange] plot [smooth,tension=1] coordinates {(a2) (lowerLeftE) (b2)};
\draw[thick,darkgreen] plot [smooth,tension=1] coordinates {(b3) (lowerRightE) (c3)};
\draw[thick,cyan] plot [smooth,tension=1] coordinates {(c1) (upperRightE) (d3)};
\draw[thick,blue] plot [smooth,tension=1] coordinates {(b1) (frontE) (d2)};
\end{tikzpicture}
}}
\end{align}
which are glued together on the triangular faces, $A \leftrightarrow A$, $B \leftrightarrow B$, $C \leftrightarrow C$, and $D \leftrightarrow D$. The lengths  $\ell_i = 4\pi b P_i$ are the total lengths of the closed curves, so each individual segment has length $\ell_i/2$.

The fixed-length path integral, at the level of the classical action, is
\begin{align}
Z_L(M, \gamma(\vecP))
&\approx
\exp\left[
-2I_L^{tet}\begin{pmatrix}
\half \ell_4 & \half \ell_5 & \half \ell_6 \\
\half \ell_1 & \half \ell_2 & \half \ell_3
\end{pmatrix}
\right]\\
&\approx
\exp\left[
-\frac{c}{3\pi}
\left( 
V_L\begin{pmatrix}
\half \ell_4 & \half \ell_5 & \half \ell_6 \\
\half \ell_1 & \half \ell_2 & \half \ell_3
\end{pmatrix}
 + \frac{1}{4}\sum_{i=1}^6 \psi_i \ell_i
 \right)
 \right]
\end{align}
In appendix \ref{app:kernels} we study the semiclassical limit of the Virasoro $6j$-symbol and derive the relation
\begin{align}\label{sixjVol}
\log
\begin{Bmatrix}
P_1 & P_2 & P_3 \\ P_4 & P_5 & P_6
\end{Bmatrix}
\sim
-\frac{c}{6\pi} \left( 
V_L\begin{pmatrix}
\half \ell_4 & \half \ell_5 & \half \ell_6 \\
\half \ell_1 & \half \ell_2 & \half \ell_3
\end{pmatrix}
 + \frac{1}{4} \sum_{i=1}^6 \ell_i \psi_i
\right) \ . 
\end{align}
Therefore, at the level of the classical action, we have confirmed that
\begin{align}
Z_L(M, \gamma(\vecP))
=  \left| \begin{Bmatrix} P_1 & P_2 & P_3 \\ P_4 & P_5 & P_6 \end{Bmatrix}\right|^2 \ . 
\end{align}
This agrees with the predictions of Virasoro TQFT and the conformal bootstrap reviewed in appendix \ref{app:cftstatistics}.

\subsection{Fixed angles}
The Virasoro TQFT graph corresponding to the geometric calculation above is
\begin{align}
\Gamma(\vecP) = \cp{\footnotesize
\begin{tikzpicture}[scale=0.8]
\centerarc[very thick](0,0)(-30:90:1);
\centerarc[very thick](0,0)(90:210:1);
\centerarc[very thick](0,0)(210:335:1);
\draw[very thick] (0,0) -- (0,1);
\draw[very thick] (0,0) -- ({cos(30)},{-sin(30)});
\draw[very thick] (0,0) -- ({-cos(30)},{-sin(30)});
\node at (.9,.9) {$1$};
\node at (-.9,.9) {$2$};
\node at (-0.2,.55) {$3$};
\node at (-.55,.0) {$4$};
\node at (0.55,0) {$5$};
\node at (0,-1.25) {$6$};
\end{tikzpicture}} \ ,
\end{align}
with amplitude
\begin{align}\label{ztets}
\Zvir(S^3, \Gamma(\vecP)) = \begin{Bmatrix} P_1 & P_2 & P_3 \\ P_4 & P_5 & P_6 \end{Bmatrix} \ . 
\end{align}
The graph $(S^3,\Gamma)$ is triangulated by two ordinary tetrahedra $\hat{\Delta}_1 = \overline{\hat{\Delta}_2}$, identical up to orientation reversal, with edges labeled as in the $6j$-symbol on the right-hand side of \eqref{ztets}. The labeling on these tetrahedra is such that $(P_1, P_2, P_3)$ meet at a vertex.  Therefore, using \eqref{tetW}, the fixed-angle path integral is
\begin{align}\label{zajcheck}
Z_A(M, \gamma(\vecP)) &= \Ztv(S^3, \Gamma(\vecP)) \\
&= \begin{Bmatrix}P_4 & P_5 & P_6 \\ P_1  & P_2 & P_3 \end{Bmatrix}^2 \\
&\approx \exp\left[ -\frac{c}{3\pi} V_A \begin{pmatrix}
\psi_4 & \psi_5 & \psi_6 \\
\psi_1 & \psi_2 & \psi_3
\end{pmatrix}
\right]
\end{align}
where in the last line we used the semiclassical approximation to the $6j$-symbol from \cite{Teschner:2012em} (reviewed in appendix \ref{app:kernels}). As expected, this agrees with the classical action of the geometry \eqref{fpres} with fixed-angle boundary conditions. 

Note the row-swap in going from \eqref{ztets} to \eqref{zajcheck}. The tetrahedron in VTQFT, which has $(P_1,P_2,P_3)$ meeting at a vertex, is dual to the tetrahedron in \eqref{fpres}, which has edges $(P_1,P_2,P_3)$ going around a face.

The relation between the fixed-angle and fixed-length amplitudes is the following exact formula for the Fourier transform of the Virasoro $6j$-symbol-squared, derived in \cite{ctv}:
\begin{align}\label{asdfJJ}
\begin{Bmatrix}
P_4' & P_5' & P_6' \\
P_1' & P_2' & P_3'
\end{Bmatrix}^2
&= 
 \int_{0}^{\infty} ( \Pi_{i=1}^6 dP_i \, S_{P_i' P_i} )
 \begin{Bmatrix}
 P_1 & P_2 & P_3 \\
 P_4 & P_5 & P_6 
 \end{Bmatrix}^2
\end{align}

\subsection{From Virasoro TQFT to geometry}\label{ss:vtqftsixj}
Let us describe the procedure to go from the VTQFT graph to the geometry in this example, following the general method of section \ref{ss:gentri}. 
The tetrahedron graph, embedded in $S^3$, can be triangulated by two finite tetrahedra:
\begin{align}\label{tetTri}
\cp{
\begin{tikzpicture}[scale=0.8]
\centerarc[thick,orange](0,0)(-30:90:1);
\centerarc[thick,red](0,0)(90:210:1);
\centerarc[thick,purple](0,0)(210:335:1);
\draw[thick,darkgreen] (0,0) -- (0,1);
\draw[thick,blue] (0,0) -- ({cos(30)},{-sin(30)});
\draw[thick,cyan] (0,0) -- ({-cos(30)},{-sin(30)});
\draw (0,0) circle (2.5);
\node at (-2,-1) {$S^3$};
\node at (-0.83,0.83) {\footnotesize 2};
\node at (0.83,0.83) {\footnotesize 1};
\node at (0,-1.2) {\footnotesize 6};
\node at (.2,.4) {\footnotesize 3};
\node at (.6,-.14) {\footnotesize 5};
\node at (-.6,-.14) {\footnotesize 4};
\end{tikzpicture}
}
\qquad = 
\qquad 
\cp{
\begin{tikzpicture}[scale=3]
\coordinate (v3) at (0, 0.3); 
\coordinate (v2) at (0.75, 0); 
\coordinate (v1) at (1, 0.5); 
\coordinate (v4) at (0.5, 1); 
\draw[thick,cyan] (v3) -- (v2) node[midway, below,black] {\footnotesize  4};
\draw[thick,purple] (v1)--(v3) node[pos=0.65,above,black] {\footnotesize 6};
\draw[white,fill=white] (0.64,0.43) circle (0.1em);
%
\draw[thick,red] (v3) -- (v4) node[midway,left,black] {\footnotesize 2};
\draw[thick,blue] (v2) -- (v1) node[midway,right,black] {\footnotesize 5};
\draw[thick,darkgreen] (v2) -- (v4) node[pos=0.7,left,black] {\footnotesize 3};
\draw[thick,orange] (v1) -- (v4) node[midway,right,black] {\footnotesize 1};
\end{tikzpicture}
}
\cp{
\begin{tikzpicture}[scale=3]
\coordinate (v3) at (0, 0.3); 
\coordinate (v2) at (0.75, 0); 
\coordinate (v1) at (1, 0.5); 
\coordinate (v4) at (0.5, 1); 
\draw[thick,blue] (v3) -- (v2) node[midway, below,black] {\footnotesize  5};
\draw[thick,purple] (v1)--(v3) node[pos=0.65,above,black] {\footnotesize 6};
\draw[white,fill=white] (0.64,0.43) circle (0.1em);
%
\draw[thick,orange] (v3) -- (v4) node[midway,left,black] {\footnotesize 1};
\draw[thick,cyan] (v2) -- (v1) node[midway,right,black] {\footnotesize 4};
\draw[thick,darkgreen] (v2) -- (v4) node[pos=0.7,left,black] {\footnotesize 3};
\draw[thick,red] (v1) -- (v4) node[midway,right,black] {\footnotesize 2};
\end{tikzpicture}
}
\end{align}
The like faces of the tetrahedra glued together, producing conical defects at the edges. This describes the light operator regime, where the Wilson lines are conical defects meeting at finite vertices. 
To describe heavy operators, we truncate the tetrahedra, then dualize all the edges.
The 1-skeleton of the first tetrahedron transforms as
\begin{align}
\Delta = \cp{
\begin{tikzpicture}[scale=0.8]
\centerarc[thick,orange](0,0)(-30:90:1);
\centerarc[thick,red](0,0)(90:210:1);
\centerarc[thick,purple](0,0)(210:335:1);
\draw[thick,darkgreen] (0,0) -- (0,1);
\draw[thick,blue] (0,0) -- ({cos(30)},{-sin(30)});
\draw[thick,cyan] (0,0) -- ({-cos(30)},{-sin(30)});
\node at (-0.83,0.83) {\footnotesize 2};
\node at (0.83,0.83) {\footnotesize 1};
\node at (0,-1.2) {\footnotesize 6};
\node at (.2,.4) {\footnotesize 3};
\node at (.6,-.14) {\footnotesize 5};
\node at (-.6,-.14) {\footnotesize 4};
\end{tikzpicture}
}
\quad \longrightarrow
\quad
\cp{
\begin{tikzpicture}[scale=1]
\begin{scope}[rotate=180]
\centerarc[thick,red](0,0)(-20:80:1);
\centerarc[thick,orange](0,0)(100:200:1);
\centerarc[thick,darkgreen](0,0)(224:317:1);
\draw[thick,purple] (0,0) -- (0,0.96);
\draw[thick,cyan] (0,0) -- ({0.96*cos(30)},{-0.96*sin(30)});
\draw[thick,blue] (0,0) -- ({-0.96*cos(30)},{-0.96*sin(30)});
\filldraw[thick,black,fill=white] (0.2,0.98)-- (0,0.7) --  (-0.2,0.98) -- cycle;
\filldraw[thick,black,fill=white] (-0.94,-0.33) -- (-0.6,-0.35) -- (-0.72,-0.68) -- cycle;
\filldraw[thick,black,fill=white] (0.94,-0.33) -- (0.6,-0.35) -- (0.72,-0.68) -- cycle;
\filldraw[thick,black,fill=white] (0,.2)-- (.2,-.1) -- (-.2,-.1) -- cycle;
\node at (-0.83,0.83) {\footnotesize 1};
\node at (0.83,0.83) {\footnotesize 2};
\node at (0,-1.2) {\footnotesize 3};
\node at (.2,.4) {\footnotesize 6};
\node at (.5,-.1) {\footnotesize 4};
\node at (-.5,-.1) {\footnotesize 5};
\end{scope}
\end{tikzpicture}
}
 = \hat{\Delta}
\end{align}
and the second tetrahedron is the mirror image.
Insertions of $C_{ijk}$ correspond to vertices in the VTQFT graph, which are faces of the dualized tetrahedra. The original face pairings become identifications of the triangular faces. Therefore the final result is a truncated tetrahedron, doubled along its truncated faces --- and this is exactly the geometry studied above in \eqref{fpres}.

\section{Discussion}\label{s:discussion}
Observables in AdS/CFT are usually calculated by gravitational path integrals with asymptotic AdS boundaries. However, the statistics of heavy operators have very little to do with the geometry near infinity. The leading-order variance is calculated from a wormhole with two asymptotic boundaries, but after normalizing by this factor, all of the corrections --- including subleading 2-point statistics, and non-Gaussian higher-point statistics --- are calculated by the gravitational path integral on a finite-volume, compact spacetime, $M$. In 3D gravity, standard ensemble-averaged CFT observables, which are products of correlation functions of local operators on Riemann surfaces, can be re-assembled by gluing in $C_0$ flares to attach $M$ to asymptotic boundaries.

In $D$ dimensions, the natural boundary conditions, for these purposes, fix the areas or conjugate angles of codimension-2 extremal surfaces on $\p M$. In AdS$_3$, we argued that the exact path integral with fixed-length and fixed-angle boundary conditions are
\begin{align}\label{discformulas}
Z_L  = |\Zvir|^2 , \qquad Z_A = \Ztv \ . 
\end{align}
See the introduction for the precise statements. The fixed-length amplitude is directly related to OPE statistics; the fixed-angle amplitude is directly related to triangulations of $M$ into generalized tetrahedra; and the transformation from one to the other is a modular $S$-transform. In TQFT language, the change of boundary conditions is realized by the exact transform between $|\Zvir|^2$ and $\Ztv$ derived in \cite{ctv}.

In the semiclassical limit, we studied two examples, and demonstrated how to build the geometries from generalized hyperbolic tetrahedra. These basic building blocks are polyhedra in $H_3$ that can be described by intersecting hyperplanes, but they are most naturally viewed as tetrahedra in the Klein space $H_3 \cup dS_3$.

Some open questions and directions for further study include:
\begin{enumerate}
\item We have worked out only two of the simplest semiclassical geometries. It would be worthwhile to have more examples, especially with internal edges or nontrivial embedding manifolds, which introduce some new ingredients. (Another interesting class of examples --- knots in $S^3$ with an attached particle --- is discussed in \cite{knotspaper}.)
\item We argued for \eqref{discformulas} indirectly -- by matching the fixed-length amplitude to known results with asymptotic boundaries, and through the Laplace transform to obtain the fixed-angle amplitude. It would be interesting and instructive to derive them by direct evaluation of the gravity path integral.
\item Some of the most interesting unanswered questions about quantum gravity in three dimensions involve contributions from off-shell topologies, i.e., topologies that do not admit a hyperbolic metric \cite{Maloney:2007ud,Cotler:2020ugk,Maxfield:2020ale,Boruch:2025ilr,deBoer:2025rct}. Any 3-manifold admits a triangulation. Can the techniques developed here by applied to non-hyperbolic topologies? One potential advantage of the CTV approach over Virasoro TQFT is that it is closer to the metric description, and the gap between TQFT and gravity is more significant on off-shell topologies because of the difference in how they treat large diffeomorphisms \cite{Collier:2023fwi,Collier:2024mgv,Yan:2025usw}.
\item How do the results of this paper reduce to 2D gravity? See \cite{Kontsevich:1992ti,Ghosh:2019rcj,Maxfield:2020ale,Saad:2022kfe,Yan:2025usw,Blommaert:2022lbh,Kruthoff:2024gxc} for hints.
\item Similar methods can by applied to dS$_3$, and perhaps even to other cosmologies or closed universes. Indeed, the main results on CTV theory derived in \cite{ctv} were based on the spin network model of dS$_3$ gravity in \cite{Barrett:2004im}. Perhaps the external edges of the graph play a role similar to that of an `observer', or an entangled auxiliary system, in recent approaches to quantum gravity in a closed universe \cite{Almheiri:2019hni,Chen:2020tes,Hartman:2020khs,Chandrasekaran:2022cip,Maldacena:2024spf,Collier:2025lux,Abdalla:2025gzn,Harlow:2025pvj,Ivo:2025yek}.
\item Can these methods be used to calculate the statistics of black hole matrix elements in higher dimensions? In principle, the statistics are encoded in the path integral on compact manifolds with fixed-area boundary conditions on a set of boundary extremal surfaces. 
\item Can we do the sum over topologies?
\item In \cite{Cotler:2020ugk,Belin:2023efa,Jafferis:2025vyp,Boruch:2025ilr,deBoer:2025rct} the authors discuss random-matrix-like models of random 2d CFTs, inspired by the JT/Random matrix duality in 2D gravity \cite{Saad:2019lba}. The random tensor model in \cite{Belin:2023efa,Jafferis:2025vyp,Jafferis:2025yxt} is designed to reproduce certain amplitudes of Virasoro TQFT, with building blocks corresponding to VTQFT tetrahedra. Another natural approach to this problem is to model the building blocks using the dual spine of the triangulated 3-manifold. For heavy states, this is the difference between the triangulation of the embedding manifold $(M_E, \Gamma)$ and the generalized triangulation of the gravity manifold $M = M_E - N(\Gamma)$. Does the dual perspective produce an interesting tensor model?
\item The fixed-length amplitude holomorphically factorizes, which is manifest in the Virasoro TQFT expression $|\Zvir|^2$. This makes it straightforward to include angular momentum in the fixed-length path integral by complexifying the lengths. However, since the CTV partition function does not holomorphically factorize, we do not currently have a way to calculate $\Zvir(M,\gamma(\vecP)) \Zvir(\bar{M}, \gamma(\bar{\vecP}))$, with independent $\vecP$ and $\bar{\vecP}$, or the OPE statistics of spinning operators, from a triangulation (except by first calculating the scalar statistics, then doing the obvious analytic continuation). Can this be done? 
\end{enumerate}
Let us propose a preliminary answer to the last question. A fixed-angle amplitude with two independent complex angles on each of $n$ cycles can be defined as
\begin{align}\label{twoangles}
Z_A(\psi, \bar{\psi}) \sim \int d^n \ell d^n \bar{\ell} e^{\frac{c}{24\pi}\sum_{i=1}^n(\psi_i \ell_i + \bar{\psi}_i \bar{\ell}_i)}Z_L(\ell, \bar{\ell}) \ . 
\end{align}
In Liouville parameterization,
\begin{align}\label{liouAA}
Z_A(M, \gamma(\vecP', \bar{\vecP}'))
\sim
\int d^n \vecP d^n \bar{\vecP} 
\left(
\Pi_{i=1}^n
S_{P_i',\half P_i}
S_{\bar{P}_i', \half \bar{P}_{i}}
\right)
Z_L(M, \gamma(\vecP, \bar{\vecP})) \ . 
\end{align}
Now the key observation is that the kernel $S_{P', \half P}$ is (up to normalization) exactly the wavefunction of an FZZ state  $|B_{P}\rangle$ in Liouville CFT, in the Ishibashi basis  \cite{Fateev:2000ik}:
\begin{align}
S_{P', \half P} \sim \langle\!\langle P' | B_P\rangle \ . 
\end{align}
Now, any VTQFT amplitude can be expressed as an overlap of an identity conformal block ${\cal F}_{\id}$ with a heavy conformal block ${\cal F}_{\vecP}$  in another channel (see section 5.2 of \cite{ctv}.). Schematically,
\begin{align}\label{zlblocks}
Z_L = |\Zvir(M_E, \Gamma(\vecP))|^2 \sim  \langle {\cal F}_{\id}|{\cal F}_{\bar{\vecP}}\rangle \langle {\cal F}_{\vecP} | {\cal F}_{\id} \rangle
 \ . 
\end{align}
The conformal blocks also depend on a choice of channel, but we suppress this in the notation.
Plugging \eqref{zlblocks} into \eqref{twoangles} and doing the holomorphic and anti-holomorphic integrals separately gives
\begin{align}\label{zafzz}
Z_A(M, \gamma(\vecP', \bar{\vecP}')) \sim \langle {\cal F}_{\id} | B_{\bar{\vecP}'} \rangle \langle 
B_{\vecP'} | {\cal F}_{\id}\rangle
\end{align}
where $|B_{\vecP'}\rangle$ is the Fourier transform of $|{\cal F}_{\vecP}\rangle$. This can be interpreted as a dual conformal block, projected in the intermediate channel onto an FZZ-like state rather than a conformal primary. Thus we propose \eqref{zafzz} as the exact gravitational path integral with fixed-angle boundaries when there are independent left- and right-moving potentials. This is no longer a CTV partition function, but the saddlepoint equations for \eqref{liouAA} are simply the complexification of section \ref{s:classicaltri} above, so the semiclassical geometries are directly related to generalized triangulations with complex lengths assigned to the edges.

The fixed-angle path integral in the form \eqref{liouAA} has some aspects very similar to \cite{Chua:2023ios,Chen:2024unp,Hung:2024gma,Bao:2024ixc,Hung:2025vgs,Geng:2025efs}, but we leave a detailed comparison to future work.

 \bigskip \ \\

\noindent\textbf{Acknowledgments}

\noindent It is a pleasure to thank Alex Belin, Jeevan Chandra, Wan Zhen Chua, Scott Collier, Gabriele Di Ubaldo, Luca Iliesiu, Yikun Jiang, Adam Levine, Nate MacFadden, Juan Maldacena, Alex Maloney, Greg Mathys, Eric Perlmutter, Julian Sonner, Douglas Stanford, Diandian Wang, and Cynthia Yan for discussions on related topics. This work is supported by NSF grant PHY-2309456.

\appendix

\section{Semiclassical crossing kernels}\label{app:kernels}
In this appendix, we discuss the semiclassical limits of the Virasoro $6j$-symbol and modular $S$-matrix. The final results in various regimes are summarized in section \ref{ss:kernelsummary}, and the calculations are done in sections \ref{ss:kernelS}-\ref{ss:kernelJ}. Results are expressed in terms of the volumes of generalized hyperbolic tetrahedra, which were reviewed in section \ref{s:polyhedra}. The volume as a function of dihedral angles is $V_A(\psi)$ and the volume as a function of edge lengths is $V_L(L)$. The central charge and conformal weights are parameterized in the standard way,
\begin{align}
c = 1+6Q^2\ ,  \qquad h = \frac{Q^2}{4} + P^2\ , \qquad Q=b+b^{-1} \ . 
\end{align}
The semiclassical limit is defined as $b \to 0$ with $b P_i$ held fixed.

Our conventions for the crossing kernels follow \cite[Appendix A]{ctv}. The Virasoro $6j$-symbol $\begin{Bsmallmatrix}P_1 & P_2 & P_3 \\ P_4 & P_5 & P_6\end{Bsmallmatrix}$ is proportional to the fusion kernel for 4-point crossing on the sphere, and enjoys tetrahedral symmetry. The modular $S$-matrix in Racah-Wigner normalization is $\hatS_{P_1P_2}[P_3]$, which is symmetric in its lower arguments and has phase $\hatS_{P_1P_2}[P_3] = e^{i\pi h_3/2} \times \mbox{(real)}$ for $P_i \in \mathbb{R}_+$. 

\subsection{Summary}\label{ss:kernelsummary}

\subsubsection*{$6j$-symbol in the defect regime}
Teschner and Vartanov \cite{Teschner:2012em} derived the semiclassical $6j$-symbol in the regime of light defects, meaning $P_i \in i(0, \frac{Q}{2})$ (i.e. $h_i \in (0, \frac{c-1}{24}))$ for $i=1\dots 6$, and 
\begin{align}\label{vertpi}
\sum_{i \in I} P_i \quad \in \quad i (0, \frac{Q}{2})
\end{align}
for $I = \{1,2,3\}$, $I = \{1,5,6\}$, $I  = \{3,4,5\}$, and $I=\{2,4,6\}$. The $6j$-symbol in this limit is \cite{Teschner:2012em}
\begin{align}\label{tvsixj}
\log \begin{Bmatrix}
P_1 & P_2 & P_3 \\
P_4 & P_5 & P_6 
\end{Bmatrix}
\sim -\frac{c}{6\pi} V_A \begin{pmatrix}
\psi_1  & \psi_2 & \psi_3 \\
\psi_4 & \psi_5 & \psi_6
\end{pmatrix}
\ , \quad 
\psi_i = 2\pi i P_i \ . 
\end{align}
The condition \eqref{vertpi} ensures that the sum of dihedral angles at each vertex of the tetrahedron satisfies $\sum \psi_i < \pi$, and therefore $V_A$ in \eqref{tvsixj} is the volume of an ordinary, non-truncated hyperbolic tetrahedron with finite vertices.

By virtue of the results in \cite{ushijimaMR2191251} and the saddlepoint analysis in \cite{Teschner:2012em}, the formula \eqref{tvsixj} continues to hold in the regime of heavy defects, where the individual weights are still below the black hole threshold but we no longer impose \eqref{vertpi}. In this case, each vertex is either finite (if $\sum\psi_i < \pi$), ideal (if $\sum \psi_i = \pi$), or truncated (if $\sum \psi_i > \pi$), and $V_A$ in \eqref{tvsixj} is the volume of a generalized tetrahedron, if one exists with these angles. The result \eqref{tvsixj} does not apply if there is no hyperbolic tetrahedron with these dihedral angles; we will not discuss this in detail in the defect regime but it is similar to the degenerate cases for heavy weights that will be discussed momentarily.

\subsubsection*{$6j$-symbol in the black hole regime}

For heavy weights, $P_i \in \mathbb{R}_+$, for $1\dots 6$, the semiclassical $6j$-symbol is calculated in section \ref{ss:kernelJ}. Parameterize the weights by $\ell_i = 4\pi b P_i$
and let $\tilde{G}$ be the Gram matrix in \eqref{sixjD}. Depending on the sign of $\det \tilde{G}$, we find
\begin{align}\label{sixjVolApp}
\log
\begin{Bmatrix}
P_1 & P_2 & P_3 \\ P_4 & P_5 & P_6
\end{Bmatrix}
\sim
-\frac{c}{6\pi} \left( 
V_L\begin{pmatrix}
\half \ell_4 & \half \ell_5 & \half \ell_6 \\
\half \ell_1 & \half \ell_2 & \half \ell_3
\end{pmatrix}
 + \frac{1}{4} \sum_{i=1}^6 \ell_i \psi_i
\right)
\qquad (\det \GL < 0)
\end{align}
with $\psi_i$ is the dihedral angle on the edge of length $\ell_i/2$, or
\begin{align}\label{sixjresDegen}
\log
\begin{Bmatrix}
P_1 & P_2 & P_3 \\ P_4 & P_5 & P_6
\end{Bmatrix}
\sim
-\frac{c}{24}\max (\ell_1+\ell_4, \ell_2+\ell_5,\ell_3+\ell_6)
\qquad (\det \GL > 0) \ . 
\end{align}
A mildly truncated generalized hyperbolic tetrahedron exists for $\det \tilde{G} < 0$, while in the opposite regime, $\det \tilde{G} > 0$, the result \eqref{sixjresDegen} can be interpreted as a degenerate case of \eqref{sixjVolApp} where the tetrahedron has zero volume but the second term contributes.

The $6j$-symbol for heavy weights \eqref{sixjVolApp} can also be obtained from the defect result \eqref{tvsixj} by analytic continuation \cite{Chen:2024unp}.

\subsubsection*{Modular $S$-matrix in the black hole regime}
The modular $S$-matrix in the heavy regime, $P_i \in \mathbb{R}_+$ for $i=1\dots 3$, is calculated in section \ref{ss:kernelS}. Parameterize the weights by $\ell_i = 4\pi b P_i$ and let $D$ be the discriminant defined in \eqref{discrimS}. For $D<0$, we find
\begin{align}\label{logSV}
\log  \hatS_{P_1P_2}[P_3] 
&\sim -\frac{c}{12\pi} \left[
V_L
\begin{pmatrix}
\frac{1}{4}\ell_3 &\frac{1}{4}\ell_3 & \ell_1 \\ \frac{1}{4}\ell_3 & \frac{1}{4}\ell_3 &\ell_2 
\end{pmatrix}
+\frac{1}{2}\sum_{i=1}^3 \ell_i \psi_i
- i \left( \frac{\pi^2}{4}+\frac{\ell_3^2}{16}\right)
\right]
\end{align}
where $(\psi_1, \psi_2, \psi_3)$ are the dihedral angles on the edges of length $(\ell_1, \ell_2, \frac{1}{4}\ell_3)$, respectively. For $D>0$, the tetrahedron degenerates and the result is instead
\begin{align}\label{logSfinalneg}
\log \hatS_{P_1P_2}[P_3] \sim   - \frac{c}{6\pi} 
\left( \frac{\pi}{4}(\ell_1+\ell_2)  -  i( \tfrac{\pi^2}{8} + \tfrac{\ell_3^2}{32}) \right)
\end{align}

\subsection{Saddlepoint analysis of the $S$-matrix}\label{ss:kernelS}

The modular S-matrix has the integral representation  \cite{Teschner:2013tqy}
\begin{align}\label{shatIntegral}
\hatS_{P_1P_2}[P_3]
&= \frac{e^{i\pi h_3/2}\sqrt{ S_b(\frac{Q}{2}\pm 2iP_2 - iP_3)S_b(\frac{Q}{2} \pm 2i P_1 + i P_3) }}{ S_b(\frac{Q}{2}+iP_3) } \\
&\qquad  \times 
\int_{-\infty}^{\infty} dP e^{-4\pi i P P_1} S_b(\tfrac{Q}{4}+ i\tfrac{P_3}{2} \pm iP_2 \pm iP)\notag
\end{align}
where $S_b$ is the double sine function. This equation is \cite[eqn (3.88)]{Eberhardt:2023mrq} rescaled to Racah-Wigner normalization.
We use the shorthand $S_b(x\pm y) = S_b(x+y)S_b(x-y)$, and similarly $S_b(x\pm y \pm z)$ is a product of 4 terms.  The integral converges for real $P_{1,2,3}$.%

The semiclassical limit is taken by rescaling the integration variable as $P = \frac{\delta}{4\pi b}$. The semiclassical limit of the double sine function is
\begin{align}\label{dilogSemi}
\log S_b(b^{-1}z) = \frac{i}{4\pi b^2}( \dilog(e^{2\pi i z}) - \dilog(e^{-2\pi i z})) + \mathcal{O}(1) \ .
\end{align}
This formula assumes $\mbox{Re}\ z \in (0, 1)$, which is the case for all of the terms in \eqref{shatIntegral}. Therefore the integral in \eqref{shatIntegral} in the semiclassical limit is
\begin{align}
 \int_{-\infty}^{\infty}d\delta \exp\left[- \frac{1}{\pi b^2} Y(\delta, \ell_1,\ell_2,\ell_3)  \right]
\end{align}

where
\begin{align}
Y(\delta, \ell_1, \ell_2, \ell_3)
&= \frac{\delta}{8}( -2\pi +2i\ell_1+i\ell_3)
+ \frac{i}{2}\big[
\dilog(-ie^{(2\delta+2\ell_2+\ell_3)/4})
+\dilog(-ie^{(2\delta-2\ell_2+\ell_3)/4}) \notag \\
&\qquad
-\dilog(ie^{(2\delta-2\ell_2-\ell_3)/4})
- \dilog(ie^{(2\delta+2\ell_2-\ell_3)/4})
\big] \ .
\end{align}
This expression has been simplified using the  identity
\begin{align}\label{dilogInverse}
\dilog(z) + \dilog(\frac{1}{z}) = -\frac{\pi^2}{6}-\frac{1}{2}(\log(-z))^2
\end{align}
to move all factors of $e^{\delta/2}$ into the numerator.  The saddlepoint equation $\frac{\p}{\p \delta}Y = 0$ is a quadratic equation for $e^{\delta/2}$. The integral is dominated by the saddlepoint at
\begin{align}\label{deltazero}
\delta_0 &=2\log\big[
\frac{-2i \cosh \frac{\ell_2}{2} \sinh \frac{\ell_1}{2} 
+ \sqrt{ -D}
}{  2\cosh( \frac{\ell_1}{2}- \frac{\ell_3}{4})}
\big]
\end{align}
where 
\begin{align}\label{discrimS}
D &=  \cosh \ell_1 \cosh \ell_2 -1-\cosh \ell_1 - \cosh \ell_2  - 2 \cosh \tfrac{\ell_3}{2} \ . 
\end{align}
The factors outside the integral in \eqref{shatIntegral} have unit magnitude, and the overall phase is $e^{i\pi h_3/2}$.  Hence the modular $S$-matrix in the semiclassical limit is
\begin{align}\label{logSfinal1}
\log \hatS_{P_1P_2}[P_3] \sim   \frac{c}{6\pi} \left(  i( \tfrac{\pi^2}{8} + \tfrac{\ell_3^2}{32})
- \mbox{Re}\, Y(\delta_0+i0^+, \ell_1, \ell_2, \ell_3)
\right) \ .
\end{align}
This is symmetric under $\ell_1 \leftrightarrow \ell_2$, though not manifestly so. The branch cuts are the standard ones, $z<0$ for $\log z$ and $z \in (1,\infty)$ for $\dilog(z)$. The symbol $\sim$ has the precise meaning $\lim_{b\to 0}\frac{\rm LHS}{\rm RHS} = 1$, and we used $\frac{1}{b^2} \sim \frac{c}{6}$. 

The behavior of this function depends strongly on the sign of the discriminant, $D$.
First consider the case $D>0$. Then Im $\delta_0 = -i\pi $, and $\mbox{Re}\ Y$ can be simplified using the identity
\begin{align}
\mbox{Im\ }\dilog(x+i0^+)= \Theta(x-1) \pi \log (x) \qquad (x \in \mathbb{R})
\end{align}
where $\Theta$ is the step function. This leads to
\begin{align}\label{reyaa}
\mbox{Re}\, Y(\delta_0+i0^+,\ell_1,\ell_2,\ell_3)
&= \frac{\pi}{8}\big[
2\ell_1+\ell_3 - 2 \delta_r
+ (2\delta_r-2\ell_2-\ell_3)\Theta(2\delta_r-2\ell_2-\ell_3) \notag\\
&\qquad 
+(2\delta_r+2\ell_2-\ell_3)\Theta(2\delta_r+2\ell_2-\ell_3)
\big]
\end{align}
where $\delta_r = \mbox{Re}\, \delta_0$. 
Using $D>0$ it is straightforward to show that the argument of the first step function is always negative, while that of the second is always positive, so the right-hand side of \eqref{reyaa} is $ \tfrac{\pi}{4}(\ell_1+\ell_2) $. 
Thus the final result is
\begin{align}\label{logSfinalnegAppBB}
\log \hatS_{P_1P_2}[P_3] \sim   \frac{c}{6\pi} \left(  i( \tfrac{\pi^2}{8} + \tfrac{\ell_3^2}{32})
- \frac{\pi}{4}(\ell_1+\ell_2) 
\right)
\end{align}
when $D>0$.

Now we turn to $D<0$, and compare to the volume of a generalized hyperbolic tetrahedron. 
Consider a generalized tetrahedron where all the vertices are hyperideal and all the truncations are mild. As in section \ref{sss:volume}, we denote the volume as a function of edge lengths by $V_L\begin{psmallmatrix}L_1 & L_2 & L_3 \\ L_4 & L_5 &L_6 \end{psmallmatrix}$, where non-adjacent edges are in the same column, and the top row contains three edges meeting at a vertex. 
Let
\begin{align}\label{tetlengths}
\begin{pmatrix}L_1 & L_2 & L_3 \\ L_4 & L_5 &L_6 \end{pmatrix}
&= 
\begin{pmatrix}
\frac{1}{4}\ell_3 &\frac{1}{4}\ell_3 & \ell_1 \\ \frac{1}{4}\ell_3 & \frac{1}{4}\ell_3 &\ell_2 
\end{pmatrix} \ .
\end{align}
We will prove the following identity:
\begin{align}\label{logSVbb}
\log \left|  \hatS_{P_1P_2}[P_3]  \right|
&\sim -\frac{c}{12\pi} \left[
V_L
\begin{psmallmatrix}
\frac{1}{4}\ell_3 &\frac{1}{4}\ell_3 & \ell_1 \\ \frac{1}{4}\ell_3 & \frac{1}{4}\ell_3 &\ell_2 
\end{psmallmatrix}
+\frac{1}{2}\sum_{i=1}^6 L_i \theta_i 
\right]\\
&\sim -\frac{c}{12\pi} \left[
V_L
\begin{psmallmatrix}
\frac{1}{4}\ell_3 &\frac{1}{4}\ell_3 & \ell_1 \\ \frac{1}{4}\ell_3 & \frac{1}{4}\ell_3 &\ell_2 
\end{psmallmatrix}
+\frac{1}{2}(\ell_1 \psi_1 + \ell_2 \psi_2 + \ell_3\psi_3)
\right]\notag
\end{align}
Here $\theta_i$ is the dihedral angle on the edge of length $L_i$ and $\psi_i$ is the dihedral angle on the edge associated to the state $P_i$. Referring to \eqref{tetlengths}, these are related by the relabeling 
$\theta_3 = \psi_1$, $\theta_6 = \psi_2$, and $\theta_1=\theta_2=\theta_4=\theta_5=\psi_3$. In \eqref{logSVbb} the angles are viewed as functions of the lengths.
The vertex Gram matrix  for the tetrahedron \eqref{tetlengths} has determinant 
\begin{align}
\det \GL = (1+\cosh(\ell_1))(1+\cosh(\ell_2)) D(\ell_1,\ell_2,\ell_3) \ . 
\end{align}
Thus $D$ and $\det \GL $ are related by a strictly positive factor. Geometrically, the limit $D \to \infty$ corresponds to one or more of the hyperideal vertices becoming ideal. The limit $D \to 0$ corresponds to the dihedral angles approach $(\psi_1 = \pi, \psi_2 = \pi, \psi_3 = 0)$. In this limit, the tetrahedron degenerates, the volume vanishes, and we see that \eqref{logSVbb} reduces to \eqref{logSfinalnegAppBB}.

\textit{Derivation of \eqref{logSVbb}:} The volume of a mildly truncated tetrahedron was found in \cite{ushijimaMR2191251}, but instead of comparing to that result, it is easier to prove \eqref{logSVbb} directly. We have already shown that \eqref{logSVbb} holds in the degeneration limit $D \to 0$, so it is enough to show that both sides of the equation have the same derivatives with respect to $\ell_{i=1,2,3}$. The Schl\"afli formula \eqref{schlafliL} for a tetrahedron with six independent edges gives the derivative with respect to lengths. Specializing to the tetrahedron \eqref{tetlengths}, this implies
\begin{align}\label{covdsDelta}
\frac{\p}{\p\ell_i}\left( V_L
\begin{psmallmatrix}
\frac{1}{4}\ell_3 &\frac{1}{4}\ell_3 & \ell_1 \\ \frac{1}{4}\ell_3 & \frac{1}{4}\ell_3 &\ell_2 
\end{psmallmatrix}
 + \frac{1}{2}\sum_{j=1}^3 \ell_j \psi_j \right)
&= \frac{1}{2}\psi_i \ , \qquad i=1,2,3 \ . 
\end{align}
The angles are expressed in terms of edge lengths in \eqref{anglesFromLengths}. For the tetrahedron \eqref{tetlengths}, this equation gives the relations
\begin{align}
\cos\psi_1 &=- \frac{1+\cosh\ell_2-\cosh\ell_1\cosh\ell_2+\cosh\frac{\ell_3}{2}}{
\cosh\ell_1+\cosh\frac{\ell_3}{2}}\\
\cos\psi_3 &=- \frac{2\cosh \frac{\ell_1}{2} \cosh \frac{\ell_2}{2} \cosh \frac{\ell_3}{4}}{\sqrt{ (\cosh\ell_1 + \cosh \frac{\ell_3}{2})(\cosh\ell_2+\cosh\frac{\ell_3}{2})}} \ ,  \notag
\end{align}
and $\cos\psi_2$ is obtained by swapping $1 \leftrightarrow 2$ in first equation. Now we compare to the $\ell_1$ and $\ell_3$ derivatives of $\log|\hatS_{P_1P_2}[P_3]|$ using \eqref{logSfinal1}. A direct calculation (which is simplified by using the saddlepoint equation to drop some of the derivatives) shows 
\begin{align}
\frac{\p}{\p \ell_1} \mbox{Re}\, Y(\delta_0, \ell_1, \ell_2, \ell_3)
&= \frac{1}{4}\psi_1 \\
\frac{\p}{\p \ell_3} \mbox{Re}\, Y(\delta_0, \ell_1, \ell_2, \ell_3)
&= \frac{1}{4}\psi_3 \ , \notag
\end{align}
and similarly for $\ell_1 \leftrightarrow \ell_2$. 
Comparing to \eqref{covdsDelta}, we see that $\mbox{Re}\, Y = \frac{1}{2}V_L\begin{psmallmatrix}
\frac{1}{4}\ell_3 &\frac{1}{4}\ell_3 & \ell_1 \\ \frac{1}{4}\ell_3 & \frac{1}{4}\ell_3 &\ell_2 
\end{psmallmatrix}
 + \frac{1}{4}\sum_{i=1}^3\ell_i \psi_i
$ and \eqref{logSVbb} follows. As a check, this formula for the volume also agrees numerically with \cite{ushijimaMR2191251}. \hfill $\square$

\subsection{Saddlepoint analysis of the $6j$-symbol}\label{ss:kernelJ}

The Virasoro $6j$-symbol has the integral representation \cite{Teschner:2012em} 
{\small
\begin{align}\label{sixjTV}
&\begin{Bmatrix}
P_1 & P_2 & P_3 \\ P_4 & P_5 & P_6
\end{Bmatrix}
= 
V_{123} V_{156} V_{246} V_{345} \\
&\qquad \times
\int_{-\frac{iQ}{4}+\mathbb{R}}\!\!\!\!dP\, 
\frac{
S_b(iP-iP_{123})
S_b(iP-iP_{156})
S_b(iP-iP_{246})
S_b(iP-iP_{345}) }
{
S_b(\tfrac{Q}{2}+iP-iP_{1245})
S_b(\tfrac{Q}{2}+iP-iP_{1346})
S_b(\tfrac{Q}{2}+iP-iP_{2356})
S_b(\tfrac{Q}{2}+iP)
}\notag
\end{align}
with $P_{ijk} = P_i+P_j+P_k$, $P_{ijkl} = P_i+P_j+P_k+P_l$, and
\begin{align}\label{Vijk}
V_{ijk} &= \sqrt{ 
S_b(\tfrac{Q}{2}+iP_k \pm i(P_i+P_j))
S_b(\tfrac{Q}{2}-iP_k \pm i(P_i-P_j) )
} \ .
\end{align}
}%
The expression \eqref{sixjTV} has manifest tetrahedral symmetry: The combinations appearing in the prefactor and the numerator are the four vertices of the tetrahedron, and the first three terms in the denominator are the 4-cycles of its 1-skeleton.

In the semiclassical limit, \eqref{sixjTV} becomes
\begin{align}\label{semiclTV}
\begin{Bmatrix}
P_1 & P_2 & P_3 \\ P_4 & P_5 & P_6
\end{Bmatrix}
\ \sim \ 
e^{\frac{1}{\pi b^2} i \alpha(\ell_1,\dots,\ell_6)}
\int_{-i\pi+\mathbb{R}} ds \exp\left( -\, \frac{1}{\pi b^2} W(s; \ell_1,\dots,\ell_6)\right)
\end{align}
where $\alpha$ is a real phase and
\begin{align}\label{sixjW}
W
&= -i U(e^{-s/2}; e^{-\ell_1/2},\dots,e^{-\ell_6/2}  ) +\frac{\pi s}{2} + i\frac{\pi^2}{2}
 - \frac{i}{8} (\ell_1 \ell_4+\ell_2\ell_5+\ell_3\ell_6)
 + \frac{\pi}{4}{ \sum_{i=1}^6} \ell_i 
\end{align}
with
\begin{align}
U(z,a,b,c,d,e,f)
&= \frac{1}{2}\big[
\dilog(z)+\dilog(abdez) + \dilog(acdfz) + \dilog(bcefz) \\
&\qquad
-\dilog(-abcz)
-\dilog(-aefz)
-\dilog(-bdfz)
-\dilog(-cdez) 
 \big] \ . \notag
\end{align}
To obtain \eqref{semiclTV}, we changed the integration variable as $P = -i\frac{Q}{2} - \frac{s}{4\pi b}$, and used \eqref{dilogSemi} and \eqref{dilogInverse}.\footnote{This is the same function $U$ defined in \cite{MR2154824, ushijimaMR2191251}. There $U$ is evaluated with arguments determined by dihedral angles whereas here the arguments are related to lengths. A similar effect for non-truncated tetrahedron was found in \cite{MR2193233} and has been explained in terms of deep truncations in \cite{kolpakov2018combinatorial,belletti2020discrete}.}

The integral is evaluated by the saddlepoint method. The saddlepoint equation $\frac{\p}{\p s} W = 0$ is a quadratic equation for $e^{-s/2}$. The integral is dominated by the saddle at
\begin{align}\label{sixjsaddle}
s_0 &= -i\pi - \sum_{i=1}^6\ell_i \\
&-2 \log \left( 
\frac{- i 
(\sinh\frac{\ell_1}{2} \sinh \frac{\ell_4}{2}
+ \sinh\frac{\ell_2}{2}\sinh\frac{\ell_5}{2}
+\sinh \frac{\ell_3}{2}\sinh\frac{\ell_6}{2}
)
+\sqrt{-\det \GL}}{
1+ e^{\ell_{123}/2}+e^{\ell_{345}/2}+e^{\ell_{246}/2}
+ e^{\ell_{156}/2}
+e^{\ell_{1245}/2}
+e^{\ell_{1346}/2}
+e^{\ell_{2356}/2}
}
\right)\notag
\end{align}
where
\begin{align}\label{sixjD}
\GL = \begin{psmallmatrix}
1 & -\cosh \frac{\ell_1}{2} & -\cosh \frac{\ell_2}{2} & -\cosh \frac{\ell_6}{2} \\
-\cosh\frac{\ell_1}{2} & 1 & -\cosh \frac{\ell_3}{2} & -\cosh \frac{\ell_5}{2} \\
-\cosh \frac{\ell_2}{2} & -\cosh \frac{\ell_3}{2} & 1 & -\cosh \frac{\ell_4}{2} \\
-\cosh \frac{\ell_6}{2} & -\cosh\frac{\ell_5}{2} & -\cosh\frac{\ell_4}{2} & 1
\end{psmallmatrix} \ . 
\end{align}
The $6j$-symbol is real; the factors outside the integral give a pure phase that cancels the phase from the integral. Therefore the semiclassical $6j$-symbol is
\begin{align}\label{sixjAnsW}
\log
\begin{Bsmallmatrix}
P_1 & P_2 & P_3 \\ P_4 & P_5 & P_6
\end{Bsmallmatrix}
\sim
-\, \frac{c}{6\pi}
{\rm Re}\, W(s_0-i\varepsilon; \vec{\ell})
\end{align}
The $i\varepsilon$ prescription (with $\varepsilon \to  0^+$) chooses the correct branch cut of dilog when the saddlepoint $s_0$ moves onto the real axis.

We recognize the matrix in \eqref{sixjD} as the vertex Gram matrix $\GL$ of a generalized tetrahedron with four hyperideal vertices and edge lengths given by
\newcommand*\Bell{\ensuremath{\boldsymbol\ell}}
\begin{align}\label{sixjlengths}
L &:= \begin{pmatrix}
L_1 & L_2 & L_3 \\
L_4 & L_5 & L_6 
\end{pmatrix}
= \begin{pmatrix}
\frac{1}{2}\ell_4 & \frac{1}{2}\ell_5& \frac{1}{2}\ell_6 \\
\frac{1}{2}\ell_1 & \frac{1}{2}\ell_2 & \frac{1}{2}\ell_3
\end{pmatrix}
\end{align}
Note the row-swap in this relation, which corresponds to dualizing the tetrahedron. The faces of the geometric tetrahedron with edge lengths $L$ correspond to the vertices of the tetrahedron whose edges are labeled by the conformal weights. 

Assume that we are in the regime where the generalized tetrahedron exists, i.e.  $\det \GL < 0$, and denote its volume by $V_L(L)$. Then we will show
\begin{align}\label{Wvol}
\mbox{Re}\, W(s_0, \ell_1,\dots,\ell_6) &= V_L(L) + \frac{1}{2}\sum_{i=1}^6 L_i \theta_i\\
&= V_L(L) + \frac{1}{4}\sum_{i=1}^6\ell_i\psi_i \ .\notag
\end{align}
Here $\theta_i$ is the dihedral angle on the edge of length $L_i$, and $\psi_i$ is the dihedral angle on the edge of length $\frac{1}{2}\ell_i$; they are related by the relabeling $(\theta_1,\theta_2,\theta_3,\theta_4,\theta_5,\theta_6) = (\psi_4,\psi_5,\psi_6,\psi_1,\psi_2,\psi_3)$.  It follows from \eqref{Wvol} that the semiclassical $6j$-symbol is
\begin{align}\label{sixjVolAppBB}
\log
\begin{Bsmallmatrix}
P_1 & P_2 & P_3 \\ P_4 & P_5 & P_6
\end{Bsmallmatrix}
\sim
-\frac{c}{6\pi} \left( 
V_L\begin{psmallmatrix}
\frac{1}{2}\ell_4 & \frac{1}{2}\ell_5 & \frac{1}{2}\ell_6 \\
\frac{1}{2}\ell_1 & \frac{1}{2}\ell_2 & \frac{1}{2}\ell_3
\end{psmallmatrix}
 + \frac{1}{4} \sum_{i=1}^6 \ell_i \psi_i
\right)
\qquad (\det \GL < 0)
\end{align}
with $\ell_i = 4\pi b P_i$.  

In the region $\det \GL > 0$, where the generalized tetrahedron does not exist, \eqref{sixjAnsW} reduces to
\begin{align}\label{sixjresDegenBB}
\log
\begin{Bsmallmatrix}
P_1 & P_2 & P_3 \\ P_4 & P_5 & P_6
\end{Bsmallmatrix}
\sim
-\frac{c}{24}\max (\ell_1+\ell_4, \ell_2+\ell_5,\ell_3+\ell_6)
\qquad (\det \GL > 0) \ . 
\end{align}
The region $\det \GL > 0$ has three disconnected components. The boundaries of these components, where $\det \GL =0$, correspond to the three different ways for a hyperideal tetrahedron to degenerate: dihedral angles $\begin{psmallmatrix} \theta_1 & \theta_2 & \theta_3 \\ \theta_4 & \theta_5 & \theta_6\end{psmallmatrix}$ approaching $\begin{psmallmatrix} \pi & 0 & 0 \\ \pi & 0 & 0\end{psmallmatrix} $, $\begin{psmallmatrix} 0 & \pi & 0 \\ 0 & \pi & 0\end{psmallmatrix}$, or $\begin{psmallmatrix} 0 & 0 & \pi \\ 0 & 0 & \pi \end{psmallmatrix}$. These are the three cases responsible for the maximization in \eqref{sixjresDegenBB}. For example, $\log(6j) \sim - \frac{c}{24}(\ell_1+\ell_4)$ in the region of parameter space whose boundary is the degenerate tetrahedron with $\theta_1=\theta_4=\pi$. 

\bigskip

\noindent \textit{Derivation of the degenerate $6j$-symbol \eqref{sixjresDegen}:}

Assume $\det \GL> 0$. The saddlepoint \eqref{sixjsaddle} has Im $s_0 = 0$, and
\begin{align}
z_0 
&:= e^{-s_0/2}
=\frac{
2e^{\frac{1}{2}\ell_{123456}}\left( s_1s_4+s_2s_5+s_3s_6 - \sqrt{\det \GL} \right)}{
1+ e^{\ell_{123}/2}+e^{\ell_{345}/2}+e^{\ell_{246}/2}
+ e^{\ell_{156}/2}
+e^{\ell_{1245}/2}
+e^{\ell_{1346}/2}
+e^{\ell_{2356}/2}
}
\end{align}
with $s_i=\sinh \frac{\ell_i}{2}$. Therefore all of the dilogs in \eqref{sixjW} are evaluated at real arguments. 
Using the identity
\begin{align}
\mbox{Im} \, \dilog(x \pm i \epsilon) = \pm i \pi \Theta(x-1) \log (x)  \qquad (x \in \mathbb{R})
\end{align}
where $\Theta$ is the step function gives
\begin{align}\label{degenstp}
\mbox{Re\ }W
&= \frac{\pi s_0}{2}
+\frac{\pi}{4}\ell_{123456}
-\frac{\pi}{4} s_0\Theta(e^{-s_0/2}-1)
 - \frac{\pi}{4}(s_0+\ell_{1245})\Theta(e^{-(s_0+\ell_{1245})/2}-1)\notag\\
&\quad
- \frac{\pi}{4}(s_0+\ell_{1346})\Theta(e^{-(s_0+\ell_{1346})/2}-1)
- \frac{\pi}{4}(s_0+\ell_{2356})\Theta(e^{-(s_0+\ell_{2356})/2}-1)
\end{align}
The saddlepoint equation satisfied by $s_0$ is
\begin{align}
\frac{
(1+e^{(s_0+\ell_{123})/2})
(1+e^{(s_0+\ell_{345})/2})
(1+e^{(s_0+\ell_{246}/2})
(1+e^{(s_0+\ell_{156}/2})}{
(1-e^{s_0/2})(1-e^{(s_0+\ell_{1245})/2})(1-e^{(s_0+\ell_{1346})/2})(1-e^{(s_0+\ell_{2356})/2})}
= 1 \ . 
\end{align}
The numerator is strictly positive for real $s_0$, so each factor in the denominator is non-vanishing. These are the same quantities appearing in the step functions in \eqref{degenstp}. It follows that the region $\det \GL > 0$ has disconnected components where each step function takes a fixed value $\pm1 $. There are three components, in which the first step function and exactly one of the others have positive arguments. Simplifying \eqref{degenstp} within each component gives \eqref{sixjresDegen}.\hfill $\square$

\bigskip

\noindent \textit{Derivation of the volume relation \eqref{Wvol}:}
 
The strategy is to take a derivative with respect to edge lengths and check the Schl\"afli identity \eqref{schlafliL}. The argument is  very similar to that around \eqref{covdsDelta} except that now we have six independent edge lengths. The result is as expected, confirming \eqref{Wvol}. We omit the details.  \hfill $\square$
 
 \section{Predictions of the CFT ensemble}\label{app:cftstatistics}

In this appendix we review the predictions of the CFT ensemble dual to pure 3D gravity that were used in the main text, following \cite{Belin:2021ryy,Belin:2021ibv,Anous:2021caj,Chandra:2022bqq,Collier:2023fwi,Belin:2023efa,Collier:2024mgv,deBoer:2024mqg}.
The result for $\overline{C_{113}C_{223}}$ was first derived in 
\cite{Collier:2024mgv} using Virasoro TQFT, which leads to a calculation equivalent to the one in section \ref{ss:ccwinding} below (although we will phrase it a bit differently). It was also derived from torus 1-point functions in \cite{deBoer:2024mqg}, which is the calculation reviewed in section \ref{ss:cctorus}. 
The non-gaussianity in the four-point function was first discussed in \cite{Belin:2021ryy} and refined in \cite{Belin:2023efa,Collier:2024mgv}. Additional background on Virasoro conformal blocks and crossing kernels can be found in the review \cite{Eberhardt:2023mrq} and in an appendix of \cite{ctv}.

In order to match the thermodynamics of the BTZ black hole, the CFT ensemble is assumed to have Cardy entropy $\rho(h,\bh) \approx \rho_0(h)\rho_0(\bh)$ at leading order in the topological expansion \cite{Strominger:1997eq,Hartman:2014oaa}. Analogously, the  4-point correlation function of heavy operators in 3D gravity is given by the Virasoro identity block (squared) at leading order \cite{Hartman:2013mia},
\begin{align}\label{g4id}
G_4 &:= \langle {\cal O}_1(0) {\cal O}_1(z,\bz) {\cal O}_2(1) {\cal O}_2(\infty) \rangle
\approx
\left| \graphS{2}{2}{1}{1}{\id} \right|^2 \ ,
\end{align}
where we use the CFT convention $|f(h)|^2 := f(h) f^*(\bh)$.  This is used to derive the OPE coefficients in the dual channel \cite{Collier:2019weq} which are interpreted as the variance of the ensemble \cite{Chandra:2022bqq}. Assuming the operators labeled 1 and 2 are distinct, the conformal block expansion in the $t$-channel  is 
\begin{align}
G_4 &= \sum_{i \neq \id} | c_{12i} |^2 \left| \graphT{2}{2}{1}{1}{P} \right|^2 \ . 
\end{align}
This can be compared to the fusion transformation of \eqref{g4id}, 
\begin{align}
G_4 \approx \left| \int dP \rho_0(P) C_0(P_1,P_2,P)\graphT{2}{2}{1}{1}{i} \right|^2 \ ,
\end{align}
where $C_0$ is a known special function that is related to the structure constants of Liouville CFT \cite{Ponsot:2000mt,Collier:2019weq}.
These agree, to this approximation, provided we set 
\begin{align}
\overline{ | c_{123}|^2} \approx | C_0(P_1,P_2, P_3) |^2 \ . 
\end{align}
We have replaced the sum by an integral using $\sum_{i\neq \id} \to \int dP d\bar{P} \rho_0(P) \rho_0(
\bar{P})$. 
After normalizing by $|C_0|^2$ as in \eqref{bigC}, the leading term is
\begin{align}
\overline{| C_{123}|^2} \approx 1 \ . 
\end{align}
This is only the leading term in a topological expansion (see the discussion in \cite[$\S$2.1 and $\S$6.5]{Chandra:2022bqq}). Corrections in the bulk are calculated by higher topologies, and corrections on the boundary are calculated by iterating the conformal bootstrap. In the remainder of this appendix we review the CFT derivation of the first correction to $\overline{CC}$ and the leading non-Gaussianity in the four-point statistics $\overline{CCCC}$.

\newcommand{\windingBlock}[5]{
\cp{
\begin{tikzpicture}[scale=0.4]
\centerarc[](-1,0)(-90:45:1.6);
\centerarc[](-1,0)(55:90:1.6);
\centerarc[](1,0)(90:225:1.6);
\centerarc[](1,0)(235:270:1.6);
\draw (-0.6,0) -- (0.6,0);
\node at (-1.5,1.5) {\footnotesize \ensuremath{#1}};
\node at (-1.5,-1.5) {\footnotesize \ensuremath{#2}};
\node at (1.5,1.5) {\footnotesize \ensuremath{#3}};
\node at (1.5,-1.5) {\footnotesize \ensuremath{#4}};
\node at (0,.4) {\footnotesize \ensuremath{#5}};
\end{tikzpicture}
}}

\newcommand{\windingBlockB}[5]{
\cp{
\begin{tikzpicture}[scale=0.4]
\centerarc[](-1,0)(-90:-55:1.6);
\centerarc[](-1,0)(-45:90:1.6);
\centerarc[](1,0)(90:125:1.6);
\centerarc[](1,0)(135:270:1.6);
\draw (-0.6,0) -- (0.6,0);
\node at (-1.5,1.5) {\footnotesize \ensuremath{#1}};
\node at (-1.5,-1.5) {\footnotesize \ensuremath{#2}};
\node at (1.5,1.5) {\footnotesize \ensuremath{#3}};
\node at (1.5,-1.5) {\footnotesize \ensuremath{#4}};
\node at (0,.4) {\footnotesize \ensuremath{#5}};
\end{tikzpicture}
}}

\subsection{Correction to $\overline{CC}$ from winding channels}\label{ss:ccwinding}
Virasoro conformal blocks are not single-valued. This means that the four-point function has not just the usual $s,t,$ and $u$ channel expansions, but an infinite number of convergent OPE channels. (See e.g.~\cite{Asplund:2014coa,Anous:2016kss,Maloney:2016kee,Anous:2017tza} for discussion of how this applies to large-$c$ CFT and holography.) Denote the chiral Virasoro four-point block obtained by winding ${\cal O}_2$ around ${\cal O}_3$ and back to its starting point by
\begin{align}
\windingBlock{3}{4}{2}{1}{P} \quad .
\end{align}
In terms of the function of cross-ratios ${\cal F}(z) = \graphS{3}{4}{2}{1}{P}$, the winding block is obtained by sending $z$ around  1, through the branch cut along $z \in (1,\infty)$: $(1-z) \to (1-z)e^{-2\pi i}$. There is another similar block defined by the opposite winding $(1-z) \to (1-z) e^{2\pi i}$. In the identity block approximation, the 4-point function of heavy operators is a sum of the identity block over winding channels \cite{Hartman:2013mia,Asplund:2014coa}
\begin{align}
G_4 \approx \left| \graphS{2}{2}{1}{1}{\id} \right|^2  + 
\left| \windingBlock{2}{2}{1}{1}{\id} \right|^2 + 
\left| \windingBlockB{2}{2}{1}{1}{\id} \right|^2 + \cdots
\end{align}
The identity block in a winding channel must appear as a sum of heavy operators in the original $s$-channel. That is,
\begin{align}\label{windingv1}
\windingBlock{2}{2}{1}{1}{\id} = \int_0^\infty dR \rho_0(R) f(P_1,P_2,R) \graphS{2}{2}{1}{1}{R}
\end{align}
for some function $f$. 
To find $f$, we start from the standard identity block and its fusion transformation:
\begin{align}
\graphS{2}{2}{1}{1}{\id} = \int_0^\infty dP \rho_0(P) C_0(12P) \graphT{2}{2}{1}{1}{P} \ . 
\end{align}
Now let us move $\O_2$ around $\O_3$ on both sides of this equation. On the left this gives a winding block while on the right it is a sequence of two braiding moves, so it produces a phase:
\begin{align}
\windingBlock{2}{2}{1}{1}{\id} &= \int_0^\infty dP \rho_0(P) C_0(12P) \graphT{2}{2}{1}{1}{P} e^{2\pi i(h_1+h_2 - h_P)} \\
&= \int_0^\infty dPdR \rho_0(P)C_0(12P) e^{2\pi i(-h_1-h_2+h_P)} \fusionF{PR}{1&1\\2&2} \graphS{2}{2}{1}{1}{R} \ . 
\label{windingv2}
\end{align}
On the second line we performed another fusion transformation to bring it back to the $s$-channel.  Comparing \eqref{windingv2} to \eqref{windingv1} gives
\begin{align}
\rho_0(R) f(P_1,P_2,R) &= \int_0^\infty dP e^{2\pi i(-h_1-h_2+h_P)} 
\fusionF{\id P}{2&1\\2&1}
\fusionF{PR}{1&1\\2&2} \ . 
\end{align}
Using \cite[(2.65)]{Eberhardt:2023mrq} we can write this in terms of the modular S-matrix as
\begin{align}
f(P_1,P_2,R) = \frac{1}{\rho_0(P_2)} C_0(11R)e^{-i\pi h_R}\modularS_{P_1P_2}[R]
= \frac{1}{\rho_0(P_2)} C_0(11R) \modularS^*_{P_1P_2}[R]
\end{align}
This is the contribution of the winding identity block to the heavy OPE statistics, and there is another conjugate contribution from the opposite winding:
\begin{align}
\overline{c_{11R}c_{22R}^*} \supset f(P_1,P_2,R) f^*(\bP_1,\bP_2,\bar{R})
 + f^*(P_1,P_2,R) f(\bP_1, \bP_2, \bar{R}) \ . 
\end{align}
Normalizing the OPE coefficients by $|C_0|^2$ as in \eqref{bigC} the result is
\begin{align}\label{tpresFour1}
\overline{C_{113}C^*_{223}}
\supset \left| \hatS_{P_1P_2}[P_3] \right|^2 + \left| \hatS^*_{P_1P_2}[P_3] \right|^2 \ .
\end{align}
This can also be expressed as
\begin{align}\label{tpresFour2}
\overline{C_{113}C^*_{223}}
\supset(1+(-1)^{\ell_3})  \left| \hatS_{P_1P_2}[P_3] \right|^2  \ .
\end{align}

\subsection{Correction to the $\overline{CC}$ from torus 1-point functions}\label{ss:cctorus}
We will now rederive the same contribution \eqref{tpresFour2} from torus 1-point functions \cite{deBoer:2024mqg}.
Consider the square of the torus 1-point function,
\begin{align}\label{torusexp1}
\langle \O_1 \rangle_{g=1} \langle \O_1 \rangle_{g=1}^*
&= 
\sum_{i,j \neq \id}
c_{ii1}c_{jj1}^* \left| \torusBlock{i}{1}\torusBlock{j}{1}\right|^2
\end{align}
The leading-order Gaussian statistics has two contractions because of the repeated indices:
\begin{align}
\overline{c_{ii1}c^*_{jj1}} &\approx 
C_0(h_i,h_i,h_1)
C_0(\bh_i, \bh_i,\bh_1)\delta_{ij} (1 + (-1)^{\ell_1}) \ ,
\end{align}
with $\ell_1 = h_1 - \bh_1$ the helicity.
Plugging this into \eqref{torusexp1} gives the average
\begin{align}\label{torusleading}
\overline{\langle \O_1 \rangle_{g=1} \langle \O_1 \rangle_{g=1}^*}
&\approx
(1+(-1)^{\ell_1})
\left| \int dP \rho_0(P) C_0(1PP) \torusBlock{P}{1} \torusBlock{P}{1}  \right|^2
\end{align}
In \eqref{torusexp1} we expanded both one-point functions in the same channel. Expanding one in the S-dual channel we have instead
\begin{align}\label{toruse2}
\langle \O_1 \rangle_{g=1} \langle \O_1 \rangle_{g=1}^*
&= 
\sum_{i,j \neq \id}
c_{ii1}c_{jj1}^* \left| \torusBlockS{i}{1}\torusBlock{j}{1}\right|^2
\end{align}
For this to reproduce \eqref{torusleading}, we must add a subleading correction to the ensemble of the form 
\begin{align}
\overline{c_{113}c_{223}^*} \supset  \alpha(P_1, P_2, P_3)\alpha(\bP_1, \bP_2, \bP_3)(1 + (-1)^{\ell_3})
\end{align}
To determine the function $\alpha$, plug this into \eqref{toruse2} 
\begin{align}\label{tcomp1}
\overline{\langle \O_1 \rangle_{g=1} \langle \O_1 \rangle_{g=1}^*}
&\approx 
(1+(-1)^{\ell_1}) \left|
\int dP dR \rho_0(P) \rho_0(R)
\alpha(R,P,P_1) \torusBlockS{R}{1} \torusBlock{P}{1} \right|^2
\end{align}
Meanwhile \eqref{torusleading} after an S-transformation becomes
\begin{align}\label{tcomp2}
\overline{\langle \O_1 \rangle_{g=1} \langle \O_1 \rangle_{g=1}^*}
\approx
(1+(-1)^{\ell_1}) \left|
\int dP dR \, \rho_0(P) C_0(1PP) \modularS_{PR}[P_1] \torusBlockS{R}{1}\torusBlock{P}{1}\right|^2
\end{align}
Comparing \eqref{tcomp1} to \eqref{tcomp2} we find
\begin{align}
\alpha(P_1, P_2, P_3) &= \frac{1}{\rho_0(P_2)} C_0(P_1,P_1,P_3)\modularS_{P_1P_2}[P_3]  \ . 
\end{align}
Therefore, in terms of normalized OPE coefficients, 
\begin{align}\label{opeshat}
\overline{C_{113}C^*_{223} }
&\supset (1+(-1)^{\ell_3}) \left| \hatS_{P_1P_2}[P_3] \right|^2
\end{align}
in agreement with \eqref{tpresFour2}.

%

%

%

\subsection{Tetrahedral non-Gaussianity}
The non-gaussian tetrahedral contribution to $\overline{C_{ijk}C_{lmn}C_{pqr}C_{stu}}$ was calculated in \cite{Belin:2021ryy,Belin:2023efa,Collier:2024mgv}. One way to derive it is to look at the squared four-point function,
\begin{align}
G_4 G_4^* &:=
\langle \O_1 \O_2 \O_3 \O_4 \rangle \langle \O_1 \O_2 \O_3 \O_4 \rangle^* \ . 
\end{align}
Expanding both four-point functions in the $s$-channel and using the Gaussian contraction gives \cite{Chandra:2022bqq}
\begin{align}\label{fpgaus}
\overline{G_4 G_4^*} &=
\sum_{i,j}
\overline{c_{12i}c_{34i} c^*_{12j}c^*_{34j}}
\left| \graphS{3}{4}{2}{1}{i} \graphS{3}{4}{2}{1}{j} \right|^2\\
&\approx \left| \int dP \rho_0(P) C_0(12P)C_0(34P)
\graphS{3}{4}{2}{1}{P} \graphS{3}{4}{2}{1}{P} \right|^2
\end{align}
A fusion transformation on the first block gives
\begin{align}\label{fourexp33}
\overline{G_4 G_4^*} &\approx
\left| 
\int dPdR \rho_0(P) C_0(12P)C_0(34P)\fusionF{PR}{3&2\\4&1}
\graphT{3}{4}{2}{1}{R}
\graphS{3}{4}{2}{1}{P}
\right|^2\notag \\
&= \big|
\int dP dR \rho_0(P)\rho_0(R)\left[ C_0(12P)C_0(34P)C_0(14R)C_0(23R)\right]^{1/2}\\
& \qquad \times
\begin{Bmatrix}P_1 & P_2 & P\\ P_3 & P_4 & R \end{Bmatrix}
\graphT{3}{4}{2}{1}{R}
\graphS{3}{4}{2}{1}{P}
\qquad \big|^2 \notag
\end{align}
On the other hand, expanding one correlator in the $t$-channel and the other in the $s$-channel gives
\begin{align}\label{fourexp2}
\overline{G_4 G_4^*}
= \sum_{i,j} \overline{ c_{14i}c_{23i} c^*_{12j}c^*_{34j} }
\left| 
\graphT{3}{4}{2}{1}{i}
\graphS{3}{4}{2}{1}{j}
\right|^2 \ . 
\end{align}
In order to reproduce  \eqref{fourexp33} we must add to the ensemble a non-gaussian contribution
\begin{align}
\overline{ C_{123} C_{156}C_{246}C_{345} }
\supset \left| \begin{Bmatrix} P_1 & P_2 & P_3 \\ P_4 & P_5 & P_6 \end{Bmatrix} \right|^2 \ . 
\end{align}

\section{Tutorial on triangulations with {\tt Orb}}\label{app:orb}
{\tt Orb} is a software program written by Heard \cite{heard2005computation}, which extends the program {\tt SnapPea} by Weeks. {\tt Orb} can be used to triangulate trivalent graphs in $S^3$, to study various aspects of the triangulation, and to find hyperbolic metrics by solving the gluing equations numerically, calculate volumes, etc. It is designed to find hyperbolic metrics with orbifold singularities on the edges of the graph, so this corresponds to CFT primary states with weights below the black hole threshold in 3D gravity. Thus the output of {\tt Orb} is the triangulation of $(M_E, \Gamma(\vecP))$. In order to study black hole states, one must go through the additional procedure in section \ref{ss:gentri} to convert this output into the generalized triangulation of $M$. 

{\tt Orb} is open source software, available at \url{https://github.com/DamianHeard/orb}. Follow the installation instructions to begin. It is nontrivial to install it directly on a modern Macbook, so a an easier option (on a Mac) is to install it in a virtual Linux environment with the following steps:  Install VMWare Fusion; create a virtual machine running Linux (Fedora 64-bit ARM); download {\tt Orb} to the virtual machine; install some necessary packages on the virtual machine ({\tt sudo dnf install gcc}; {\tt sudo dnf install gcc-c++}; {\tt sudo dnf install redhat-rpm-config}); then finally compile {\tt Orb} following the installation instructions ({\tt qmake -o Makefile Orb.pro}; {\tt make}).

 In this brief tutorial, we will use {\tt Orb} to triangulate the following knotted theta graph,
 \begin{align}\label{regtref}
\Gamma \qquad =\qquad 
\cp{\includegraphics[width=1in]{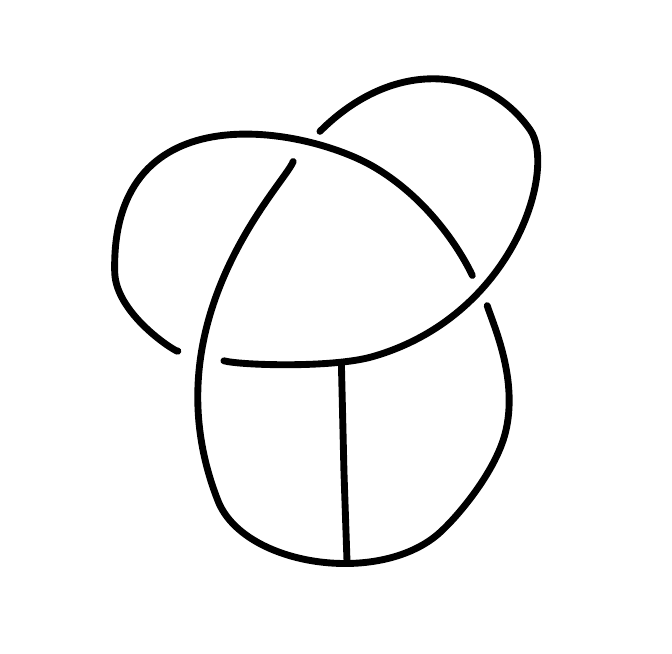}}
\end{align}
embedded in $S^3$, and to find the hyperbolic volume with orbifold singularities on the edges of the graph.

Run {\tt Orb}. Click the `Draw' button to create a new graph, use the point-and-click interface to draw the graph, and when it is complete, click the magnifying glass to analyze it. Under `Labels', assign an orbifold order to each edge; let's assign `7.00' to all three edges, which corresponds to defects with total angles $2\pi/7$. 

Click the arrow next to the {\tt tri} button and turn on `Verbose' mode, and then click {\tt tri} to view the triangulation. For the graph \eqref{regtref}, the output is
\begin{verbatim}
tri :
SolutionType geometric_solution
vertices_known

  1  3 7.000  1  2  1vu  2wu  2uv  1xv
  2  0 1.000  2  2  1wu  2ux  1wv  2vx  2xw  1xw
  3  1 7.000  2  1  1xu
  4  2 7.000  2  1  2wv
\end{verbatim}
Each row describes one edge of the triangulation, in the format
\begin{verbatim}
  n e L  v1  v2   link
\end{verbatim}
where {\tt n} is the edge number; {\tt e} is the index of the edge if it belongs to $\Gamma$  (as it appears in the `Labels' column) and otherwise 0; {\tt v1} and {\tt v2} are the vertices at the endpoints of the edge; and {\tt link} is the link of this edge. The `link' of an edge $E$ is the closed cycle of edges that go around $E$. For example, the link {\tt 1vu  2wu  2uv  1xv} is a cycle formed by the edges {\tt vu} on tetrahedron 1, {\tt wu} on tetrahedron 2, etc. 
The vertices of each tetrahedron are labeled as follows:
\begin{align}
\cp{
\footnotesize
\begin{tikzpicture}[scale=0.8]
\centerarc[ultra thick](0,0)(-30:90:1);
\centerarc[ultra thick](0,0)(90:210:1);
\centerarc[ultra thick](0,0)(210:335:1);
\draw[ultra thick] (0,0) -- (0,1);
\draw[ultra thick] (0,0) -- ({cos(30)},{-sin(30)});
\draw[ultra thick] (0,0) -- ({-cos(30)},{-sin(30)});
\node at (0,1.2) {$u$};
\node at (-1.1,-.6) {$v$};
\node at (1,-.6) {$w$};
\node at (0,-.25) {$x$};
\end{tikzpicture}
}
\end{align}
The edge $E$ consists of all the tetrahedral edges that are directly `across' from the edges in the link. Thus edge number 1, whose link is {\tt 1vu  2wu  2uv  1xv}, consists of the edges {\tt 1wx} $\equiv$ {\tt 2xv} $\equiv$ {\tt 2xw} $\equiv$ {\tt 1wu}, which are all identified. The triangulation that we read off from the {\tt Orb} output above has two tetrahedra, four distinct edges, and two distinct vertices. The two tetrahedra are shown in the following figure, 
\begin{align}
\cp{
\begin{tikzpicture}[scale=0.8]
\centerarc[thick,blue](0,0)(-30:90:1);
\centerarc[thick,darkgreen](0,0)(90:210:1);
\centerarc[thick,orange](0,0)(210:335:1);
\draw[thick,darkgreen] (0,0) -- (0,1);
\draw[thick,blue] (0,0) -- ({cos(30)},{-sin(30)});
\draw[thick,darkgreen] (0,0) -- ({-cos(30)},{-sin(30)});
\end{tikzpicture}
}
\cp{
\begin{tikzpicture}[scale=0.8]
\centerarc[thick,darkgreen](0,0)(-30:90:1);
\centerarc[thick,darkgreen](0,0)(90:210:1);
\centerarc[thick,darkgreen](0,0)(210:335:1);
\draw[thick,red] (0,0) -- (0,1);
\draw[thick,blue] (0,0) -- ({cos(30)},{-sin(30)});
\draw[thick,blue] (0,0) -- ({-cos(30)},{-sin(30)});
\end{tikzpicture}
}
\end{align}
where each color is a distinct edge, and the faces are glued in the only way compatible with the edge colorings. The green edge is internal, and the other three belong to $\Gamma$. More information on the gluing pattern can be found by clicking the {\tt tet} button.

To calculate the volume, click `Update'. This example has $ V \approx 4.726$.

\addcontentsline{toc}{section}{References}
\bibliographystyle{utphys}
{\small
\bibliography{triangulations}
}

\end{document}